\documentclass[krantz1,ChapterTOCs]{krantz} %See documentation for other class options
\usepackage{fixltx2e,fix-cm}
\usepackage{amssymb}
\usepackage{amsmath}
\usepackage{graphicx}
\usepackage{subfigure}
\usepackage{makeidx}
\usepackage{multicol}
\usepackage{float}
\usepackage{geometry}
\usepackage{listings}
\DeclareMathOperator*{\argmin}{\arg\!\min}

\newcommand{\y}{\widetilde{y}}

\newcommand{\ES}[1]{\textnormal{\tiny {#1}}}

\begin{document}

\frontmatter

\title{Handbook of Multiple Comparisons\\[8mm] 
{\Large Chapter: Identifying important predictors in large data bases - multiple testing and model
selection}} %This is a placeholder titlepage, it will not be final.

\author{Malgorzata Bogdan, Florian Frommlet}
\maketitle

%\include{frontmatter/dedication}
%\cleardoublepage
%\setcounter{page}{7} %previous pages will be reserved for frontmatter to be added in later.
%\tableofcontents
%\include{frontmatter/foreword}
%\include{frontmatter/preface}
\listoffigures
\listoftables

\mainmatter

%\part{General Methodology}
%\include{chapters/chapter1/ch1}

%\chapterauthor{Author Name}{Malgorzata Bogdan, Florian Frommlet}
\chapter{Identifying important predictors in large data bases - multiple testing and model selection}

This chapter will consider a variety of model selection strategies in a high-dimensional setting, where the number of potential predictors $p$ is large compared to the number of available observations $n$. A typical example of this situation are genome-wide association studies (GWAS). These may include several hundred thousand genetic variants (SNPs) which are used as genetic markers for DNA regions. GWAS are performed to find those variants which are related to some trait. This could be a binary trait like disease risk or a quantitative trait like height. GWAS are most often analysed by performing statistical tests for each individual marker combined with some correction for multiple testing. Often a Bonferroni corrected significance level  $5 \times 10^{-8}$ is recommended. This type of analysis is still prevailing although from a statistical perspective it has many severe drawbacks \cite{GWAS, mono, renaux2020hierarchical}. 

A major assumption underlying the rationale of performing GWAS is the common disease/common variant assumption. Accordingly the risk of common diseases should depend on a relatively large number of fairly common genetic variants. In statistical terms this corresponds to models which include moderate numbers of genetic markers as regressors. It has been shown  that under this assumption testing each marker individually results in a severe loss of power to detect important SNPs \cite{GWAS}. Furthermore the order of p-values from the individual marker tests may no longer reflect the actual importance of genetic variants and consequently also the chance of false positive findings is increased. Using model selection approaches to detect important genetic variants can help to overcome these shortcomings. For practical applications with real GWAS data see for example \cite{DBF14, hofer2017bayesian}

For the ease of presentation the focus will be on the linear model 
\begin{equation} \label{LinearModel}
y \!=\! X \beta + \epsilon \; ,
\end{equation}
where $y \in {\mathbb R}^n$, $\beta \in {\mathbb R}^p$,  $X \in {\mathbb R}^{n \times p}$ and the error terms are independent Gaussian random variables, $\epsilon \sim N_n(0, \sigma^2 I)$. The basic ideas easily extend to more general regression settings, like generalized linear models or generalized linear mixed models. Many high-dimensional model selection strategies make use of penalized likelihood methods, which can be written for example in the following form   
\begin{equation} \label{PenLik}
- 2 \log(\mathcal{L}(\beta)) + \mbox{Pen}(\beta) \; .
\end{equation}
Here $\mathcal{L}(\beta)$ denotes the likelihood function. In case of the linear model (\ref{LinearModel}) with known $\sigma$ the first term in (\ref{PenLik}) is up to a constant  $\|y - X\beta\|^2 / \sigma^2$, that is the residual sum of squares divided by the variance term. There exists a wide range of penalty functions $\mbox{Pen}(\beta)$  for high-dimensional model selection. This article will focus on  $L_0$ penalties as well as certain weighted $L_1$ penalties. 

In general model selection might serve two different purposes, identification of the actual data generating model or finding a model which is good for prediction. Depending on the application in mind the former or the latter goal might be more important and the most suitable selection strategies might be different. For example, in the context of genetic association studies one can make use of variable selection methods to identify causal mutations. Correct model identification then corresponds to correctly identifying causal mutations without detecting too many false positives \cite{mono}. In statistical terms one needs a consistent variable selection procedure to achieve this goal.

Section \ref{Sec:orthogonal} will set the stage by considering the simple setting of an orthogonal design matrix $X$ and known error variance $\sigma^2$. In that case estimates of the regression coefficients $\beta_j$ do not depend on the other components of the vector $\beta$ and  model selection becomes equivalent to multiple testing. Simple results for the classical selection criteria AIC and BIC will illustrate that these are not suitable for model selection when the number of potential regressors $p$ is getting large compared with $n$. Instead some $L_0$ penalties which are modifications of AIC and BIC will be introduced  which are designed to control the number of false discoveries. This means that for predictors which are of no relevance type I error control strategies will be applied which are known from multiple testing. Specifically selection procedures are introduced which control either the family wise error rate (FWER) or the false discovery rate (FDR). The simulation results from Section \ref{Sec:SimL0} then show that these criteria also perform really well  when regressors are stochastically independent (but not orthogonal) or strongly correlated.

Using again the framework of orthogonal designs Section \ref{Sec:Optimality} will first discuss some optimality properties of the introduced penalties in terms of model identification, followed by optimality results in terms of prediction. These theoretical results will indicate that it is often preferable to use methods which control the FDR since these can adapt to the typically unknown level of sparsity. 
While the $L_0$ penalties have superb theoretical properties, their practical application leads to a most challenging optimization problem which is known to be NP-hard. For that reason there has been a strong interest in $L_1$-penalties like the LASSO \cite{LASSO} which can be tackled via convex optimization. However, with a fixed penalty weight for all regressors entering the model LASSO can be compared with a fixed threshold rule in multiple testing. In view of the theoretical results from Section \ref{Sec:Optimality} and the good performance of the Benjamini Hochberg rule it would be desirable to have more flexible choices of penalties.
Section \ref{Sec:L1} will introduce SLOPE, where a specific choice of weighted $L_1$-penalties provides another FDR-controlling model selection procedure.   Afterwards Section \ref{Sec:Advanced} briefly discusses some advanced variable selection procedures controlling FDR, first a Bayesian version of SLOPE and then a procedure which uses the idea of knock offs. In Section \ref{Sec:RealData} the different selection methods are applied to two real data sets.  R scripts are available on-line  which provide the code to perform these analyses.

\section{Model selection under an orthogonal design}\label{Sec:orthogonal}

Consider the situation where the columns of the design matrix $X$ are orthogonal and scaled such that $X^T X = n I_p$. Apart from models using wavelets this will rarely be the case in practice. However, this simple setting allows to see the parallels between multiple testing and model selection. It also provides  the basic intuition for the behavior of $L_0$ penalties in high dimensions. The most important consequence of the orthogonal design is that the estimates of the different components of the coefficient vector $\beta$ become independent of each other. Denoting the $p$ columns of $X$  by $X_j$ one simply obtains $\hat \beta_j = \frac1n X_j^T y$ and it is fairly easy to see that in case of known $\sigma$ these estimates are statistically independent and normally distributed, $\hat \beta_j \sim N(\beta_j, \sigma^2/n)$. One can test each coefficient using a z-Test with the statistic $Z_j := \sqrt n \hat \beta_j / \sigma, j \in \{1,\dots,p\}$. Model selection thus reduces to a multiple testing problem.

Alternatively one can study the properties of model selection based on information criteria.  To this end a particular model is  characterized by the index set $M$ corresponding to non-zero coefficients of $\beta$. The notation $k_M = \|\beta\|_0$ is used for the corresponding model size.  Table \ref{Tab:L0Penalties} provides an overview over some important $L_0$ penalties discussed in more detail below.  Historically the first selection procedures of this type were developed in the 70ies, the Akaike Information Criterion AIC \cite{Aka} and the Bayesian Information Criterion (BIC) by Schwarz \cite{Schw}. AIC uses as penalty $\mbox{Pen}(\beta) = 2 k_M$, whereas BIC has the penalty $\mbox{Pen}(\beta) = k_M  \log n$ which becomes more stringent than the AIC penalty for $n > 7$.  A vast literature exists about their statistical properties (see for example \cite{BA02}). In particular AIC has some optimality properties in terms of prediction and BIC is consistent as long as the number of potential regressors is relatively moderate. However, it will soon become clear that both criteria are not really useful in a high-dimensional setting.

\begin{table}[t]
	%\noautomaticrules
	\tabletitle[Different $L_0$-penalties for high-dimensional model selection]{Different $L_0$-penalties and their corresponding properties for high-dimensional model selection. AIC and BIC are classical criteria but not suitable for high dimensions. The four modifications of AIC and BIC, respectively, are the main focus in this presentation. The other mentioned criteria are related, where this list  is by no means comprehensive.}
	\label{Tab:L0Penalties}
	\begin{tabular}{r|ll}
		\tch{Name}    &\tch{Pen$(k_M;n,p)$} &\tch{Properties} \\\hline 
		AIC \cite{Aka} & $2\ k_M   $ & Not suitable for $p > n$\\
		BIC \cite{Schw} & $\log n\ k_M $ & Not suitable for $p > n$\\[1mm] \hline 
		mBIC \cite{mBIC} & BIC + $ 2 \log (p / 4)\ k_M$ & Controls FWER  at level $\alpha < n^{-1/2}$\\
		mAIC \cite{bigstep} & AIC + $ 2 \log (2p)\ k_M$ & Controls FWER  at level $\alpha < 0.05$\\
		mBIC2 \cite{mBIC2} & mBIC $ - 2 \log k_M!$ & Controls FDR at level $\alpha < n^{-1/2}$\\
		mAIC2 \cite{bigstep} & mAIC $ - 2 \log k_M!$ & Controls FDR  at level $\alpha < 0.05$\\[1mm] \hline \vspace{1mm}
		EBIC \cite{CC08} & BIC $ + 2 \log \binom{p}{k_M}^{1 - \kappa}$ & Similar to mBIC2 for $\kappa \approx 1$\\
		RIC \cite{RIC} & $2 \log p\ k_M $ & Minimal inflation of predictive risk\\
		& & like mAIC with a different constant\\
		%\emph{ABDJ}
		\cite{ABDJ} &  $  2 k_M \log (p/k_M)$ & Minimax optimality\\
		& & similar to mAIC2\\
		%\emph{BM}
		\cite{BM2001} &  $  c k_M \log (p/k_M),  c > 2$ &  Bounds on quadratic risk\\
	\end{tabular}
\end{table}

It is well known that performing model selection using AIC in our simple setting is equivalent to performing z-tests. This can easily be seen by considering the fact that under orthogonality it holds that 
$$
- 2 \log(\mathcal{L}(\hat\beta)) = const + \|y -  \sum\limits_{j=1}^p  \hat\beta_j X_j\|^2 / \sigma^2
$$ 
where the maximum likelihood estimates and the least squares estimates of the coefficients coincide. It follows that adding regressor $X_j$ to the model reduces the log-likelihood term of AIC by  $n  \hat\beta_j ^2 /\sigma^2$ and increases the penalty  by 2, no matter which other regressors have already entered the model. So adding $X_j$ decreases AIC if and only if $|Z_j| > \sqrt 2$, which is equivalent to performing the z-test  considered previously. Hence model selection with AIC under an orthogonal design is equivalent to performing a z-test for each coefficient at the significance level $\alpha = 1 - F_{\chi^2}(2) \approx 0.157$.    

Similar  considerations hold for BIC, but here the penalty depends on the sample size $n$.  For $n = 8$ selection with BIC corresponds to performing a z-test for each coefficient at a 15\% significance level. Due to the $\log n$ penalty the significance level $\alpha_n$ decreases with increasing sample size $n$. More specifically it holds that  $\alpha_n = o(n^{-1/2})$. This is essential for the consistency property of BIC. Mathematically this follows immediately from the well known tail bounds of the normal distribution
\begin{equation} \label{Eq:NormalTail}
\frac{2 \phi(c)}{c} (1 - c^{-2}) \leq P(|Z_j| > c) \leq  \frac{2 \phi(c)}{c} \; .
\end{equation}
Model selection with BIC under orthogonality corresponds to the comparison $|Z_j| > \sqrt{\log n}$ which gives according to (\ref{Eq:NormalTail}) a type I error probability of $\alpha_n \leq \frac{\sqrt 2}{\sqrt \pi} (n \log n)^{-1/2}$.

However, neither AIC nor BIC provide any correction for multiple testing and it is immediately clear that with growing $p$ the number of type I errors will increase. 
In a high-dimensional context one is typically interested in sparse models and it follows that under sparsity both AIC and BIC will massively overfit the data (see for example \cite{BS01, FN16}).  
In particular BIC will be no longer a consistent selection procedure when considering an asymptotic regime where $p$ grows faster than $\sqrt n$. This is particularly problematic in applications where one is more interested in correct model identification than in prediction.

A first remedy is provided by the risk inflation criterion (RIC), which was  introduced by Foster and George \cite{RIC} and has the penalty $2 \log p\ k_M$.  In the orthogonal setting this relates to a z-test of the form $|Z_j| > \sqrt{\log p^2}$ with type I error rate controlled at   $\alpha_p \leq  \frac{\sqrt 2}{\sqrt \pi} p^{-1} (2 \log p)^{-1/2}$. Hence this penalty is closely related to the Bonferroni rule in multiple testing, where the nominal $\alpha$ level is divided by the number of tests $p$.  Note that RIC is consistent only in the sense that for increasing $p$ the expected number of false detections decreases at the very slow rate $(2 \log p)^{-1/2}$. RIC is not consistent in the classical sense that the  probability of identifying the true model will converge to 1 with increasing $n$. Furthermore if one is interested in model identification then RIC has still a rather large rate of false detections as long as $p$ is not exceptionally large. For small $p = 10$ the  $\alpha_p$ bound indicates that RIC will control FWER only at approximately $0.35$ and for $p=1000$ FWER is still at approximately $0.2$.

\subsection{Modifications of AIC and BIC}\label{Sec:mXIC}

We will now introduce systematically a set of modifications of BIC and AIC which are suitable for high-dimensional variable selection.
The first of these criteria called mBIC was introduced by Bogdan et al. \cite{mBIC} in the context of QTL mapping. The motivation for this criterion was based on arguments concerning the prior distribution of regressors for Bayesian model selection. According to  asymptotic arguments in its classical derivation BIC neglects the model prior. This is equivalent to giving each possible model $M$ exactly the same prior probability. While such a prior is non-informative for the model, it is highly informative for the model dimension which can be seen by a simple combinatorial argument. There are only $p$ models of size 1, $\binom{p}{2}$ models of size 2, but there are $\binom{p}{p/2}$ models of size $p/2$. Consequently BIC will have a strong bias towards choosing models of intermediate size. Now if $p$ is large and one is interested in sparse models then BIC will have a tendency to overestimate the model size. This is a Bayesian explanation of the overfitting problem of BIC in high dimensions which is complementary to the multiple testing perspective given above.

To overcome this problem mBIC was derived by using i.i.d priors for the $p$ regressors \cite{mBIC}. This is of course a classical choice in Bayesian variable selection and results in a binomial prior for the model size. The resulting criterion has a penalty of the form
\begin{equation} \label{mBIC}
\mbox{mBIC:} \quad \mbox{Pen}(k_M;n,p) = \log n\ k_M + 2 \log (p /  E)\ k_M \;.
\end{equation}
Here the BIC penalty is combined with the penalty from RIC.
Clearly if $p$ is large the second penalty term will dominate the $\log n$ term.   However,  the $\log n$ term yields a criterion which is consistent in the usual sense. In fact using similar arguments like given above for BIC  it is easy to see that mBIC controls the FWER of false detections under orthogonality at a level $\alpha_n = o(n^{-1/2})$. On the other hand a similar penalty without the $\log n$ term will control FWER essentially at a constant level. If one is interested in classical consistency then one should keep the $\log n$ term. 

The constant $E$ corresponds to the a-priori expected number of regressors which enter the model.  From a frequentist point of view 
it can be used as a tuning parameter to calibrate the level $\alpha_n$.   In case of no prior knowledge on the model dimension a choice of $E = 4$ is recommended, which guarantees that  for $n=150$ the family wise error rate is controlled approximately at a level $0.1$ for $p \geq 10$ and the bound drops already to $0.065$ for $p = 1000$. For $n=500$ FWER is below $0.05$ for $p \geq 10$ and below $0.035$ for $p = 1000$.  

We have seen that the RIC criterion controls FWER at a constant level with respect to $n$ which is fairly large. FWER decreases with $p$ but at an extremely slow rate. It would take some $p \approx 10^{50}$ to bring down the FWER to 5\%. So  `consistency' with respect to $p$ is really fairly theoretical. As an alternative we introduce the mAIC criterion as a modification of the classical AIC criterion
\begin{equation} \label{mAIC}
\mbox{mAIC:} \quad \mbox{Pen}(k_M;p) = 2 k_M + 2 \log (p / \mbox{const})\ k_M \;.
\end{equation}
Choosing Euler's number $e$ as the constant this coincides with RIC. To control at the more familiar level $\alpha = 0.05$ for small $p = 10$ (and at $0.035$ for $p = 1000$) one can use $\mbox{const} = 0.5$ which is the recommended choice for our mAIC criterion. As a consequence the criteria mAIC and mBIC coincide for sample size $n = 473$.  Using the constant  1 in (\ref{mAIC}) yields roughly an $\alpha$ level of $0.11$ for $p = 10$ ($0.07$ for $p = 1000$).

%%%%%%%%%%%%%%%%%%%%%%%%%%%%%%%%%%%%%%%%%%%%%%%%%%%%%%%%%%%%%%%%%

According to the theoretical results from Section \ref{Sec:Optimality_inference} it is desirable to have selection criteria which control the false discovery rate and not the family wise error rate. This is achieved by the following modifications of BIC and AIC. The first criterion mBIC2 controls FDR at a level which again depends on the sample size like $\alpha_n \propto (n \log n)^{-1/2}$:
\begin{equation} \label{mBIC2}
\mbox{mBIC2:} \quad \mbox{Pen}(k_M;n,p) = \log n\ k_M + 2 \log (p /  E)\ k_M - 2 \log k_M! \;.
\end{equation}
In accordance with the definition of mBIC we recommend the choice of $E = 4$, though FDR levels are then slightly higher than the FWER levels for mBIC (see the simulation results below for details).  The additional  penalty term $- 2 \log k_M!$ relaxes the penalty of mBIC and is closely related to the Benjamini Hochberg (BH) procedure, hence the control of the false discovery rate. In fact  $- 2 \log k_M!$ is a first order approximation of the penalty 
\begin{equation} \label{BH_penalty}
\mbox{Pen}_{BH}:=\sum_j q_N^2(\alpha j / 2 p) \;,
\end{equation}
where $q_N$ denotes the quantile of the normal distribution. This penalty was introduced by Abramovich et al. \cite{ABDJ} in their seminal paper on minimax optimality of FDR controlling model selection rules. Details of the derivation of mBIC2 and its theoretical properties are provided in \cite{FCMB11} which also considers a second order approximation mBIC1 of $\mbox{Pen}_{BH}$. However, for all practical purposes mBIC2 performs just as well and is much easier to compute.  The extended Bayesian Information Criterion (EBIC) from Chen and Chen \cite{CC08} provides another family of BIC modifications suitable for high-dimensional variable selection. It depends on a parameter $\kappa$ which varies between $0$ and $1$. EBIC with $\kappa = 0$ coincides with the original BIC, whereas for  $\kappa$ being large EBIC behaves very similar to mBIC2. However,  there is not such an immediate interpretation of the parameter $\kappa$ in terms of controlling the FDR level. 

Augmenting the mAIC criterion with $- 2 \log k_M!$ yields criteria which control FDR roughly at a fixed level:
\begin{equation} \label{mAIC2}
\mbox{mAIC2:} \quad \mbox{Pen}(k_M;p) = 2 k_M +  2 \log (p / \mbox{const})\ k_M - 2 \log k_M! \;.
\end{equation}
To control FDR at a level close to $\alpha = 0.05$ we recommend once again to use $\mbox{const} = 0.5$.  The penalty of mAIC2 is extremely similar to the penalty $2 k_M \log (p/k_M)$ suggested by Abramovich et al.\cite{ABDJ} as an approximation of $\mbox{Pen}_{BH}$. The difference (up to a constant) between  $k_M \log k_M$ and $\log k_M!$ is due to Sterling's approximation and for small values of $k_M$ mAIC2 is actually closer to $\mbox{Pen}_{BH}$ than $2 k_M \log (p/k_M)$. Similar penalties of the form $c k_M \log (p/k_M)$ with $c > 2$ have been studied by Birge and Massard \cite{BM2001}.

\begin{figure}
	\begin{center}
		\subfigure[\label{Fig:Crit_n200}]{\includegraphics[angle=90,width=5.1cm,angle=-90]{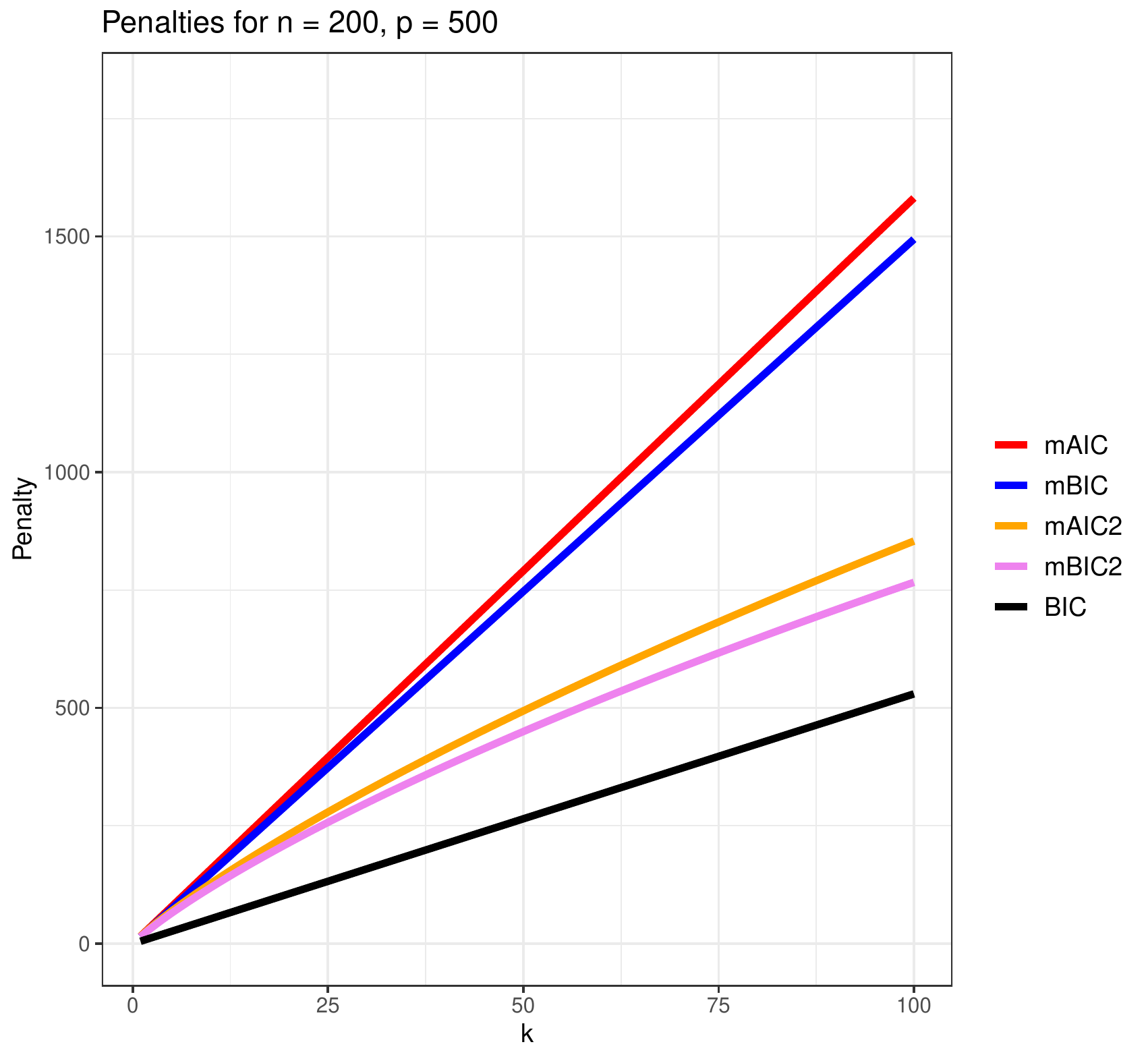}} 
		\subfigure[\label{Fig:Crit_n1000}]{\includegraphics[angle=90,width=5.1cm,angle=-90]{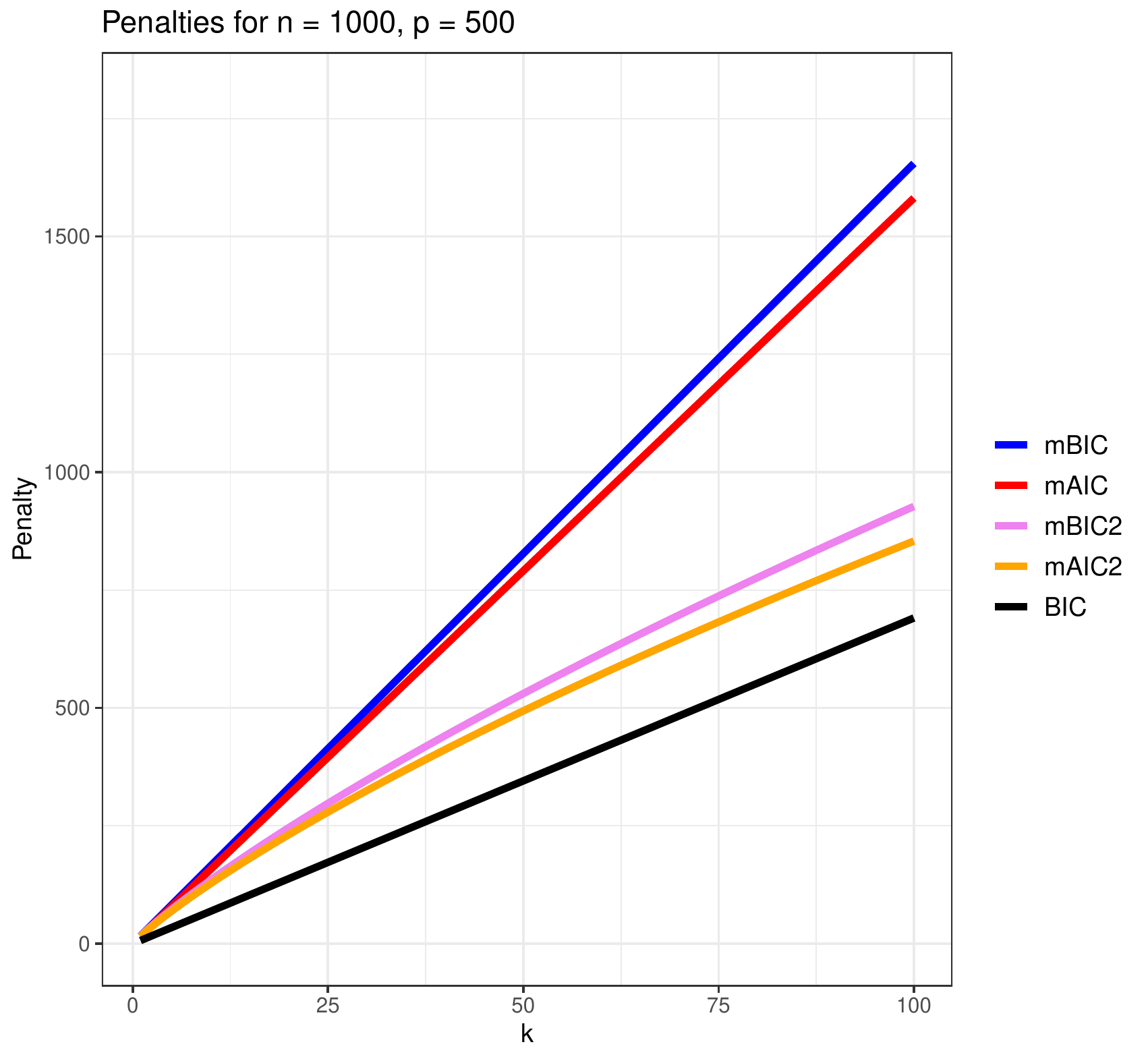}}
	\end{center}
	\caption[$L_0$-Penalty functions]{Different $L_0$-Penalties for $n = 200$ and $n = 1000$ as a function of $k$. }\label{Fig:Crit}
\end{figure}

Figure \ref{Fig:Crit} illustrates the functional form of different $L_0$ penalties. The two penalties $mBIC$ and $mAIC$ which control the FWER are much more severe than the other penalties. For $n = 200$ mAIC (mAIC2) penalizes stronger than mBIC (mBIC2), for $n = 1000$ the opposite is true. The FDR - controlling criteria mAIC2 and mBIC2 have been designed for sparse model selection and should only be used for $k < p/4$. For larger $p$ they can actually penalize less than BIC.  It is also important that $k \ll n$. Otherwise one might run into the problem of getting saturated models where the log-likelihood is converging to infinity. In high-dimensional applications (with $p > n$) $L_0$  penalties will typically have their global minimum at $k = n$. What one is looking for in practice is a local minimum with $k \ll n$.

\section{Simulation study for $L_0$-penalties}\label{Sec:SimL0}

The following simulations illustrate the properties of the four $L_0$ penalties (\ref{mBIC}) -  (\ref{mAIC2}) introduced above in comparison with Schwarz BIC. Data are simulated according to the linear model (\ref{LinearModel}). The first part is concerned with independent regressors and illustrates that in that case  type I error control of the different criteria is very similar to the orthogonal setting. The second part studies a specific scenario with correlated regressors and illustrates that even in such a setting our $L_0$ penalties perform quite well as long as correlations between regressors are not getting excessively large. Analysis was performed with the R package bigstep which is available at CRAN \cite{bigstep}.

\subsection{Independent regressors}

In the scenarios considered here both the columns of $X$ and the error term $\epsilon$ are i.i.d. standard normal. Scenario 0 is concerned with the type I error rate (number of false discoveries) under the assumption that there are no regressors associated with the dependent variable (that is $k = 0$). The other three scenarios consider sparse data generating models, where the total number of regressors $p$ behaves differently with growing sample size $n$. In the first scenario $p$ remains constant with growing $n$; in the second scenario $p$ is proportional to $\sqrt n$ and in the final scenario $p$ equals $n$.  In Scenario 1 the number of regressors in the model $k$ remains also constant whereas in the other two scenarios  $k$ is mildly growing. Table \ref{Tab:SimScenConsisteny} provides more details.

\begin{table}[h]
	%\noautomaticrules
	\tabletitle{Characteristics of the four simulation scenarios}%
	\label{Tab:SimScenConsisteny}
	\begin{tabular}{r|cc|cc|cc|cc}
		&\multicolumn{2}{|c|}{ \tch{Scen 0}}&\multicolumn{2}{|c|}{ \tch{Scen 1}}&
		\multicolumn{2}{|c|}{ \tch{Scen 2}}&\multicolumn{2}{|c}{ \tch{Scen 3}} \\
		$n$ & $p$&$k$ & $p$&$k$  & $p$&$k$  & $p$&$k$  \\\hline 
		49  & 49 & 0  & 49 & 5  & 49 & 5 & 49 & 5 \\
		100 & 49 & 0  & 49 & 5  & 70 & 7 &100 & 7 \\ 
		225 & 49 & 0  & 49 & 5  &105 &10 &225 &10 \\
		529 & 49 & 0  & 49 & 5  &161 &13 &529 &15 \\ 
		1024& 49 & 0  & 49 & 5  &224 &16 &1024&20 \\
		2048& 49 & 0  & 49 & 5  &
	\end{tabular}
\end{table}

All coefficients from $\beta$ were set to 0.4 for those regressors which enter the data generating model. In Section  \ref{Sec:Optimality_inference} we will pay more attention to the effect sizes which can actually be detected with different model selection criteria. Here our main focus is rather on the type I error rates where the simulations are supposed to illustrate the control rates claimed above. 

To estimate FWER and FDR for each scenario 1000 simulation runs were performed. Regressors selected by some criterion are counted as true positives (TP) if they are part of the data generating model, otherwise they are counted as false positives (FP). FWER is then estimated as the average number of simulation runs with at least one FP detection. FDR is defined as the average over simulation runs of the proportion of false discoveries $\#FP / \max(1,\#FP +\#TP)$.

\begin{figure}
	\begin{center}
		\subfigure[\label{Fig:Sim0_FWER}]{\includegraphics[angle=90,width=5.1cm,angle=-90]{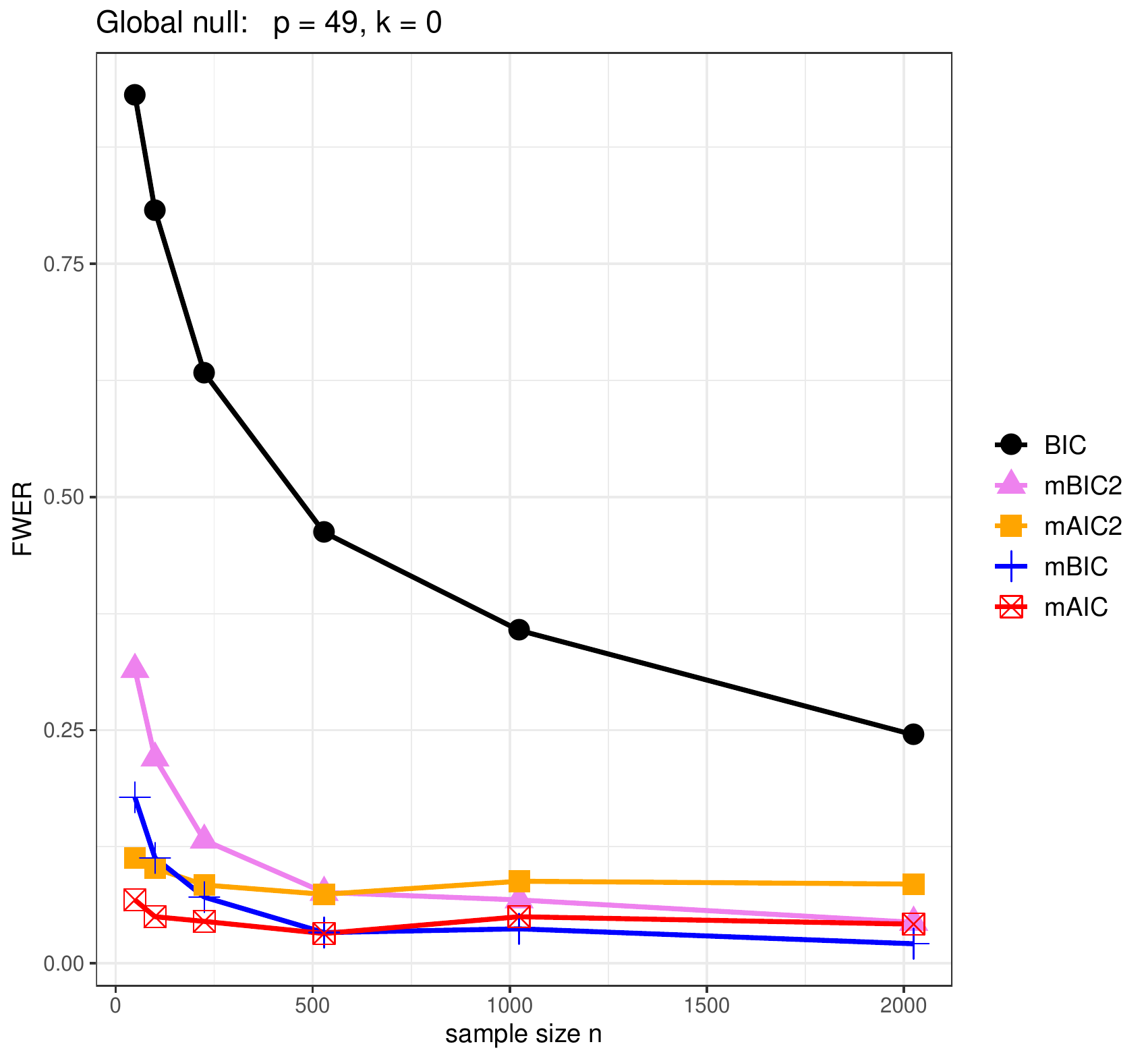}} 
		\subfigure[\label{Fig:Sim1_Power}]{\includegraphics[angle=90,width=5.1cm,angle=-90]{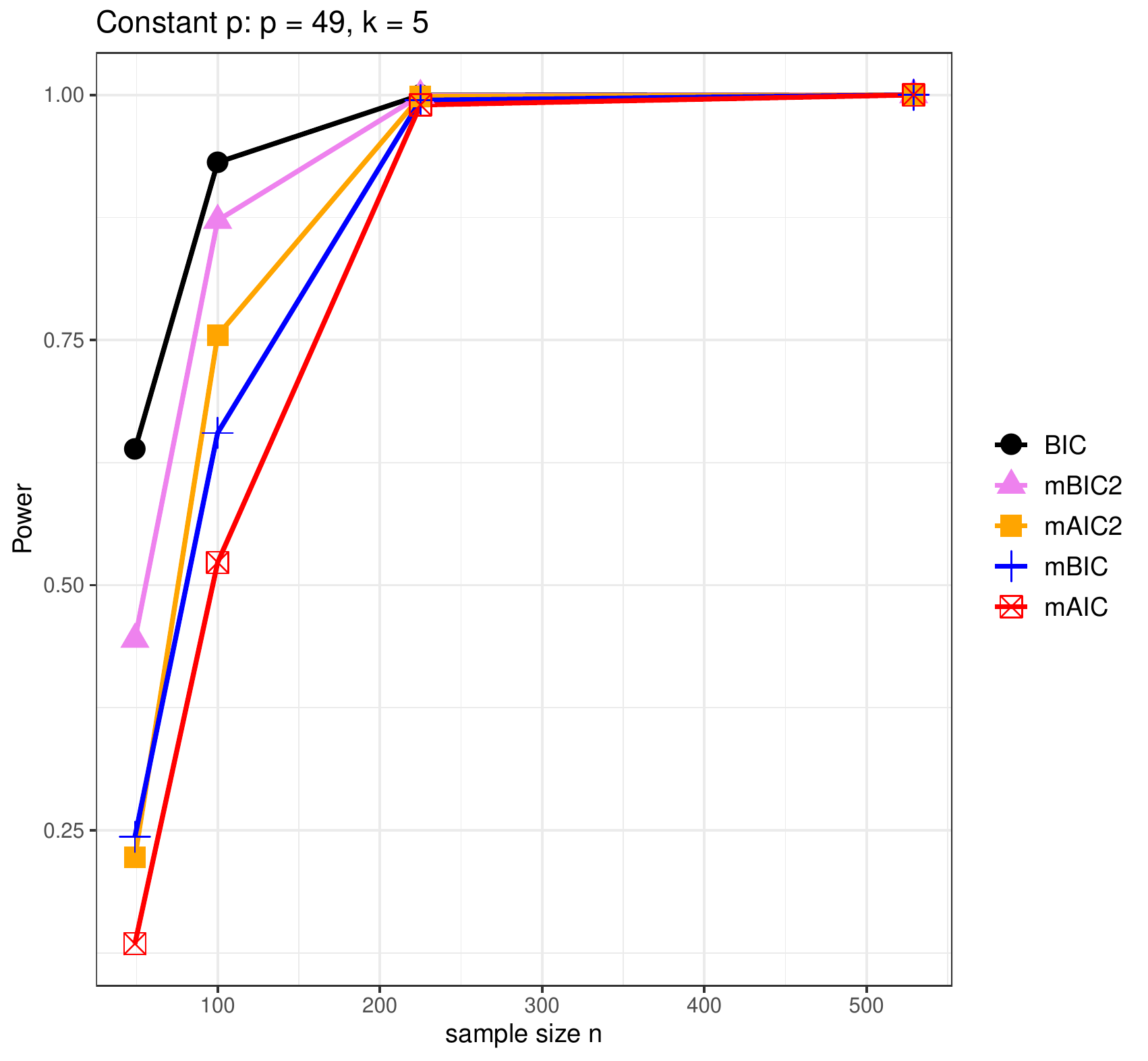}}
	\end{center}
	\caption[Simulation with independent regressors: FWER under global null and example of power]{\textbf{Panel a}: Family wise error rate for different $L_0$-penalties depending on sample size $n$ under the global null model of Scenario 0. \textbf{Panel b}: Power depending on $n$ for Scenario 1 with constant $p=49$ and constant $k=5$. }\label{Fig:Sim0}
\end{figure}

Figure \ref{Fig:Sim0_FWER} shows the dependence of FWER on the sample size. Clearly BIC has a much larger type I error rate than the other four criteria. One can see that mAIC nicely controls FWER at the nominal level 0.05, while under the global null mAIC2 has a FWER (and thus FDR) closer to 0.08. mBIC has larger FWER than mAIC for $n < 500$ and smaller type I error for $n > 500$. The same relationship hold for mAIC2 and mBIC2.

\begin{figure}
	\begin{center}
		\subfigure[\label{Fig:Sim1_FWER}]{\includegraphics[angle=90,width=5.1cm,angle=-90]{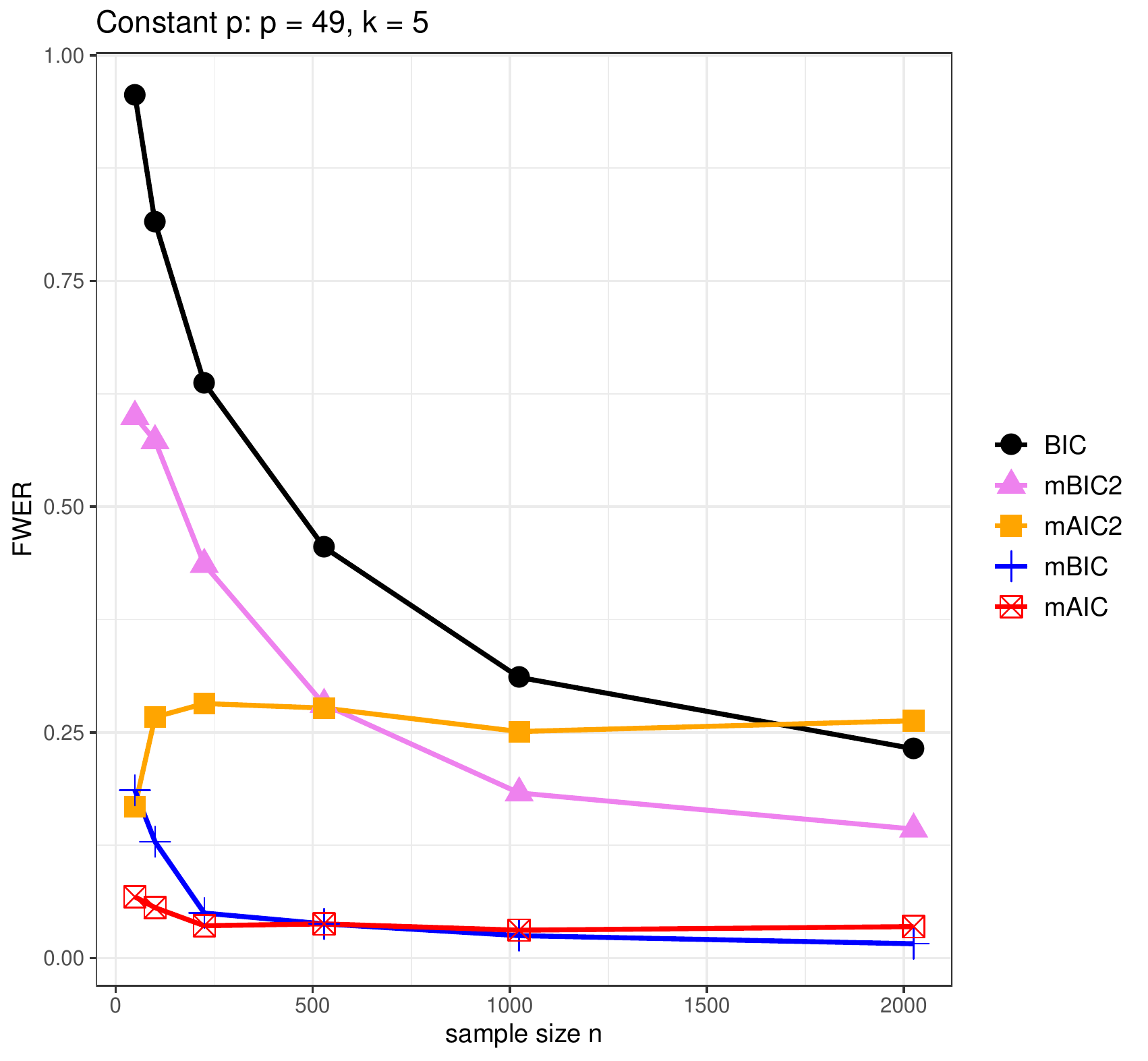}} 
		\subfigure[\label{Fig:Sim1_FDR}]{\includegraphics[angle=90,width=5.1cm,angle=-90]{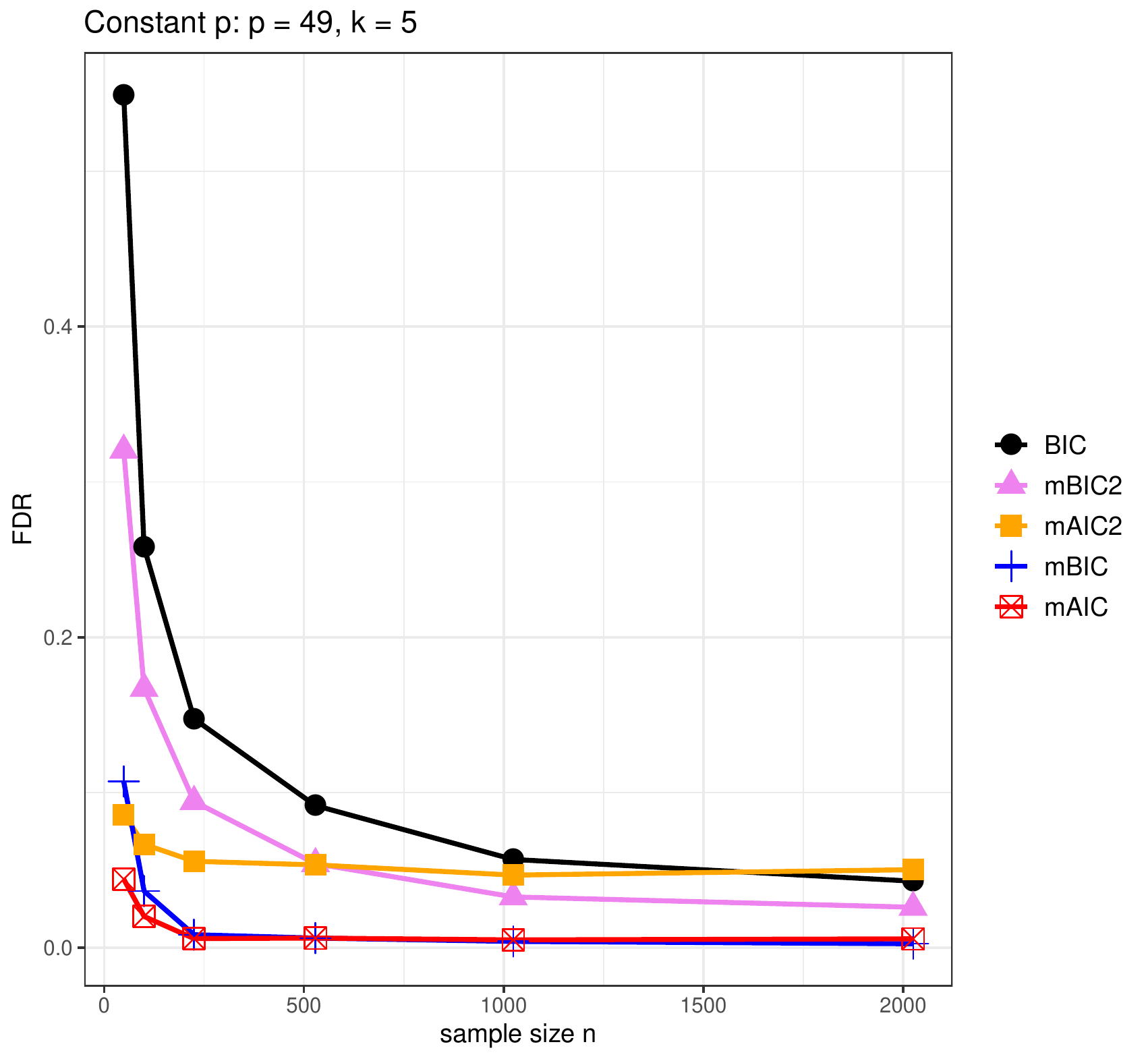}}
		\subfigure[\label{Fig:Sim2_FWER}]{\includegraphics[angle=90,width=5.1cm,angle=-90]{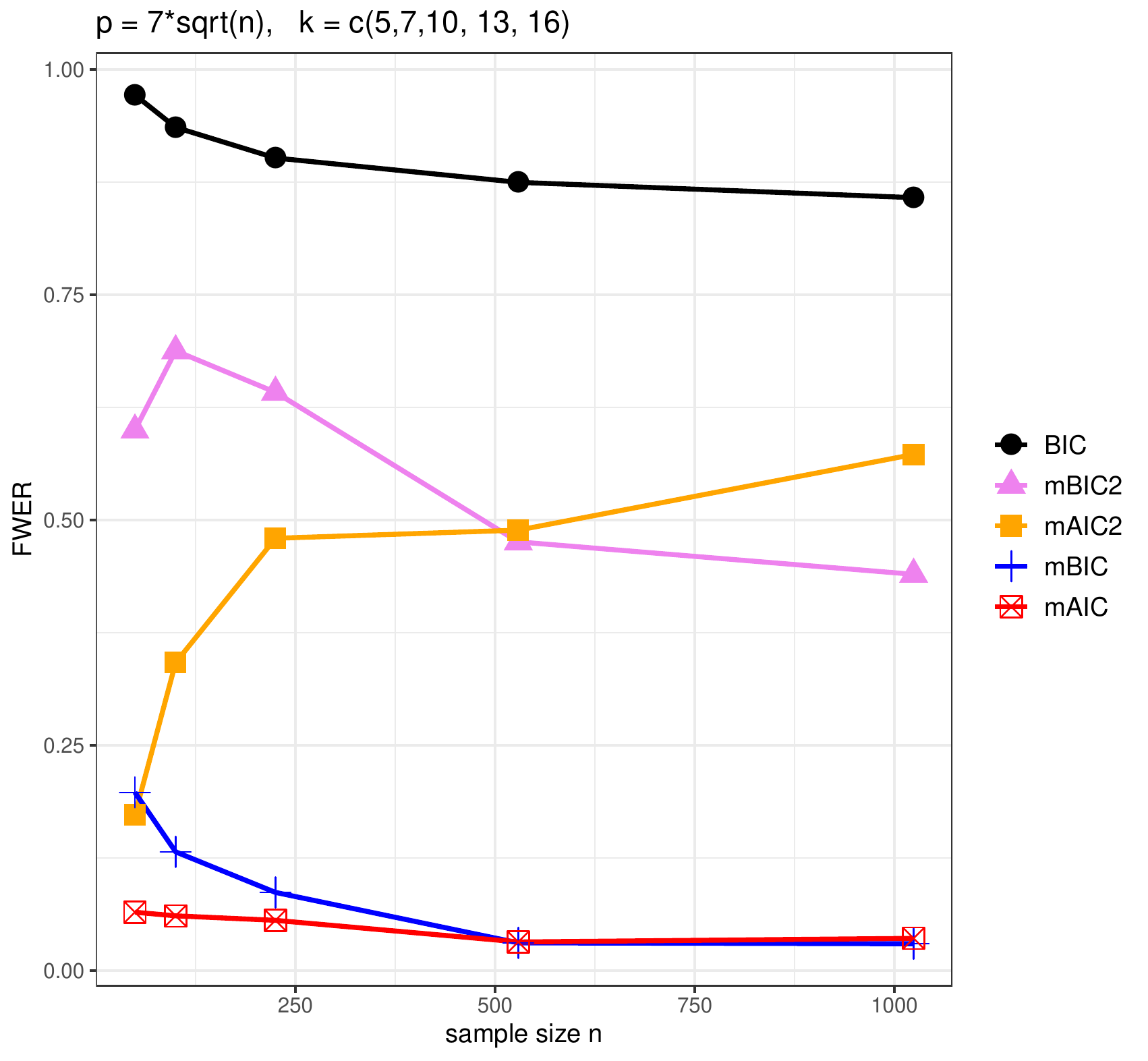}} 
		\subfigure[\label{Fig:Sim2_FDR}]{\includegraphics[angle=90,width=5.1cm,angle=-90]{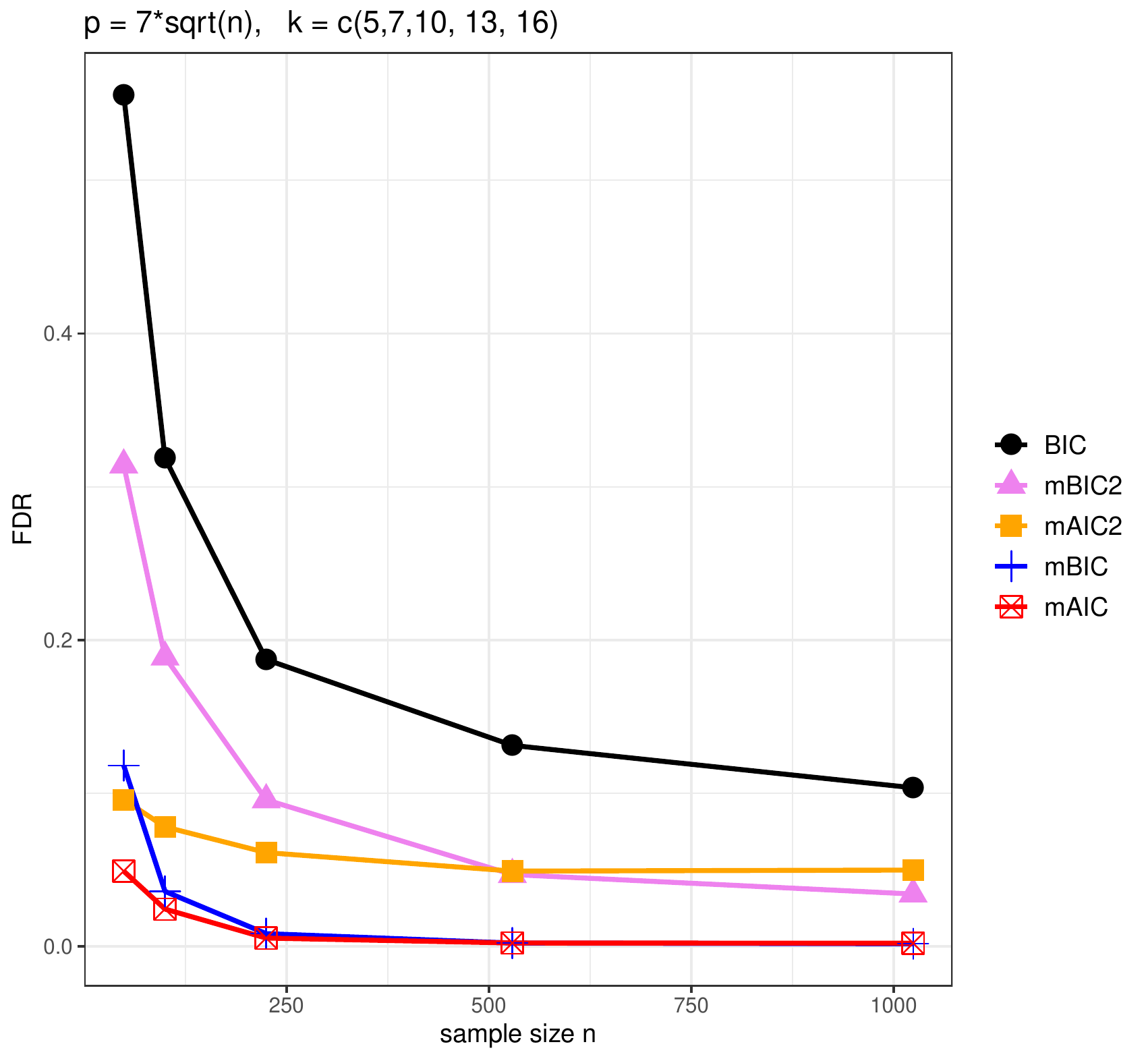}}
		\subfigure[\label{Fig:Sim3_FWER}]{\includegraphics[angle=90,width=5.1cm,angle=-90]{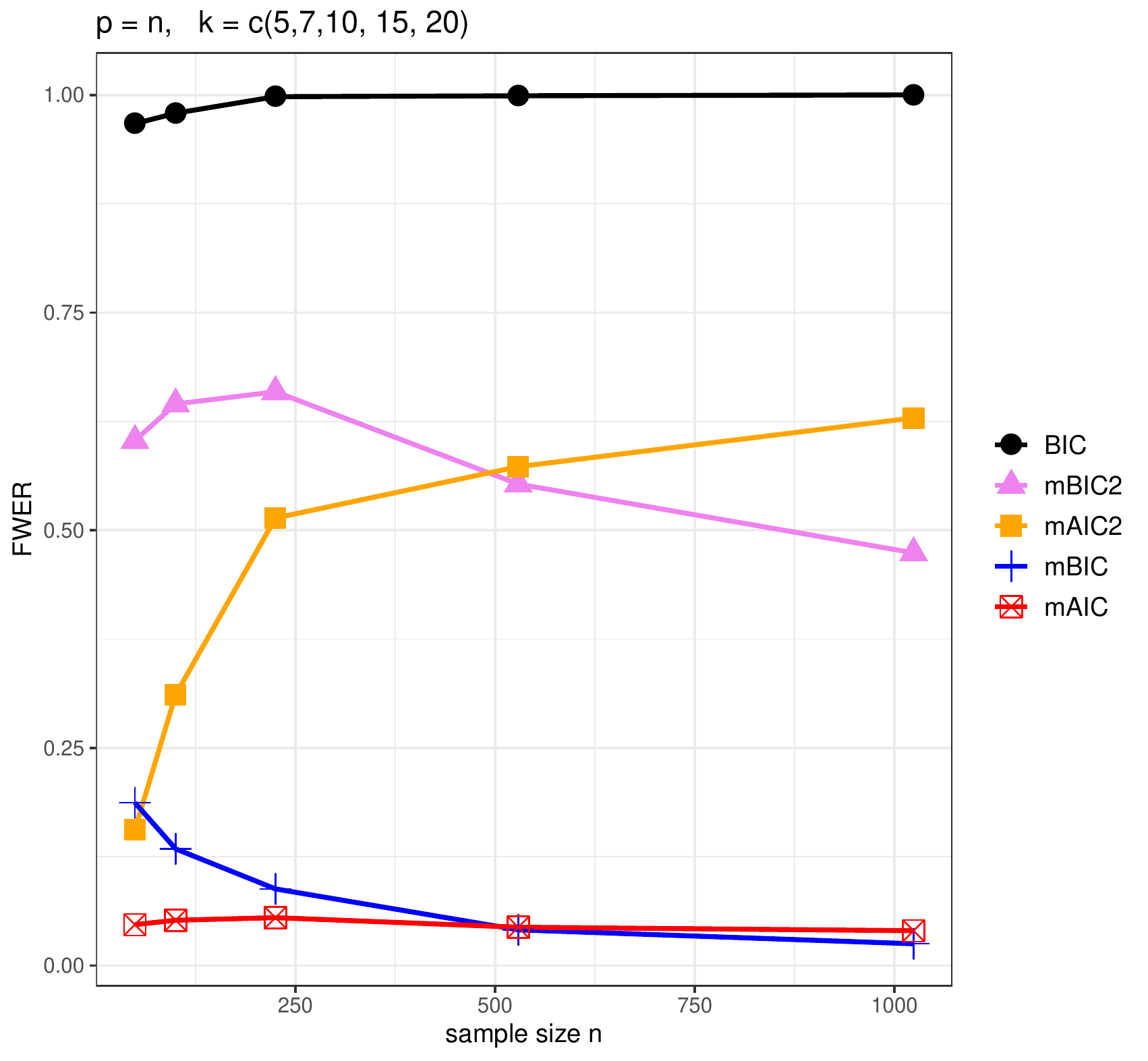}} 
		\subfigure[\label{Fig:Sim3_FDR}]{\includegraphics[angle=90,width=5.1cm,angle=-90]{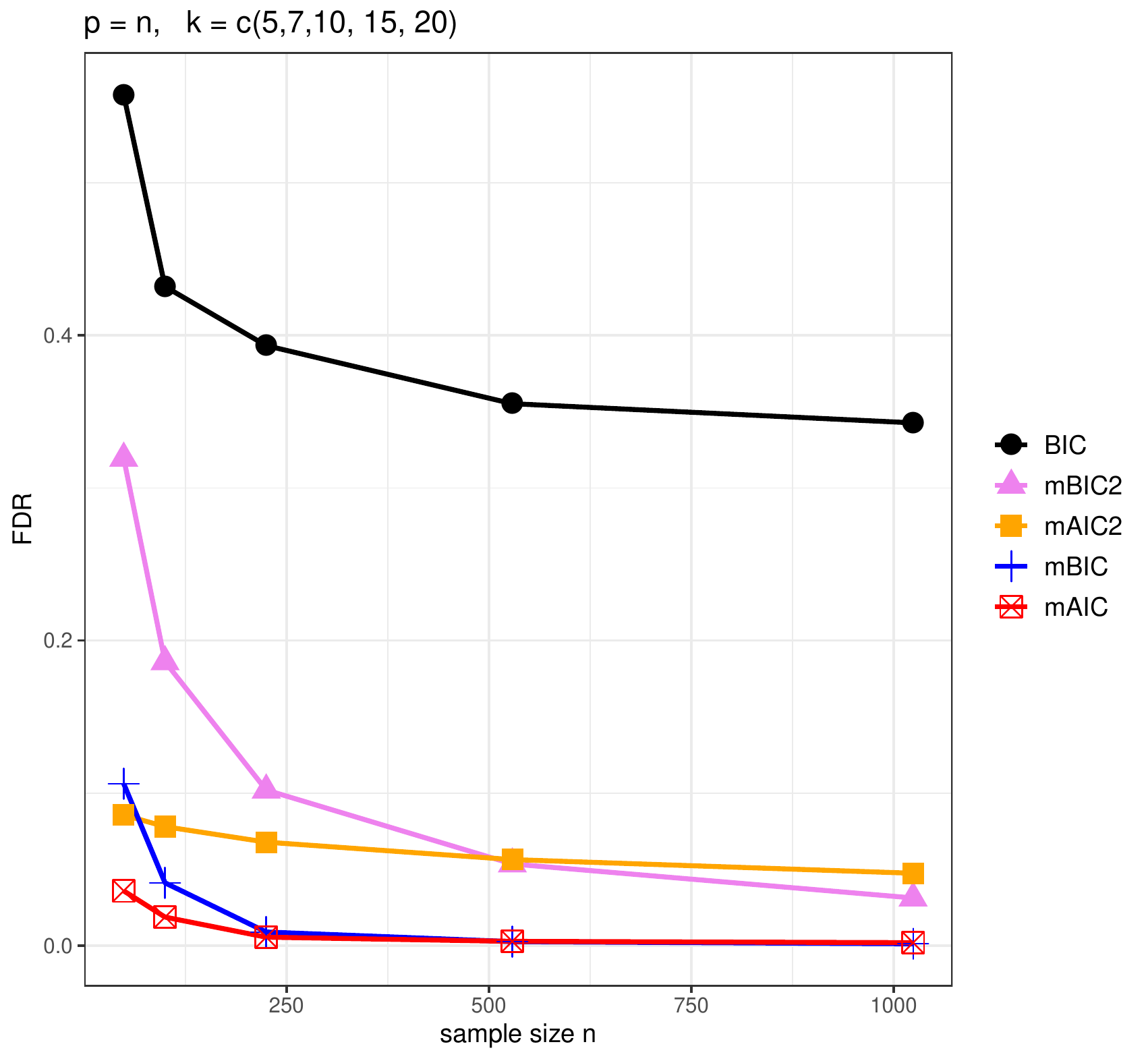}}
	\end{center}
	\caption[Simulation with independent regressors: FWER and FDR for three simulation scenarios]{Family wise error rate and false discovery rate depending on $n$ for three different simulation scenarios. \textbf{Panel a and b}: Scenario 1 with constant $p=49$ and constant $k=5$;  \textbf{Panel c and d}: Scenario 2 with $p \propto \sqrt n$ and $k$ growing mildly with $n$; \textbf{Panel e and f}: Scenario 3 with $p = n$ and $k$ growing mildly with $n$;   }\label{Fig:Sim123}
\end{figure}

Figure \ref{Fig:Sim1_Power} provides the power for Scenario 1, which is defined here as the percentage of correctly detected regressors from the data generating model. The corresponding plots for Scenario 2 and Scenario 3 look fairly similar and are not presented.  BIC has the largest power followed by mBIC2, but already for $n = 500$ all criteria achieve a power of 1. In terms of consistency it is therefore for these scenarios of primary importance to look at the type 1 error rates depicted in Figure \ref{Fig:Sim123}.

The three left panels of  Figure \ref{Fig:Sim123} show FWER depending on the sample size. For constant $p$ it is known that  BIC is consistent and consequently FWER keeps on decreasing with increasing $n$. However, even for $n = 2000$ the average number of FP detections is still at 0.27. In comparison FWER of mBIC and mAIC are really small already for quite moderate sample size with values which are in accordance with the results for the orthogonal design. Note  that  the FWER of mAIC remains at about 0.04 even for large $n$ and is actually not expected to get smaller for arbitrary large $n$. Just like AIC, mAIC is not consistent, whereas mBIC is. 

Looking at the plots \ref{Fig:Sim2_FWER} and \ref{Fig:Sim3_FWER} shows that when $p$ is growing with $n$ BIC is no longer consistent at all. For $p \propto \sqrt n$ the average number of false detections still decreases very slowly but for $p = n$ it actually keeps  on growing with $n$. So even for $p = n$ BIC is already completely unsuitable as a selection criterion if one is interested in model identification and for $p > n$ things are only getting worse. In contrast both mAIC and mBIC are doing a very good job in controlling FWER for all our simulation scenarios and they work similarly well in case of $p > n$.  The FWER estimated from simulations for  independent regressors are remarkably close to the theoretical values from the orthogonal design.

The three right panels of Figure \ref{Fig:Sim123} provide the FDR. One can observe in all three scenarios that at least for larger $n$ mAIC2 nicely controls FDR at the level 0.05. Similarly the FDR of mBIC2 drops with growing $n$ even below 0.05.  mAIC and mBIC tend to have extremely small FDR and in view of the discussion of Section \ref{Sec:Optimality_inference} are therefore potentially too conservative for many high-dimensional applications.

\subsection{Correlated regressors} \label{Subsec:Corr}

The previous simulations were performed for statistically independent regressors.
For highly correlated predictor variables it usually becomes difficult to distinguish between correctly identified predictors from an underlying data generating model and variables which are only highly correlated. Nevertheless mBIC and mBIC2 have been repeatedly  applied successfully in the context of genetic association studies \cite{GWAS, memetic, DBF14, admixtures}.  The following simulation scenario taken from \cite{F11} gives an impression to which extent model selection based on our modifications of AIC and BIC still performs well in case of correlated regressors. 

%{\bf THIS FIGURE NEEDS TO BE WORKED AT IN TERMS OF FORMATTING}

\begin{figure}
	\begin{center}
		%\subfigure[\label{Fig:CorrStruct1}]{\includegraphics[angle=90,width=6.1cm,angle=-90]{CorrStruct.pdf}} 
		\includegraphics[width=6.5cm]{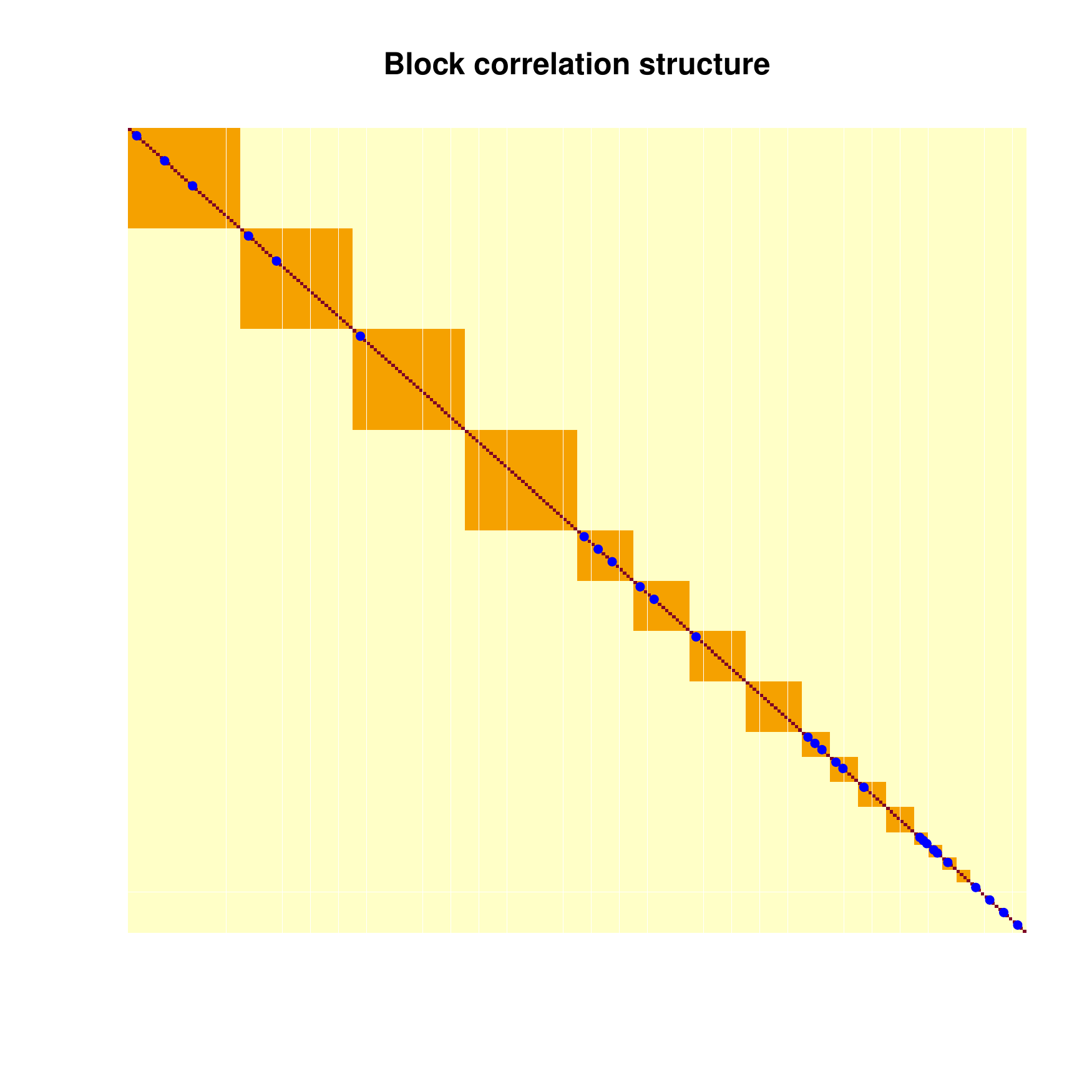}
	\end{center}
	\caption[Simulation with correlations: Correlation structure of the matrix $X$.]{Block correlation structure of the matrix $X$.  Four different block sizes and four blocks per size, where within each block all variables have pairwise correlation $\rho$. Simulations are run with $\rho$ ranging from  0 to 0.6. Blue dots indicate the causal variables from the data generating model.}\label{Fig:CorrStruct}
\end{figure}

Consider 256 potential regressor variables with a specific block correlation structure sketched in Figure \ref{Fig:CorrStruct}. The first four blocks include each 32 variables, then come four blocks with 16 variables, four blocks with 8 variables and four blocks with 4 variables, respectively. Within each block one has compound symmetry with  variance 1 and correlation $\rho$, where depending on the simulation run $\rho$ is ranging from 0 to 0.6. Otherwise the blocks of variables are independent from each other and there are some additional 16 variables which are also independent. This block structure is inspired by correlation patterns one might find in genome-wide association studies, though apparently it is a simplified setting.

Regressors from the data generating model are referred to as `causal' in this example.  For each block size there is one block with three, one with two and one with one causal variable, respectively. Additionally there are four causal variables among the 16 independent variables. In summary our data generating model thus has $k^* = 4*3 + 4*2 + 4*1 + 4 = 28$ causal variables. Effect sizes are randomly drawn from a normal distribution with mean 0 and variance 0.5. For this scenarios 2000 data sets are simulated to assess the performance of the different selection criteria.

\begin{figure}
	\begin{center}
		\subfigure[\label{Fig:SimCorr_FWER}]{\includegraphics[angle=90,width=5.1cm,angle=-90]{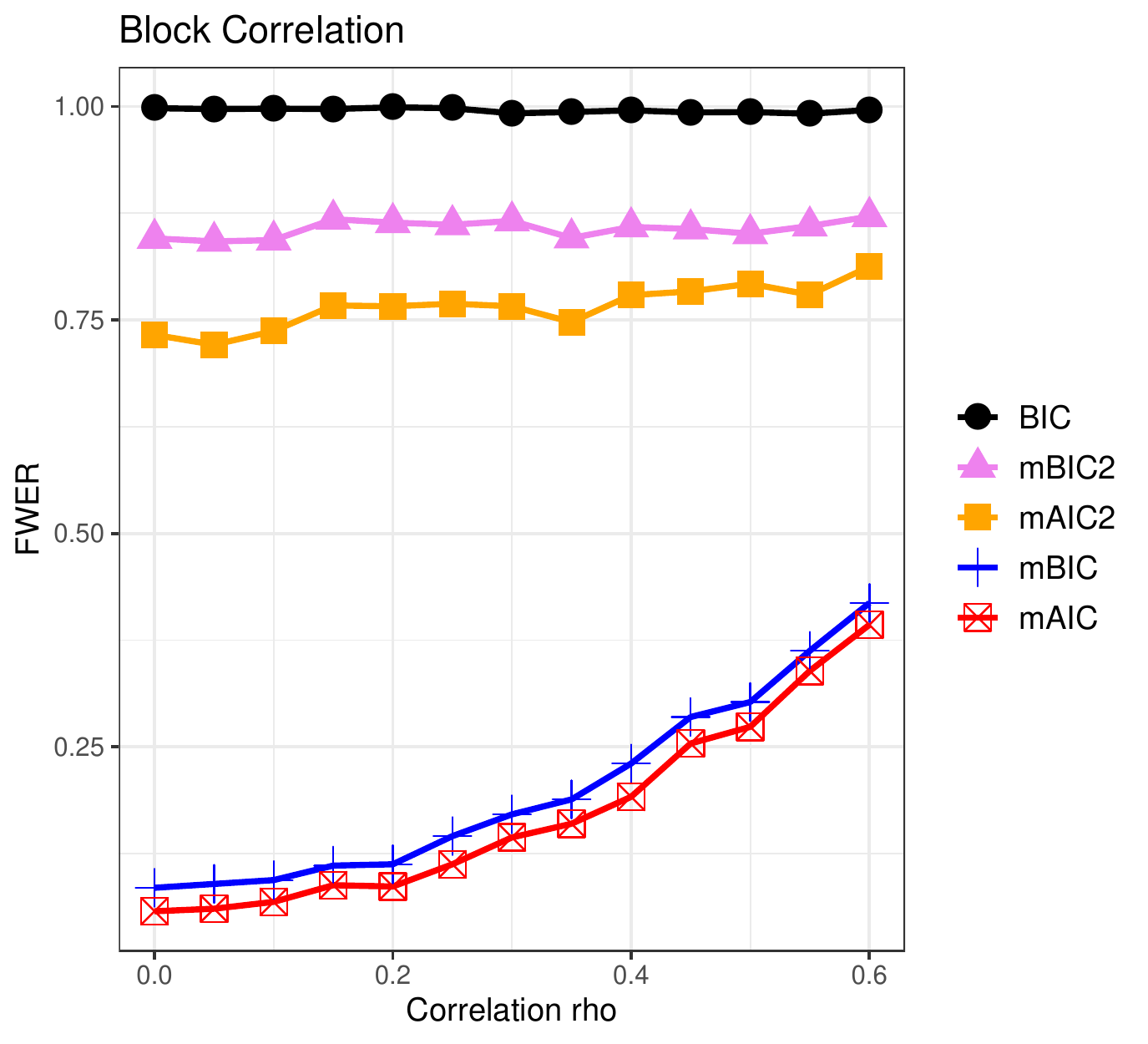}} \subfigure[\label{Fig:SimCorr_FDR}]{\includegraphics[angle=90,width=5.1cm,angle=-90]{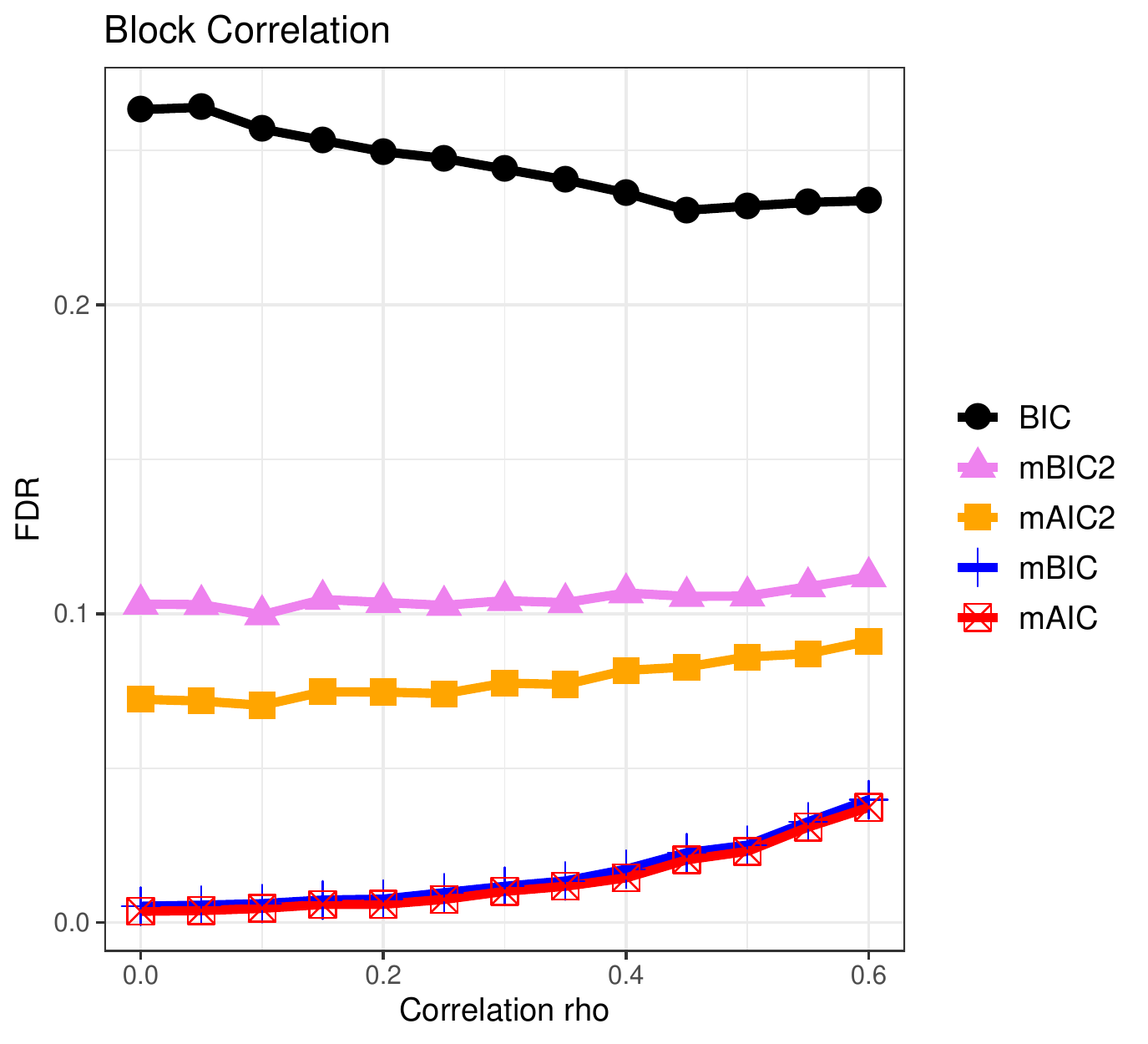}}
		\subfigure[\label{Fig:SimCorr_Power}]{\includegraphics[angle=90,width=5.1cm,angle=-90]{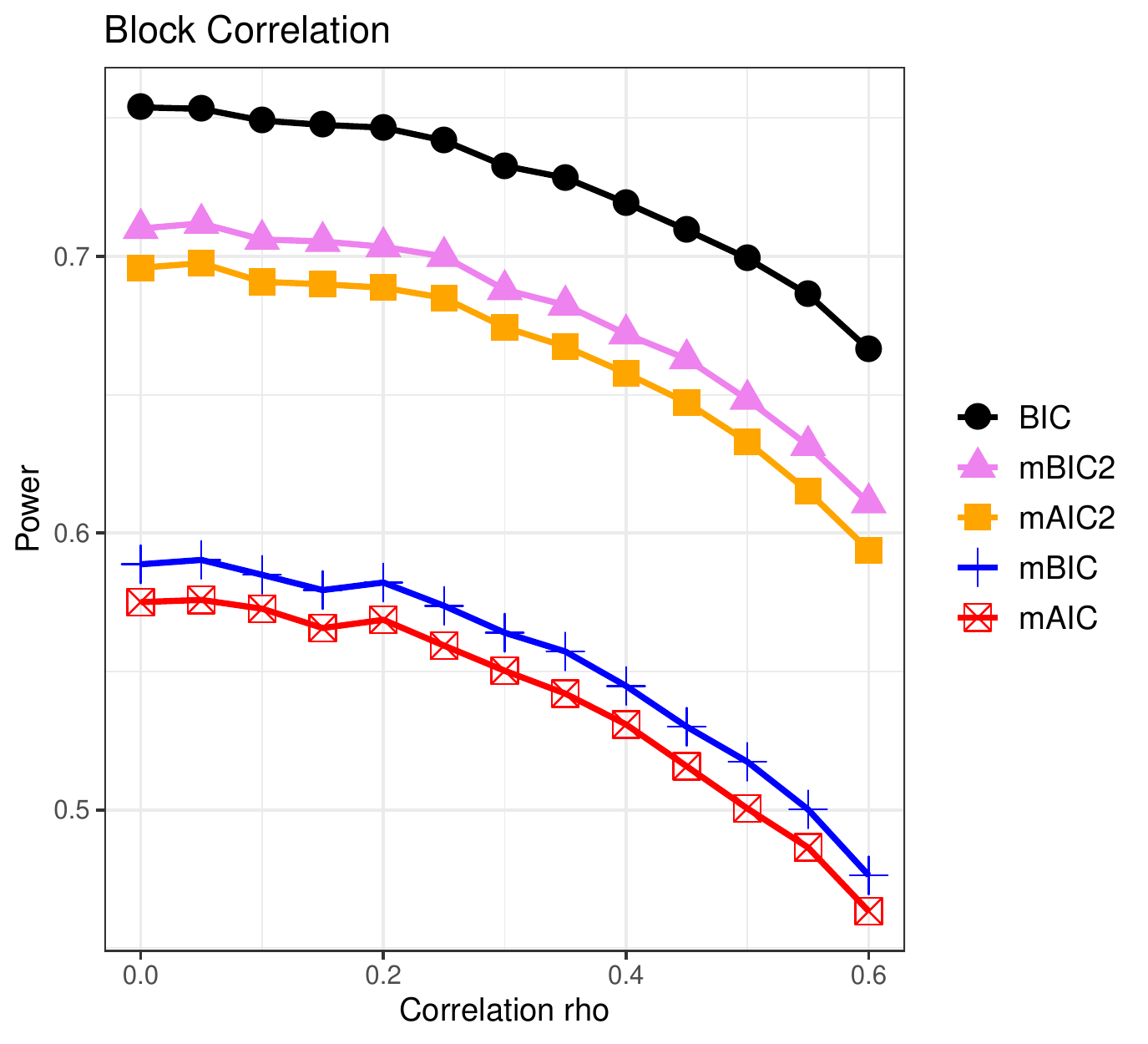}} \subfigure[\label{Fig:SimCorr_Mis}]{\includegraphics[angle=90,width=5.1cm,angle=-90]{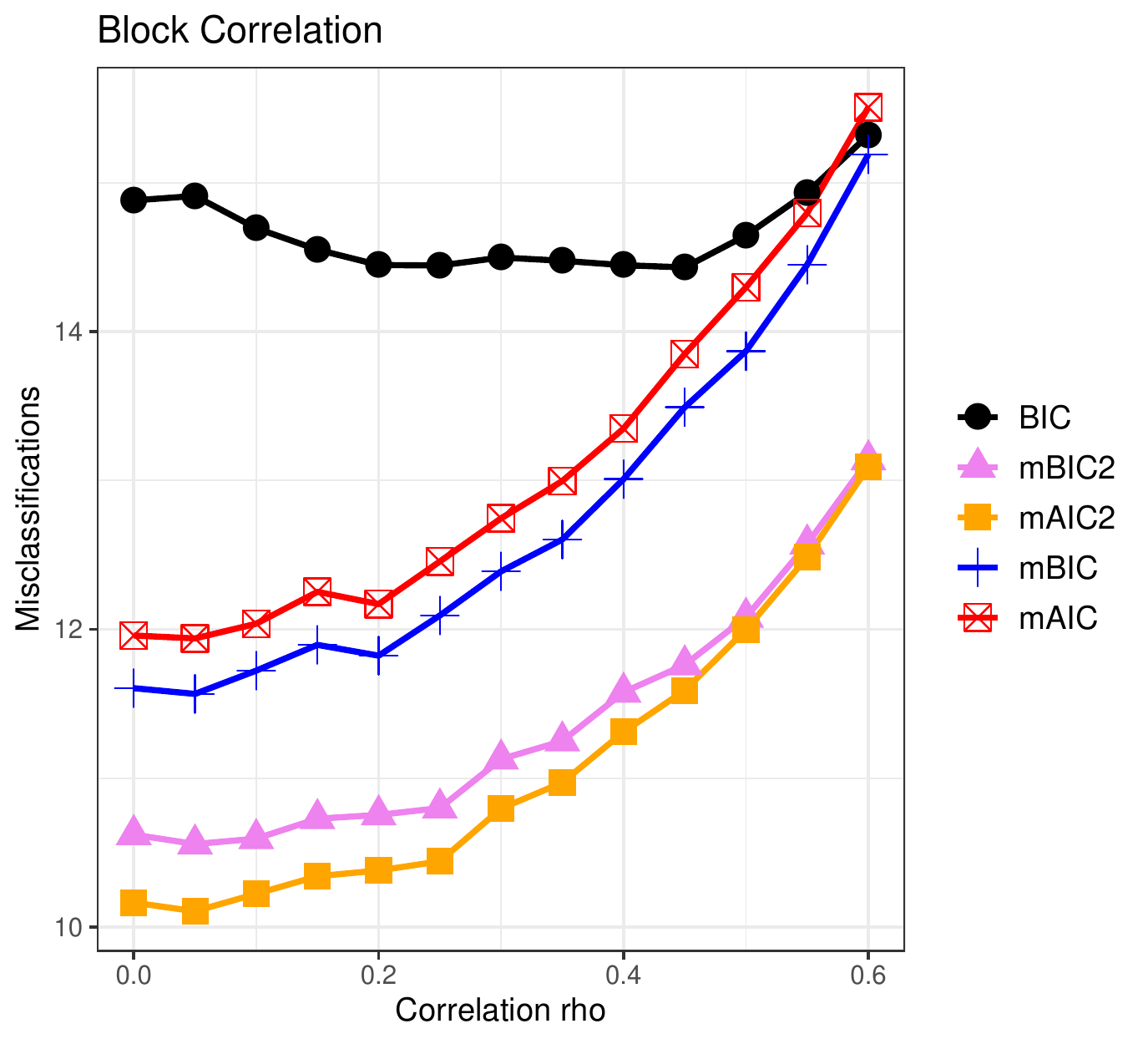}} 
	\end{center}
	\caption[Simulation results for correlated variable]{Power, number of misclassifications, family wise error rate and false discovery rate depending on within block correlation $\rho$ for  different $L_0$ penalties.   }\label{Fig:SimCorr}
\end{figure}

Figure \ref{Fig:SimCorr} presents the results of the simulation study for correlated regressors. Apart from Power, FDR and FWER it also shows the average number of misclassifications obtained with each selection criterion.  For $\rho = 0$ the potential regressor variables are all independent and the type I error rates are similar to those from Scenario 3 of the previous simulation study. In particular mAIC and mBIC are controlling FWER at levels 0.06 and 0.08, respectively. Furthermore mAIC2 and mBIC2 are controlling FDR at levels 0.07 and 0.1, whereas BIC has  FWER close to 1 and FDR at 0.26. 

BIC has with 75\% the largest power but due to the large number of false positives also the largest number of misclassifications. Note that this scenario is not particularly sparse and hence BIC is performing not that bad. Increasing the number of potential regressors $p$ while keeping the same number of causal variables would provide scenarios where BIC would perform much worse compared with the other criteria.  Note that mAIC2 and mBIC2 have the lowest number of misclassifications which corresponds to the optimality properties discussed in Section \ref{Sec:Optimality_inference}. Controlling FDR gives the best trade off between controlling the number of false positives and  having sufficient power to detect true positives.

Increasing the correlation within blocks has the following effects. For all criteria the power decreases. This is natural because for larger $\rho$ it becomes more likely that a causal variable is substituted by a highly correlated variable in the model. Eventually for very large $\rho$ correlated variables will become more or less indistinguishable and the concept of true and false positives breaks down for variables within the same block. For the very same reason mAIC and mBIC suffer from an increasing  number of false positive    detections with increasing $\rho$ and are no longer controlling FWER. Now interestingly the average number of false positives detected  by mAIC2 and mBIC2 remains fairly stable with increasing $\rho$ while for BIC the number of false positives is actually decreasing.  FDR of mAIC2 and mBIC2 are increasing only slowly with increasing $\rho$ and they keep on having the lowest misclassification rate.

\section{Optimality results}\label{Sec:Optimality}

In the last simulation scenario the FDR controlling criteria mAIC2 and mBIC2 performed best in terms of the number of misclassifications. The following section briefly recaps theoretical results from multiple testing which corroborate this observation. Subsequently some results from the literature are presented which show that mAIC2 is asymptotically optimal with respect to prediction.

\subsection{Optimality for model identification}\label{Sec:Optimality_inference}

Bogdan et al. \cite{ABOS}
introduced the concept of Asymptotic Bayes Optimality under Sparsity (ABOS) for multiple testing in the context of normal mixture distributions. The optimality results obtained there were later extended to more general multiple testing settings \cite{ABOS2, ABOS3} and in case of orthogonal designs can be directly transferred to the model selection criteria which were introduced above \cite{FCMB11}. Here we will only sketch the ABOS results for multiple testing to provide an intuition about the advantages of using FDR controlling model selection criteria in a high-dimensional setting.

%%%%%%%%%%%%%%%%%%%%%%%%%%%%%%%%%%%%%%%%%%%%%%%%%%%%%%%

Consider a set of normally distributed populations $T_j | \mu_j \sim N(\mu_j, \sigma^2)$, $j=1,\ldots,p$. A two groups normal mixture model 
\begin{equation}\label{Eq:NormalModel}   
T_j \sim  (1-\eta) N(0,\sigma^2)  +  \eta N(0,\sigma^2 + \tau^2) \;
\end{equation}
can be used for testing the hypotheses
\begin{equation}\label{Eq:NormalHyp}
H_{0j}: \mu_j=0 \ \ \mbox{ against } \ \  H_{Aj}: \mu_j \sim N(0,\tau^2) \;.
\end{equation}
This is similar to the classical two-sided test, but like in the simulation scenario of Section \ref{Subsec:Corr} the  effect size under the alternative  is a random variable. The parameter $\eta$ gives the proportion of true alternative hypotheses in the population. It is assumed to be small and will be referred to as sparsity parameter.

The concept of ABOS is based on a decision theoretical framework. For each individual test a type I error imposes a loss of $\delta_0$  and a type II error a loss of $\delta_A$.  The total loss is then defined as the additive loss over all individual tests. This is a common choice in classification tasks.  
For a specific multiple testing procedure let $t_{1j}$ and $t_{2j}$ be the probabilities of type I and type II errors for the $j$-th hypothesis, respectively. The corresponding  Bayes risk is then defined as the expected total loss,
\begin{equation}\label{Eq:risk}    
R = p \left(  (1-\eta) t_{1j} \delta_0 + \eta t_{2j} \delta_A\right) \;.
\end{equation}
In case of $\delta_0 = \delta_A = 1$, this is simply the expected number of misclassifications.

Assuming that the $p$ tests based on model (\ref{Eq:NormalModel})  are independent and that $\sigma$ is known it is easy to explicitly obtain the risk $R_{opt}$ of the Bayes classifier  which minimizes the Bayes risk (\ref{Eq:risk}). In \cite{ABOS} certain asymptotic regimes were studied and a multiple testing procedure in this setting was classified as ABOS if the ratio between its Bayes risk and the risk of the Bayes classifier converges to one, i.e. $R/R_{opt} \rightarrow 1$.

Without including all the technicalities of \cite{ABOS} the main result can be illustrated by considering the asymptotic setting where the number of tests $p$ converges to infinity and the sparsity parameter $\eta$ is decreasing with $p$ according to $\eta \propto p^{-\beta}$ for some $\beta \in (0,1]$. Here $\beta$ describes the asymptotic level of sparsity. Given some other technical conditions it turns out that Bonferroni correction is ABOS only when $\eta \propto 1/p$. This is the assumption of  ``extreme" sparsity, under which the expected number of true signals does not increase with $p$. Instead, Benjamini-Hochberg correction is ABOS for any $\beta \in (0,1]$. When the number of true signals is very small then the results of BH are not substantially different from those obtained using the Bonferroni correction. However, BH can adapt to the unknown sparsity level and is ABOS for a wide range of scenarios under which $\eta \rightarrow 0$.

For a loss with $\delta_0 = \delta_A = 1$ it is also necessary that with increasing $p$ the FDR level $\alpha$ of BH converges to 0 at a certain rate to obtain ABOS. In \cite{ABOS3} model (\ref{Eq:NormalModel}) was extended to consider the situation of tests based on random samples of size $n$ and conditions are given under which BH with FDR level $\alpha \propto n^{-1/2}$ is ABOS. In terms of model selection criteria this rate of decrease directly corresponds to mBIC2 as we have seen in Section \ref{Sec:orthogonal}.

\subsection{Optimality for prediction}\label{Sec:Optimality_prediction}

Abramovich et al. \cite{ABDJ} analyzed the properties of BH for estimating a vector of random variables with expected values $\mu$.  Specifically, they consider the hard-thresholding estimator 
\begin{equation}
\tilde \mu_j=
\left\{
\begin{array}{cc}
X_j &\mbox{if BH rejects}\;\; H_{0j}:\mu_j=0\\
0&\mbox{otherwise}
\end{array}
\right.
\end{equation}

The optimality of BH is analyzed with respect to the estimation risk over ``sparse" balls in the space $\mu \in R^p$. In \cite{ABDJ} different notions of sparsity are considered and highly technical results are proven. In essence it is shown that the hard-thresholding rule based on BH is asymptotically minimax over a wide range of sparsity levels or, in other words, it adapts to the unknown sparsity and optimally selects those components of $\mu$ for which the signal strength substantially exceeds the variance of the estimation error. 

In Wu and Zhao \cite{Wu_13} these results are extended to the class of estimators of the form:
\begin{equation}\label{Wu}
\hat \mu = \argmin_{\mu} \left\{||Y-\mu||^2 + \sigma^2 Pen \left(||\mu||_0\right)\right\}\;\;,
\end{equation}
where $Pen\left(||\mu||_0\right)$ is the penalty for the number of non-zero elements of $\mu$. 
Specifically, in \cite{Wu_13} it is shown that $\hat \mu$ is asymptotically minimax if  the penalty grows like $2k \log(p/k)$.
It is easy to check that mAIC2 is asymptotically equivalent to such a penalty. 
Note also that under orthogonality with $X'X=I$ a multiple regression model can be represented as
$$\hat \beta=X'Y=\beta+\tilde\epsilon\;\;,$$
where $\tilde\epsilon = X'\epsilon\sim N(0,\sigma^2 I)$.
Thus, the results of \cite{Wu_13}  prove also asymptotic optimality of mAIC2 for minimizing the estimation and prediction error in  a multiple regression model when $\sigma$ is known and the design matrix is orthogonal and normalized such that $X'X=I$.

\section{Model selection with the Sorted L-One Norm}\label{Sec:L1}

The last section has shown that there is a lot of theoretical underpinning for FDR controlling model selection criteria like mBIC2 or mAIC2. The main difficulty in applications with high dimensional data is the computational complexity. Identifying the model that minimizes any of these criteria is an NP-hard problem. In the context of genetic data,  which exhibit rather low range spatial correlations, very good results were obtained by certain modifications of a simple step-wise search \cite{GWAS, DBF14, admixtures}. Available software includes the C++ program {\it MOSGWA}\cite{DBF14} which is particularly designed to analyze GWAS data and the more generally applicable R package {\it bigstep} \cite{bigstep}. Another interesting possibility is to use some adaptive Ridge regression to obtain good models according to the modified information criteria \cite{FN16}.
However, all these  optimization algorithms give no guarantee that actually the optimal model has been identified. 

In contrast convex optimization problems have a unique solution which can be solved efficiently by a number of different algorithms. There is a strong interest in considering penalized likelihood methods of the form
\begin{equation}\label{con1}
\hat \beta=\rm{argmin}_{\beta} \{-\log \mathcal{L}(Y|X,\beta) +\|\beta\|\}\;\;,
\end{equation}
where 
$\|\cdot\|$ is some norm, because for classical generalized linear models (GLM) $\hat \beta$ then becomes a convex function of the parameter vector $\beta$. Note that in contrast to (\ref{PenLik}) we are no longer multiplying the log-likelihood term with a factor 2. This is quite common when working with $L_1$ penalties.

The most popular model selection procedure of this type is the LASSO (Least Absolute Shrinkage and Selection Operator, \cite{Santosa86, chen1994basis,LASSO}), which  uses the standard $L_1$ norm multiplied by a tuning parameter $\lambda$:
\begin{equation}\label{LASSO}
\hat \beta^L=\rm{\rm{argmin}}_{\beta} \left\{ \frac12 \|Y-X\beta\|_2^2 + \lambda \|\beta\|_1 \right\}\;\;,
\end{equation}
with
$\|\beta\|_1=\sum_{j=1}^p |\beta_j|$.
It is easy to check that in case of 
$X'X=I$ it holds that
$$\hat \beta^L_j = 0\;\;
\mbox{
	if and only if
}\;\;|X_j'Y|\leq \lambda\;,$$
and furthermore one has $X_j'Y \sim N(\beta_j,\sigma^2)$. Hence a Bonferroni like tuning parameter 
$\lambda=\lambda_{Bon}=\sigma \sqrt{2 \log p} (1+o_p)$ is needed to control FWER.
This provides the intuition why most of the theoretical results on  consistency and optimality of LASSO require that 
$\lambda$ is proportional to $\sqrt{\log p}$. 

Thus, similarly to mBIC or mAIC,  under orthogonality LASSO can be interpreted as a fixed threshold multiple testing  procedure. The theoretical results for multiple testing under sparsity show that procedures based on decaying sequences of thresholds (like Benjamini-Hochberg) perform better than fixed threshold rules (like Bonferroni). Furthermore we have seen that for high-dimensional model selection  mAIC and mBIC are outperformed by the non-linear penalties mAIC2 and mBIC2.  Therefore it is quite natural to consider replacing the single tuning parameter $\lambda$ from LASSO with a decaying sequence of tuning parameters.

This idea was used by Bogdan et al. 
\cite{SLOPE, SLOPE2} to propose the  SLOPE (Sorted L-One Penalized Estimation) procedure.
For any non-zero and non-increasing sequence 
$\lambda_1\geq \ldots \geq \lambda_p\geq 0$  the SLOPE estimator is given by 
\begin{equation}\label{SLOPE}
\hat \beta^{SL}=\rm{argmin}_{\beta} \left\{ \frac12 \|Y-X\beta\|_2^2 + J_{\lambda}(\beta)\right\}\;\;,
\end{equation}
where
$J_{\lambda}(\beta)=\sum_{j=1}^p \lambda_j |\beta|_{(j)}$,
and
$|\beta|_{(1)} \geq \ldots \geq |\beta|_{(p)}$ is the vector of sorted absolute values of elements of 
$\beta$. 
It is easy to check that the function $J_{\lambda}(b)=\sum_{j=1}^p\lambda_j\big|b\big|_{(j)}$
is a norm (see \cite{SLOPE,SLOPE2}) and hence (\ref{SLOPE}) can be solved with convex optimization tools.

\begin{figure}[t!]
	\begin{minipage}[t]{0.3\textwidth}
		\includegraphics[scale=0.3]{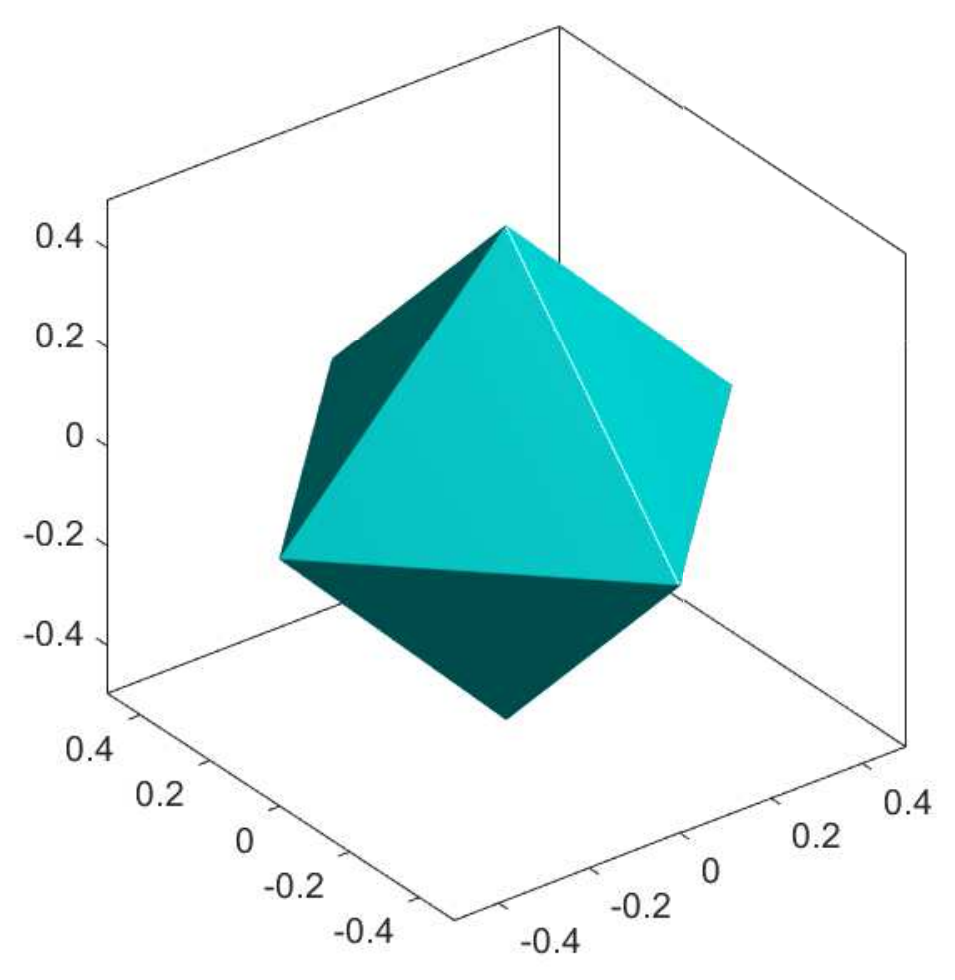}
		$\lambda=(2,2,2)$
	\end{minipage}
	\hspace{0.2cm}
	\begin{minipage}[t]{0.3\textwidth}
		\includegraphics[scale=0.3]{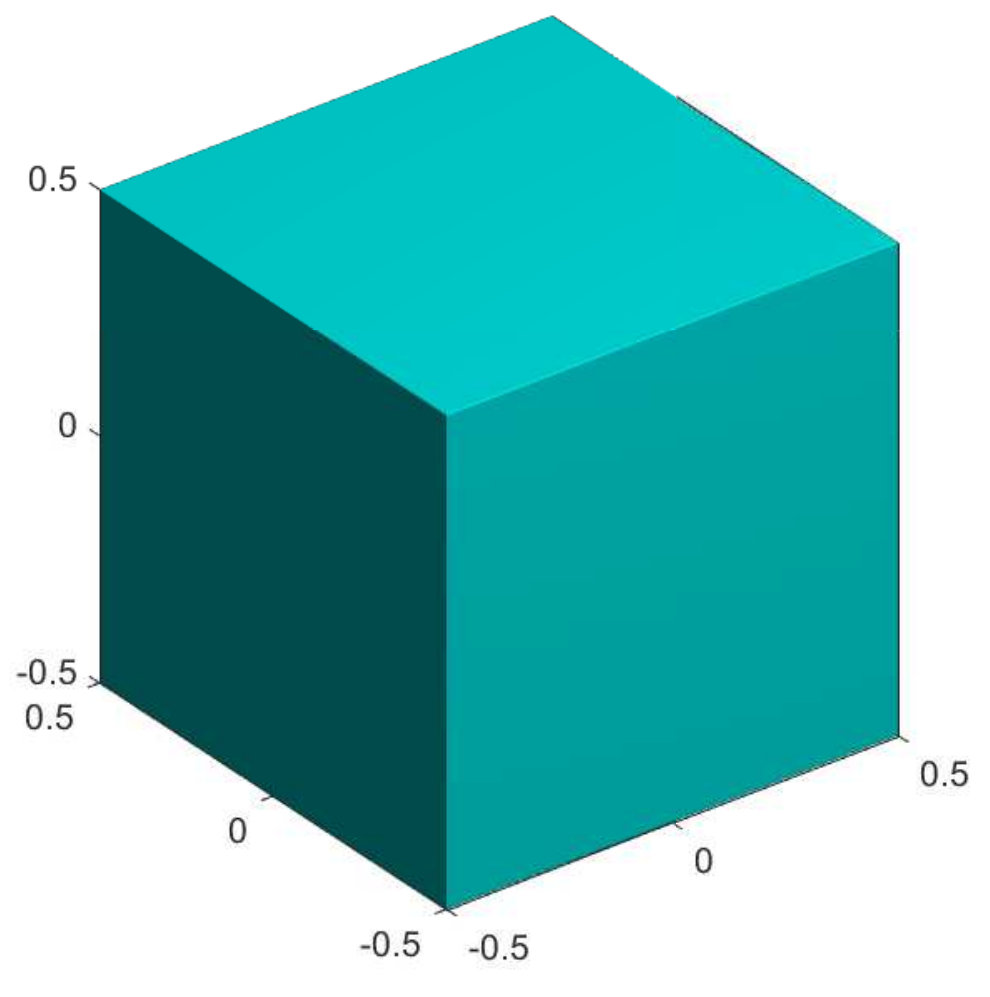}
		$\lambda=(2,0,0)$
	\end{minipage}
	\hspace{0.2cm}
	\begin{minipage}[t]{0.3\textwidth}
		\includegraphics[scale=0.3]{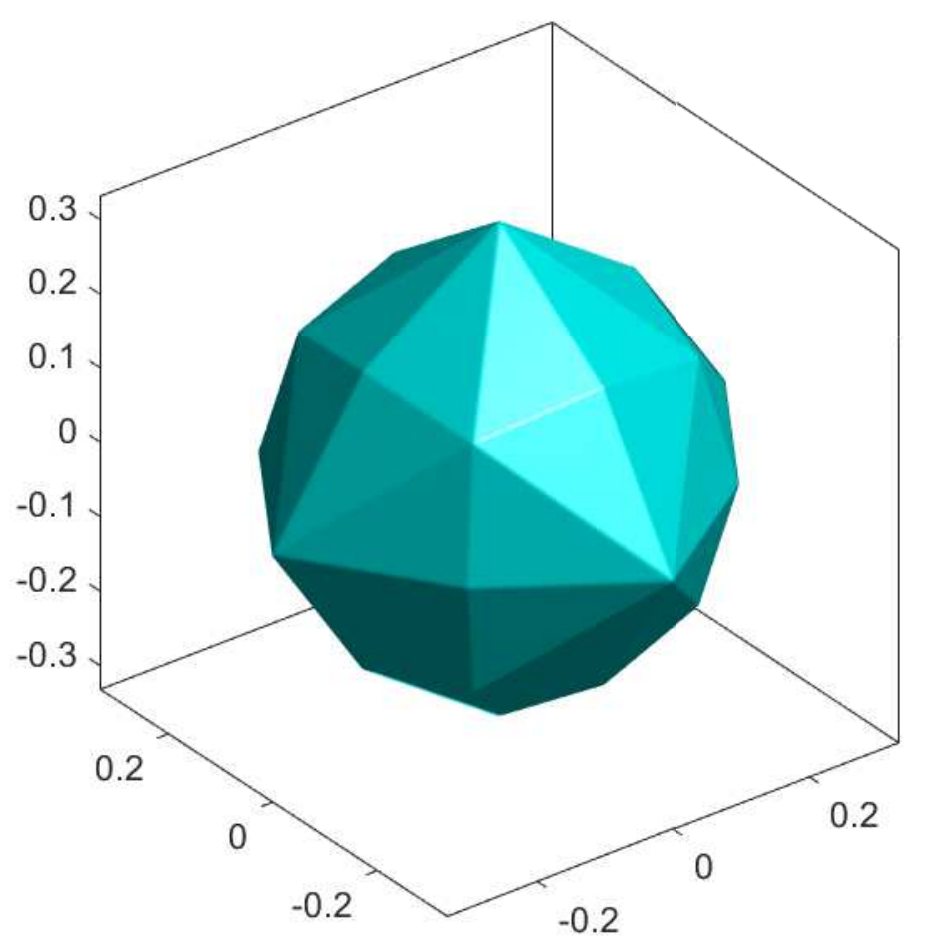}
		$\lambda=(3,2,1)$
	\end{minipage}
	\caption{Shapes of different SLOPE balls}
	\label{balls}
\end{figure}

Figure \ref{balls} illustrates different shapes of the unit balls corresponding to different versions of the Sorted L-One Norm. Since the solutions of SLOPE tend to occur on the vertices of respective balls, Figure 
\ref{balls} demonstrates large flexibility of SLOPE with respect to dimensionality reduction.
For $\lambda_1=\ldots=\lambda_p$ SLOPE coincides with LASSO and reduces dimensionality by shrinking the coefficients to zero. In contrast for  $\lambda_1>\lambda_2=\ldots=\lambda_p=0$ the reduction of dimensionality is performed by shrinking coefficients towards each other (since the vertices of the $l_{\infty}$ ball correspond to vectors $b$ such that at least two coefficients are equal to each other). When the sequence of thresholding parameters is monotonically decreasing SLOPE reduces the dimensionality both ways: it shrinks them towards zero and towards each other. Thus it returns sparse and stable estimators.

We are particularly interested in decaying parameter sequences which results in FDR controlling selection procedures. This is once again achieved by translating the thresholds from the BH multiple testing procedure into penalty parameters. The BH penalty (\ref{BH_penalty}) from Abramovich et al.  corresponds to the SLOPE parameter sequence
\begin{equation}\label{lBH}
\lambda^{BH}_j(c,q) = c\ \Phi^{-1}\left(1-\frac{jq}{2p}\right),\;\;j\in\{1,\ldots,p\},\;\;q\in(0,1) \; ,
\end{equation}
where $c$ is some tuning parameter to be discussed later and $q$ corresponds formally to the FDR level in BH. Note however that the  nominal FDR level of SLOPE will depend both on $c$ and $q$ which is the reason why we change here notation and do not use $\alpha$ in (\ref{lBH}). Also note that for the limit $q = 0$ the parameter sequence $\lambda^{BH}_j(c,0)$ is constant and the procedure turns into LASSO with tuning parameter $\lambda = c\ \Phi^{-1}(1)$.

A second order approximation, which was also used in \cite{ABDJ}, yields a similar sequence of tuning parameters of the form
\begin{equation}\label{lBH2}
\lambda_j \propto \sqrt{2\log (p/j)},\;\;j\in\{1,\ldots,p\} \;.
\end{equation}
The following two sections present different properties of SLOPE using these BH parameter sequences. We will first focus on prediction and then on model identification. In particular we will see that different choices of the tuning parameter $c$ in (\ref{lBH}) are necessary to achieve these different goals.

\subsection{Prediction properties of SLOPE}\label{predSLOPE}

To evaluate the estimation and prediction properties of SLOPE we consider two different mean squared errors.
With the notation $\hat \mu=X\hat \beta$  and $\mu=E(Y)=X\beta$, the mean squared prediction error is defined as
\begin{equation}\label{MSPE}
MSE(\hat \mu)=E ||X \hat \beta - X \beta||_2^2\;\;,
\end{equation} 
whereas for the coefficients we consider 
\begin{equation}\label{MSE}
MSE(\hat \beta)= E ||\hat \beta - \beta||_2^2\;\;,
\end{equation} 

For the convergence of $MSE(\hat \beta)$ some theoretical results for high dimensional linear and logistic regression are available  which were proven for SLOPE with the sequence (\ref{lBH2}), see e.g. \cite{su2016, Bellec, Felix}. 
Specifically under some assumptions on the sparsity of $\beta$ and the structure of the design matrix $X$, SLOPE 
achieves an asymptotic minimax rate  $k\log\left(\frac{p}{k}\right)$, where $k=||\beta||_0$ is the number of non-zero coefficients of $\beta$. Since the optimal rate of convergence of the LASSO estimator with a fixed tuning parameter $\lambda$ is only $k\log p$ one would expect that SLOPE can outperform LASSO in terms of the estimation rate for larger values of $k$. According to the following simulation study similar theoretical results should also hold for the mean squared prediction error. \\

\noindent \textbf{Simulation:} In the following simulation study the estimation and prediction properties of SLOPE and LASSO are compared. 
The sample size and number of potential predictors is $n=p=1000$. The rows of the design matrix are simulated as independent random vectors from a multivariate normal distribution $N(0, \frac{1}{n} \Sigma)$. In the first scenario predictors are {\it independent}, that means  $\Sigma = I$. In the second scenario {\it correlated} predictors are  simulated using a compound symmetry matrix with  $\Sigma_{i,i}=1$ and $\Sigma_{i,j}=0.5$ for $i\neq j$. Values of $Y$ are generated according to the linear model (\ref{LinearModel}) with 
\begin{displaymath}
\left\{
\begin{array}{l}
\beta_{1} = \dots = \beta_{k} = \sqrt{2 \log \left( \frac{1000}{k} \right)} \\
\beta_{k+1} = \dots = \beta_{1000} = 0\;\;,
\end{array} \right.
\end{displaymath}
with $\sigma=1$. For the model size two different values $k \in \{20, 100\}$ are considered. 
Estimation of $MSE(\hat \mu)$ and $MSE(\hat \beta)$ is based on 100 independent replicates of the whole experiment.

\begin{figure}[h!]
	\begin{minipage}[t]{0.5\textwidth}
		\centering
		independent, $k=20$
		\includegraphics[scale=0.7]{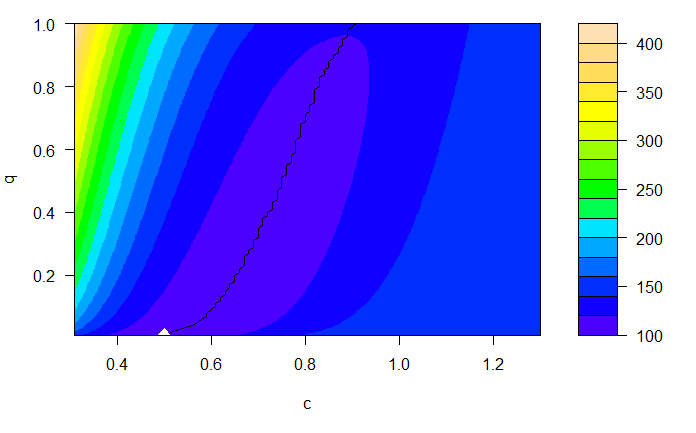}
	\end{minipage}
	\hspace{0.2cm}
	\begin{minipage}[t]{0.5\textwidth}
		\centering
		independent, $k=100$
		\includegraphics[scale=0.7]{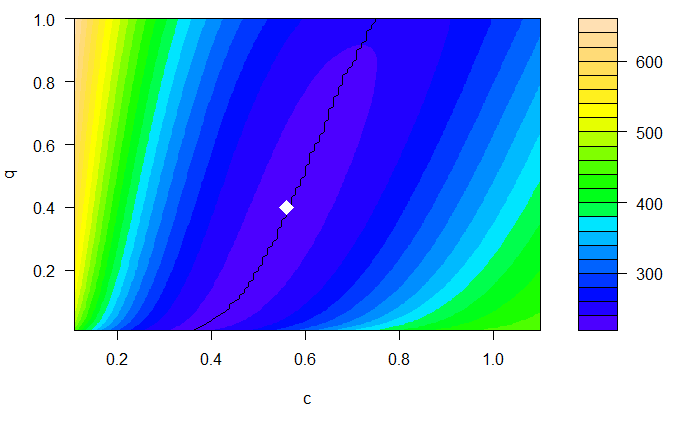}
	\end{minipage}
	\hspace{0.2cm}
	\begin{minipage}[t]{0.5\textwidth}
		\centering
		correlated, $k=20$
		\includegraphics[scale=0.7]{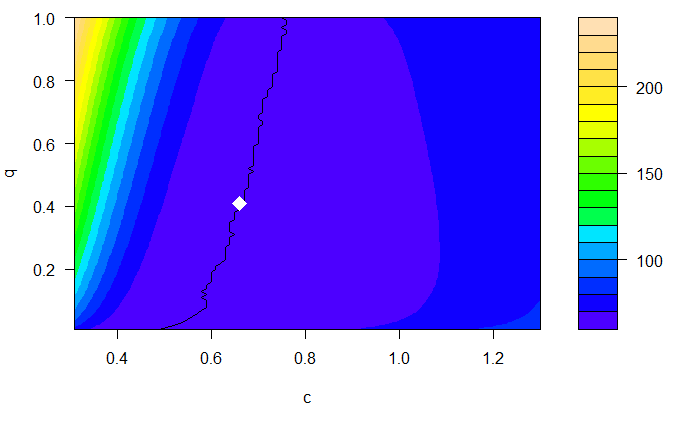}
	\end{minipage}
	\hspace{0.2cm}
	\begin{minipage}[t]{0.5\textwidth}
		\centering
		correlated, $k=100$
		\includegraphics[scale=0.7]{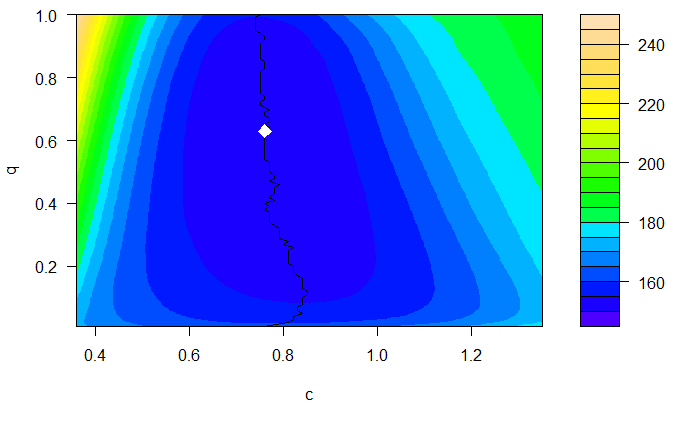}
	\end{minipage}
	\caption{Hypsometric maps of $MSE(\hat \mu)$ defined in (\ref{MSPE})  for SLOPE with the sequence of tuning parameters (\ref{lBH}) and different values of parameters $c$ and $q$. The white triangle marks the positions with the minimal $MSE(\hat \mu)$. For $q=0$ one obtains $MSE(\hat \mu)$ from LASSO. 
		Here $n=p=1000$ and $k$ denotes the number of nonzero elements in $\beta$. }
	\label{hyps}
\end{figure}

Figure \ref{hyps} presents  hypsometric maps of the prediction error of SLOPE with parameter sequence  (\ref{lBH})
using a range of values for $c$ and $q$.
The white triangle marks the combination of $(c,q)$ values for which $MSE(\hat \mu)$ is optimal.  For $\Sigma=I$ and $k=20$  
the optimal $q$ is very close to zero. This illustrates that LASSO has good prediction properties  when the  regressors are independent and $\beta$ is very sparse. 
However, when $k=100$ the optimal value of $q$ is close to 0.4 and  the prediction error of the optimal version of SLOPE is substantially smaller than the prediction error of the optimal version of LASSO. The advantage of SLOPE over LASSO is even more pronounced when regressors are correlated. Here SLOPE offers a much lower $MSE(\hat \mu)$ even when $k=20$. For $k=100$ the difference between these two methods becomes very large.

\begin{figure}[t]
	\begin{minipage}[t]{0.3\textwidth}
		\centering
		$\;\;\;\;\;\;\;\;$ independent, $k=100$
		\includegraphics[scale=0.7]{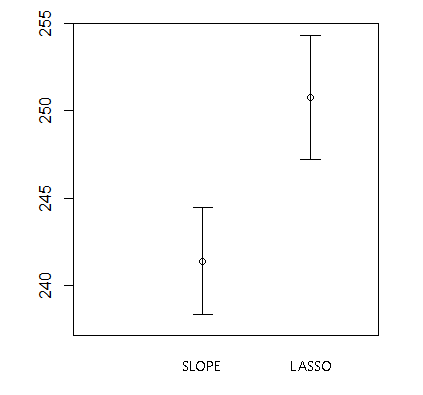}
	\end{minipage}
	\hspace{0.2cm}
	\begin{minipage}[t]{0.3\textwidth}
		\centering
		$\;\;\;\;\;\;$ correlated, $k=20$
		\includegraphics[scale=0.7]{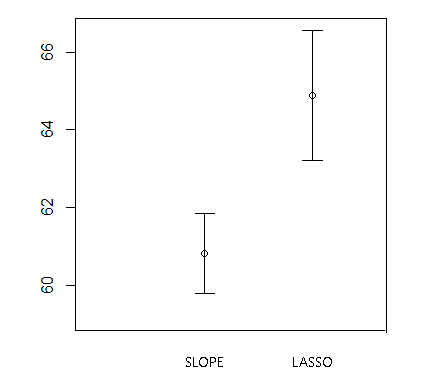}
	\end{minipage}
	\hspace{0.2cm}
	\begin{minipage}[t]{0.3\textwidth}
		\centering
		$\;\;\;\;\;\;$ correlated, $k=100$
		\includegraphics[scale=0.7]{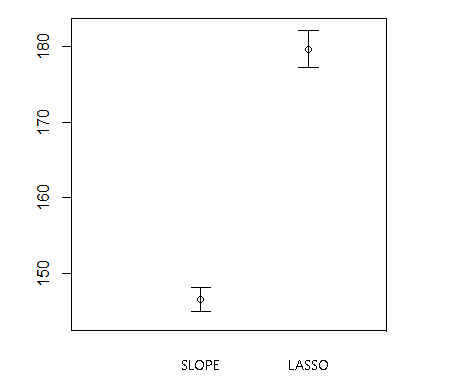}
	\end{minipage}
	\caption{95\% confidence intervals for MSE for optimal versions of LASSO and SLOPE based on 100 independent experiment replicates. }
	\label{CI}
\end{figure}

Figure \ref{CI} compares confidence intervals of $MSE(\hat \beta)$ for LASSO with optimal $\lambda$ and SLOPE with  optimal choice of $(c,q)$. Results for $k=20$ and independent regressors are not shown because in that case LASSO is more or less identical with SLOPE. For all other scenarios SLOPE is performing substantially better than LASSO.

These simulations illustrate the great potential of SLOPE, particularly when the signal is relatively strong and the number of true regressors is moderately large or when predictors are strongly correlated.  For the practical application of SLOPE with real data there remains to be explored to which extent the optimal values of $c$ and $q$ can be obtained via  cross-validation,  which can provide almost unbiased estimators of prediction error (see e.g., \cite{burman:89, yanagihara:06, fushiki:11}).

The good performance of SLOPE under correlated designs confirms the conjectures of Bondell and Reich \cite{Bondell2008}, who developed  OSCAR (Octagonal Selection and Clustering Algorithm for Regression), a predecessor of SLOPE. They proposed to use a linearly decaying sequence of tuning parameters to improve the properties of LASSO. According to results from extensive simulation studies (not reported here) SLOPE with the sequence $\lambda^{BH}$ (\ref{lBH}) usually performs substantially better than OSCAR. Here also the work of Zeng and Figueiredo \cite{Zeng2014} should be mentioned, who independently developed a version of SLOPE (called OWL) as an extension of OSCAR to deal with correlated designs.

\subsection{Model identification properties of SLOPE}

Like in case of other selection procedures it is to be expected that cross validation for SLOPE with $\lambda^{BH}(c, q)$   will result in  values of $c$ and $q$  which will  give too large models. This means the corresponding selection procedure will include too many false positive regressors. It has been shown  that under an orthogonal design with known error variance the choice of $c = \sigma$ provides FDR control at the level $q$  \cite{SLOPE,SLOPE2}. Unfortunately, this is no longer true if the inner products between columns of the design matrix are different from zero, which will almost always be the case in practical applications. 

LASSO suffers from a very similar problem and to better understand what is going on here it is worth looking at the general solution of the LASSO procedure:
\begin{equation}\label{restrict}
\hat \beta_j^L = \eta_\lambda(\beta_j + X_j' \epsilon  + v_j)\;\;,
\end{equation}
where $$\eta_\lambda(t) = sign(t) (|t| - \lambda)_+\;\;$$
and
$$ v_j = \left<X_j, \sum_{l \neq j} X_l (\beta_l - \hat \beta^L_l)\right>\;\;.$$

Only for an orthogonal design matrix 
$X$ it holds that $v_j = 0$, otherwise this term contributes to the variance of the estimator of $\beta$. The magnitude of this additional noise depends on the model sparsity, the bias of large regression coefficients and the inner products between columns of the design matrix. As long as the correlations and the number of non-zero elements in $\beta$ are small enough, this additional variance can be controlled by increasing the value of the tuning parameter $\lambda$. However, increasing $\lambda$ leads to increased bias and the whole process gets out of control when the number of non-zero elements in $\beta$ exceeds some limiting value. 

This phenomenon is captured e.g. by Theorem 2 of Wainwright \cite{wainwright2009}, which says that no matter how large the non-zero regression coefficients of the data generating model are, the probability that LASSO can identify the true model is smaller than 0.5 unless a stringent {\it irrepresentability} condition is satisfied  \cite{zhaoyu:06, conditions:2009, wainwright2009}. This condition in principle sets a limit on the sparsity of $\beta$, which depends on the correlations between columns in $X$. A thorough discussion of this condition, with examples of {\it irrepresentability} curves for different design matrices, can be found in \cite{Patrick}. These issues were also thoroughly analyzed in \cite{SLOPE, FDR_LASSO} for design matrices with i.i.d. standard normal columns. Specifically,  Bogdan et al. \cite{SLOPE} used the theory of Approximate Message Passing Algorithms  \cite{BM12} to  provide sparsity limits needed for FDR control when LASSO is used with an arbitrary but fixed $\lambda$.    Su et al. \cite{FDR_LASSO}  discuss the trade-off between FDR and Power provided by LASSO when $\lambda$ is chosen adaptively based on the data.  

The theory describing the limitations of SLOPE for model identification still needs to be fully developed, but its behavior under orthogonality and the cited results for LASSO suggest that SLOPE can efficiently control FDR if $\beta$ is sparse enough and the regressors are roughly independent. There are some theoretical results available which point in that direction.  Kos \cite{Kos2} considered SLOPE with a sequence of tuning parameters $(1+\delta_n) \lambda^{BH}$, where $\delta_n$ is slowly converging to zero. and showed  that it asymptotically controls FDR at the level $q$ when the design matrix is random with uncorrelated predictors and $p$ is fixed while $n$ diverges to infinity.
Kos \cite{Kos2} and Kos and Bogdan \cite{Kos1} prove also that if the columns of the design matrix are i.i.d random variables from a Gaussian distribution then SLOPE with the sequence of tuning parameters 
\begin{equation}\label{ldelta}
(1+\delta) \lambda^{BH}\;\;, \delta>0 \;,
\end{equation}
has FDR converging to zero and power converging to one if the number of true nonzero regression coefficients $k$ satisfies
\begin{equation}\label{as2}
\frac{k^2 \log p}{n}\rightarrow 0\;\;,
\end{equation}
and the magnitude of these nonzero coefficients is large enough.

To improve FDR controlling properties of SLOPE, Bogdan et al. \cite{SLOPE}  used equation (\ref{restrict}) to derive a heuristic adjustment of the $\lambda^{BH}$ sequence which is well justified for design matrices with i.i.d normally distributed columns:
\begin{equation}\label{heuristics}
\lambda^{ad}_i(q) = \Bigg \{\begin{array}{ccc}
\sigma \Phi^{-1}(1-q/2p) & if & i=1\\
\min\left(\lambda_{i-1}, \sigma\Phi^{-1}(1-qi/2p)\sqrt{1+\frac{\sum_{j<i}\lambda_j^2}{n-i-2}}\right) & if & i>1\;\;.
\end{array}
\end{equation}

\begin{figure}[h!]
	
	\centering
	\includegraphics[width=0.47\textwidth]{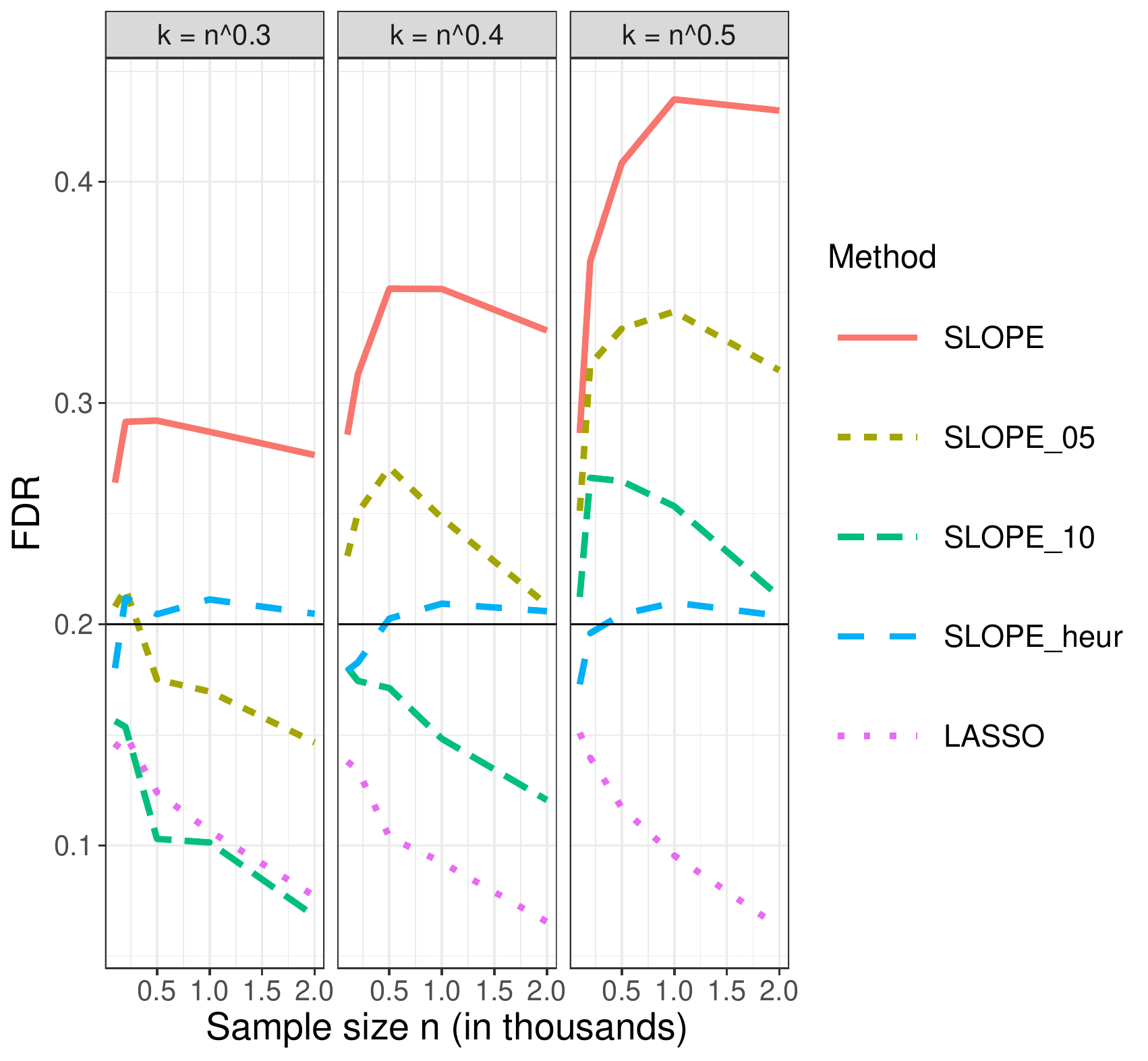} 
	\includegraphics[width=0.47\textwidth]{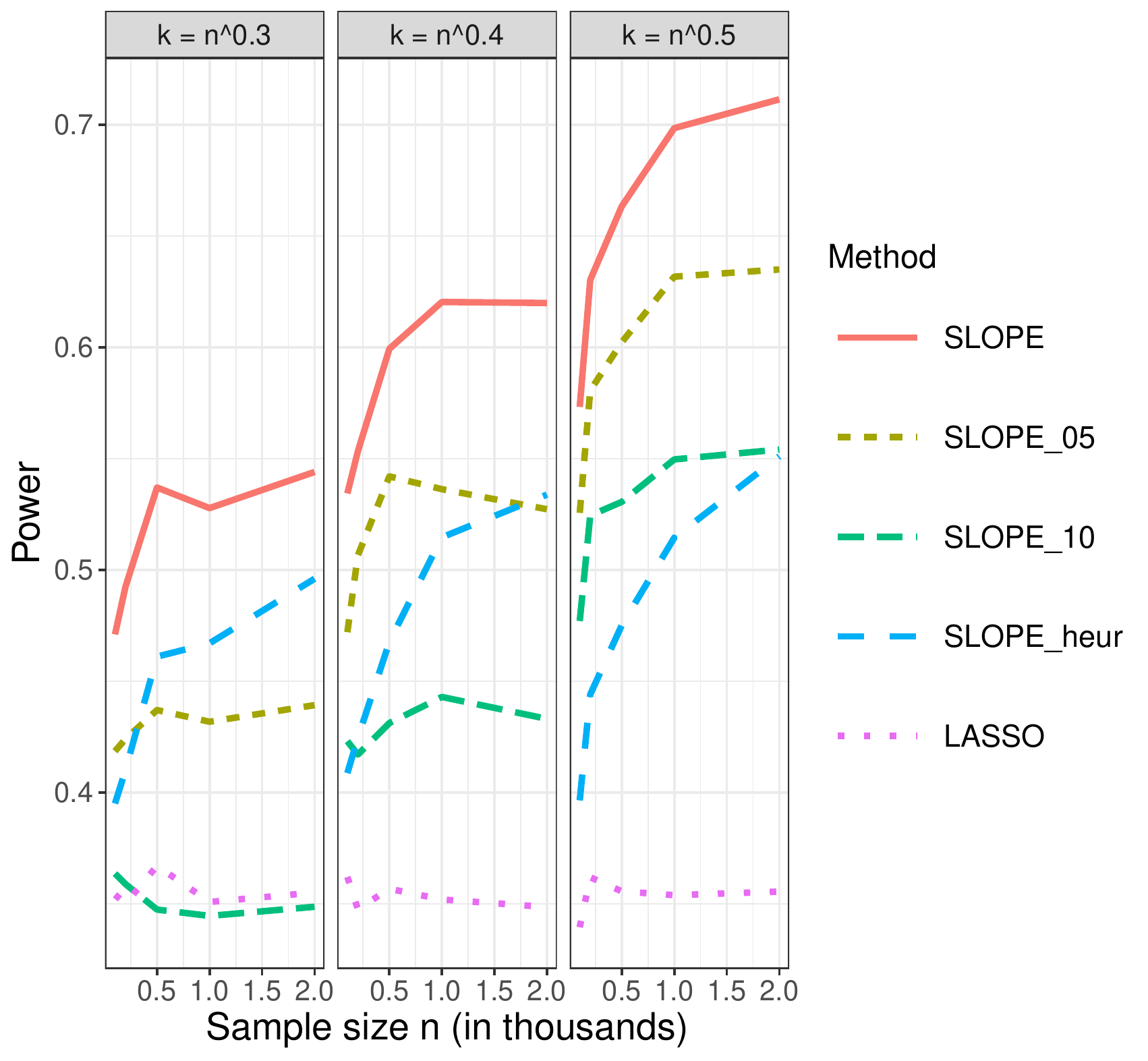} 
	
	\caption{Comparison of LASSO and four different SLOPE versions. In the figure labels SLOPE refers to the tuning sequence $\lambda^{BH}(\sigma, q)$, SLOPE\_05 and SLOPE\_10 to the adjusted sequence (\ref{ldelta}) with $\delta = 0.05$ and  $\delta = 0.1$, respectively and SLOPE\_heur  to the heuristic sequence (\ref{heuristics}).   FDR and Power are given as a function of $n$ for $q=0.2$ and for three different sparsity levels $k=round(n^{\alpha})$. Estimates were obtained by averaging the false or true positive rates over  500 independent simulation replicates.}
	\label{FDR_small}
\end{figure}

\ \\

\noindent  \textbf{Simulation:}  We want to compare the performance of LASSO and SLOPE for model identification in different simulation scenarios. For SLOPE we use four different tuning parameter sequences: $\lambda^{BH}(\sigma, q)$ according to (\ref{lBH}), then  $(1 + \delta) \lambda^{BH}(\sigma, q)$ according to (\ref{ldelta}) with $\delta = 0.05$ and  $\delta = 0.1$, respectively; and finally  $\lambda^{ad}(q)$ according to (\ref{heuristics}). As tuning parameter for LASSO, the Bonferroni threshold $\lambda = \Phi^{-1}\left(1-\frac{q}{2p}\right)$  was chosen.

Simulations were performed as before using a linear model with design matrix $X$ having independent Gaussian columns. The signal magnitude to generate $Y$ was $\beta_1=\ldots=\beta_k=0.9\sqrt{2 \log p}$. To study the asymptotic behavior of different procedures simulations were performed for sample size $n \in \{100, 200, 500, 1000, 2000\}$. The number of potential regressors was set accordingly to $p=0.05 n^{1.5}$. Three scenarios of different sparsity levels were studied by setting the model size to $k=round(n^{\alpha})$, with  $\alpha \in \{0.3, 0.4, 0.5\}$.

Figure \ref{FDR_small} presents FDR and  Power as a function of $n$ for $q=0.2$. 
For $\alpha = 0.3$ and $\alpha = 0.4$  assumption (\ref{as2}) is satisfied for the two SLOPE versions with $\delta>0$, and it is noticeable that in those cases FDR  seems to converge and is close to or below the nominal level $q=0.2$ for the whole range of considered values of $n$. For $\alpha = 0.5$ the assumption is violated FDR is still a decreasing function of $n$ but the rate of this decrease is rather slow and it is difficult to predict if it would converges to 0 with increasing $n$.  

Apparently larger values of $\delta$ lead  to more conservative versions of SLOPE. For $\delta=0$ one has the original $\lambda^{BH}$ sequence and FDR is above the nominal level. For $\alpha<0.5$ FDR is slowly decreasing with $n$  but it remains to be checked if it actually converges to $q=0.2$. When $\delta=0$ and $\alpha=0.5$ FDR seems to stabilize substantially above the nominal value of 0.5, which suggests that condition (\ref{as2}) is indeed necessary  for the asymptotic FDR control with the original $\lambda^{BH}$ sequence.  

Figure \ref{FDR_small}  shows that LASSO with Bonferroni tuning parameter is substantially more conservative than all the different versions of SLOPE. In all three scenarios its FDR converges to zero, which in case of such moderate signals leads to a substantial decrease in power compared with SLOPE. 
Interestingly, SLOPE with the heuristic choice of tuning parameters provides a stable FDR control for all considered scenarios.
This suggests that the upper bound on $k$ provided in assumption (\ref{as2}) could be relaxed when working with this heuristic sequence. 
In practical applications SLOPE with the heuristic sequence (\ref{heuristics}) has been shown to control FDR when the design matrix contains genotypes of independent or strongly correlated SNPs and the number $k$ of non-zero elements in the sequence $\beta$ is sufficiently small \cite{SLOPE,SLOPE2}.

\subsection{Extensions of SLOPE}

\noindent  \textbf{Clustered regressors:}  In \cite{geneSLOPE} SLOPE was combined with an algorithm for clustering of correlated SNPs to control FDR for spatially correlated GWAS data. After a preliminary selection of representatives of groups of correlated genetic markers, SLOPE is used to identify significant representatives. It is proven that the suggested method of identifying representatives does not impair FDR control. The method has been implemented in the publicly available package {\it geneSLOPE}, and the simulations presented in 
\cite{geneSLOPE} show good properties of {\it geneSLOPE} in terms of FDR control and power to identify relevant genes. 

It is possible to go one step further and use  SLOPE to select groups of predictors \cite{grpSLOPE}.  Let
$I=\{I_1,\ldots,I_m\}$ be a partition of the set 
$\{1,\ldots,p\}$ and rewrite the multiple regression model (\ref{LinearModel}) in the form
\begin{equation}
\label{gmodel}
y=\sum_{i=1}^mX_{I_i}\beta_{I_i}+ \epsilon \; ,
\end{equation}
\noindent where 
$X_{I_i}$ is a sub-matrix of 
$X$ consisting of columns with indices from the set 
$I_i$, 
and
$\beta_{I_i}$ consists of elements of
$\beta$  with indices from 
$I_i$.
Then the norm 
$\|X_{I_i}\beta_{I_i}\|_2$ will serve as a measure for the influence of the
$i$-th group on the response variable. We will say that the
$i$-th group has an impact on 
$Y$ if and only if 
$\|X_{I_i}\beta_{I_i}\|_2>0$.
Thus, the task of identifying significant groups of variables boils down to identifying the support of the vector 
$$[[\beta]]_I:= \big(\|X_{I_1} \beta_{I_1}\|_2, \ldots, \|X_{I_m} \beta_{I_m}\|_2\big)^\mathsf{T}\;\;.$$

For any non-negative and non-increasing sequence   
$\lambda_1,\ldots,\lambda_m$ and any positive sequence of weights 
$w_1,\ldots,w_m$ the group SLOPE (gSLOPE) estimator is defined as
\begin{equation}\label{gSLOPE}
\beta^\ES{gS}: = argmin_b\ \ \Big\{\frac 12\big\|y-Xb\big\|_2^2+\sigma J_{\lambda}\big( W[[b]]_I\big)\Big\},
\end{equation}
where 
$W$ is a diagonal matrix with elements 
$W_{i,i}:=w_i$.

FDR control can then be obtained by selecting a sequence 
$\lambda$ compatible with chi-square distribution quantiles  \cite{grpSLOPE}. A conservative selection of this sequence allows FDR control when the columns of the design matrix belonging to different groups are mutually orthogonal. In that case asymptotic optimality of estimation of $[[\beta]]_I$ has  been proved.
In addition, a heuristic adaptation of the sequence $\lambda$ has been proposed 
which allows for FDR control when variables in different groups are mutually independent. 

$gSLOPE$ has been successfully applied to the problem of gene localization, where groups consist of two variables corresponding to the additive effect and the dominance effect of a given gene. The proposed method works particularly well for identifying  so-called rare recessive variants, where the dominance effect is of particular importance. \\

\noindent  \textbf{Outlier detection:} 
One approach to outlier detection and robust estimation of regression coefficients is the mean-shift model  \cite{mean_shift2,mean_shift1,mean_shift3,ipod}:
\begin{equation}
Y = X\beta + I \mu + \varepsilon\;\;,
\label{model}
\end{equation}
where $I$ is the $n\times n$ identity matrix, $\mu=(\mu_1,\ldots,\mu_n) \in R^n$ and $\mu_i\neq 0$ means that observation $i$ is an outlier.
In \cite{outliers} an extension of SLOPE is used to estimate $\beta \in R^p$ and $\mu=(\mu_1,\ldots,\mu_n)$ according to 
\begin{equation}
\label{pen}
(\hat \beta, \hat\mu) =argmin_{\beta \in R^p, \mu \in R^n} \bigg\{\Vert y - X \beta - \mu \Vert_2 ^2 + 2\rho_1 J_{\tilde{\lambda}}(\beta)  + 2\rho_2 J_\lambda (\mu) \bigg\}\;\;,
\end{equation}
where $\rho_1$ and $\rho_2$ are two positive constants. It is shown that under a proper adaptation of the Restricted Eigenvalue condition \cite{bickel:09, geer:08, yezhang:10}
SLOPE with a sequence of tuning parameters proportional to $\lambda^{BH}$ or to the sequence with the elements $\lambda_i=\sigma \sqrt{\log \Big(\frac{2n}{i}\Big)}$ satisfies
\begin{eqnarray}
\Vert\hat{\beta}-\beta^\ast\Vert_2^2 + \Vert\hat{\mu}-\mu^\ast\Vert_2^2 &\leq&C_1 \left(k \log\left(\frac{p}{k}\right)+s\log \left(\frac{n}{s}\right)\right),
\end{eqnarray}
where $s$ is the number of outliers. Thus,  when $p>n$ and the number of outliers $s$ is smaller than the number of non-zero coefficients of $\beta$, the version (\ref{pen}) of SLOPE  for the mean-shift model allows to obtain a minimax rate $k \log\left(\frac{p}{k}\right)$ to estimate $\beta$. Moreover, it was shown that SLOPE with the sequence (\ref{ldelta}) asymptotically controls FDR with respect to outlier detection if 
\begin{equation*}
\frac{\left(s\log (n/s) \vee k\log (p/k)\right)^2}{n} \rightarrow 0.
\end{equation*}
Here FDR control can be obtained even when the columns of the design matrix $X$ are strongly correlated. This is due to the fact that  in the mean-shift model (\ref{model}) the vector $\mu$ is accompanied by the identity matrix $I$.   \\

\section{Advanced methods for model identification and prediction}\label{Sec:Advanced}

Convex optimization methods, like LASSO or SLOPE, are shrinking estimates of regression coefficients towards zero. 
Model size is reduced when the values of tuning parameters $\lambda$ are getting large enough that coefficients are shrunk to zero. However, large values of $\lambda$ result in shrinkage of all parameters and consequently in a large bias of the estimators of the most important regression coefficients. It follows that estimation and prediction properties can be rather poor as we have seen in the real data example.

Therefore it is practically impossible to tune LASSO or SLOPE in such a way that one obtains both good prediction and selection properties at the same time. One practical solution suggested  consist of applying a two stage procedure \cite{SLOPE, grpSLOPE}.  \textit{i)} Use LASSO or SLOPE to detect significant predictors; \textit{ii)} apply  standard least-squares methods for the selected predictors to estimate coefficients. This two stage procedure allows to correct for bias of LASSO and SLOPE estimates under a variety of scenarios, but it still does not solve the problem of deterioration of the model selection properties of these methods, which occurs when the number of non-zero coefficients in the true vector of regression coefficients is sufficiently large. A brief look at the term $v_j$ from equation(\ref{restrict}) is enough to see that this additional noise will typically increase with $\lambda$, the correlation between columns of the design matrix $X$ and the sparsity of the true vector of regression coefficients. When the noise variable $v_j$ becomes too large then LASSO or SLOPE are not capable of recovering the true order of the magnitude of regression coefficients and will therefore not classify true and false discoveries correctly. This in turn results in a deterioration of the model selection properties. 

Two different solutions have been developed to solve this problems for LASSO. In the adaptive or reweighted LASSO \cite{zou2006adaptive, candesreweighted} the values of the tuning parameters become different for different variables and depend on some initial estimators of the vector of regression coefficients. Large expected values of regression coefficients are assigned smaller weights, which results in debiasing the large regression coefficients and decreasing the value of the noise variable $v_j$. This allows the adaptive LASSO to recover the true model for a much wider range of realistic scenarios than the regular LASSO. The second solution relies on applying LASSO with a relatively small value of the tuning parameter $\lambda$, such that the ordering of false and true discoveries is optimal. False discoveries are then eliminated by using an appropriate threshold or some model selection criterion. Specifically, in \cite{pokmiel:15} the consistency of such a two-step procedure with thresholding based on the Generalized Information Criterion (GIC, \cite{GIC}) is proved and very good properties of the Extended BIC or Modified BIC are shown via simulations. We have also seen in our real data example that this is a viable strategy. Good model selection properties of adaptive and thresholded LASSO are reported in \cite{Patrick, Wojtek}. In case when one aims at FDR control then thresholding can also be performed by using the knockoff filter \cite{ko,knockrandom}, which provably controls FDR.

The rest of this chapter is devoted to have a look at these advanced methods.
First we will briefly describe the Adaptive Bayesian Slope, which uses the Bayesian framework for the selection of weights in reweighted SLOPE. Then we will introduce the knockoff methodology. A final short simulation study will compare these different methods of convex optimization with respect to  model identification and estimation properties.  

\subsection{Adaptive Bayesian SLOPE}

To address the described problems with  model identification and estimation properties of SLOPE,  a new synergistic procedure called adaptive Bayesian SLOPE (ABSLOPE) was proposed \cite{wei}. ABSLOPE effectively combines SLOPE  
with the Spike-and-Slab LASSO method of \cite{spikeslabLASSO}, which provides a Bayesian version for LASSO reweighting. ABSLOPE  performs simultaneous variable selection and  parameter estimation based on  data which can contain missing values. 
As with the Spike-and-Slab LASSO,  the regression coefficients are regarded as arising from a hierarchical model consisting of two groups: (1) the spike for the nonactive variables or negligibly small signals and  (2) the slab for  large signals. In contrast to the Spike-and-Slab LASSO, the ''ABSLOPE'' spike prior is fixed and relies on the sequence $\lambda^{BH}$ of the SLOPE tuning parameters in order to control FDR.  Other prior parameters like the signal sparsity  or the ''average'' magnitude of the large signals are treated as  latent variables and are iteratively updated in the spirit of a Stochastic Approximation EM algorithm (SAEM). The algorithm can handle  missing data under the  Missing at Random (MAR) assumption and estimates the variance of the error term.   
The prior is designed in such a way that the Maximization step of the algorithm is performed by invoking a  reweighted SLOPE, with weights dependent on the current conditional probability that a given variable is a large signal. According to simulation results in \cite{wei}, ABSLOPE allows to control FDR under a much wider range of scenarios than SLOPE. One also obtains good model identification  and estimation properties at the same time.

\subsection{Model selection with Knockoffs}\label{Sec:Knockoff}

In  \cite{ko,knockrandom} the, so called, {\it knockoff} methodology was proposed to control the number of false discoveries. The method can be used with almost any measure of importance for the explanatory variables, like e.g. the marginal correlation with the response variable or the estimate of regression coefficients provided by any regularization method. The main idea is to create a matrix of {\it fake} explanatory variables in such a way that its correlation structure corresponds to the correlation structure of $X$. More specifically, in case of a random design matrix $X$ with independent rows, swapping any set of columns $X$ with the same columns in $\tilde X$ should not affect the distribution of the extended matrix and the fake variables should be conditionally independent of $Y$, given $X$. The knockoff matrix $\tilde X$ is attached to $X$ and the method to evaluate the importance of explanatory variables is run on the extended design matrix. 

Knockoffs are then used to define a measure of importance $W_j$, $j\in \{1,\ldots,p\}$ in such a way that the signs of $\{W_j:\beta_j = 0\}$ are i.i.d.~coin flips. Here, it is important to note that this property is usually not satisfied for statistics calculated based on the matrix $X$ only, since in this case the sign of $\beta_j$ will depend on the correlations between $X_j$ and true predictors.
To construct our importance measure, suppose that a statistic 
\begin{equation}\label{eq:tstat}
T=(U, \widetilde{U}) = (U_1,...,U_p,\widetilde{U}_1, ..., \widetilde{U}_p)
\end{equation}
is computed from $(Y,X, \tilde X)$, where $T$ has the natural property that swapping the $j$ and $j+p$ columns in $\mathbb{X}$ results in swapping the corresponding components of $T$. 
Suppose  that the feature importance statistics are formed as
\begin{equation}
W_j = f(T_j, T_{j+p}),\ \ \ \ j=1,...,p
\end{equation}
where $f$ is an anti-symmetric function; for example, we can take $W_j = T_j - T_{j+p}$. 
Then it is easy to see that the statistic $W = (W_1,...,W_p)$ has a {\it flip-sign} property, namely, swapping the $j$ and $j+p$ columns in $\mathbb{X} = (X, \tilde X)$ has the effect of changing the sign of $W_j$.
Then, according to the results of \cite{ko, knockrandom}  the knockoff filter defined as
\begin{equation}
\widehat{\mathcal{S}} = \{j: W_j\geq \hat t\}\;\;,
\end{equation}
where
\begin{equation}\label{eq:that}
\hat t = \min\left\{ t>0: \frac{1+\#\{j:W_j\leq -t\}}{\#\{j:W_j\geq t\}} \leq q \right\}\;\;,
\end{equation}
controls FDR at the level $q$.

Knockoff thresholding allows for FDR control with SLOPE or LASSO for any choice of  tuning parameters. Here the tuning parameter should be selected to provide a proper ranking of explanatory variables rather than to execute their selection.

\subsection{Simulation study}\label{Sec:SimAdvanced}

In this section we provide results of a simulation study comparing different convex optimization methods with mBIC2 for selection of important variables and estimation of parameters in the multiple regression model \ref{LinearModel}. In all these simulations $n=p=500$, the error $\epsilon$ is iid normally distributed $N(0,I)$  
and the number $k$ of nonzero regression coefficients in the vector $\beta$ takes values from the set $k\in \{10,20,40,60,80,100\}$.
We consider weak signals with
\begin{equation}\label{weak}
\beta_1=\ldots=\beta_k=1.3 \sqrt{2 \log p}
\end{equation}
and strong signals with
\begin{equation}\label{strong}
\beta_1=\ldots=\beta_k=2 \sqrt{2 \log p}.
\end{equation}
The rows of the design matrix are generated as independent random vectors from a multivariate normal distribution $N\left(0,\frac{1}{n}\Sigma\right)$. We consider two scenarios, one with {\it independent} regressors, where the correlation matrix $\Sigma=I$, the other one with {\it correlated} regressors, where   $\Sigma$ is the compound symmetry matrix with $\Sigma_{i,j}=0.5$ for $i\neq j$. 
\\[2mm]
We compare five different estimation and model selection methods: 
\begin{itemize}
	\item{\bf mBIC2}: mBIC2 with an advanced stepwise search procedure implemented in the {\it bigstep} package,
	\item {\bf SLOPE}: SLOPE with the vector of tuning parameters (\ref{heuristics}) with $q=0.2$ for the {\it independent} scenario and the regular $\lambda^{BH}$ sequence (\ref{lBH}) with $c=1$ and $q=0.2$ for the {\it correlated} scenario,
	\item {\bf SLOBE}: a simplified version of ABSLOPE described in Section 3.4  of \cite{wei}  with $q=0.1$,
	\item {\bf Lcv}: LASSO with $\lambda$ selected by 10-fold cross-validation aimed at minimizing the prediction error,
	\item {\bf knLcv}: model free knockoffs \cite{knockrandom} based on estimates of LASSO applied to the augmented design matrix $[X,\tilde X]$ and with the tuning parameter $\lambda$ selected by cross-validation. For the independent case the iid rows of the knockoff matrix are generated from $N(0, I_{p\times p})$ distribution. For the correlated case the knockoff matrix is generated using the equicorrelated construction from Section 3.4.2 of \cite{knockrandom} with the parameter $s$ equal to the minimal eigenvalue of $\Sigma$. According to our simulations this choice of $s$ allows to achieve a higher power in the considered example than the choice suggested in \cite{knockrandom}.
\end{itemize}

The performance of the different methods to identify models correctly is assessed like in the previous simulation study by the FDR and Power. These are estimated based on 200 simulation runs. Additionally we consider here the relative estimation and prediction errors, defined as
$$MSE=\frac{MSE(\hat \beta)}{||\beta||^2}$$ 
and
$$MSP= \frac{MSE(\hat \mu)}{||X\beta||^2}\;.$$

\begin{figure*}
	\centering
	\begin{minipage}[t]{0.4\textwidth}
		\includegraphics[scale=0.3]{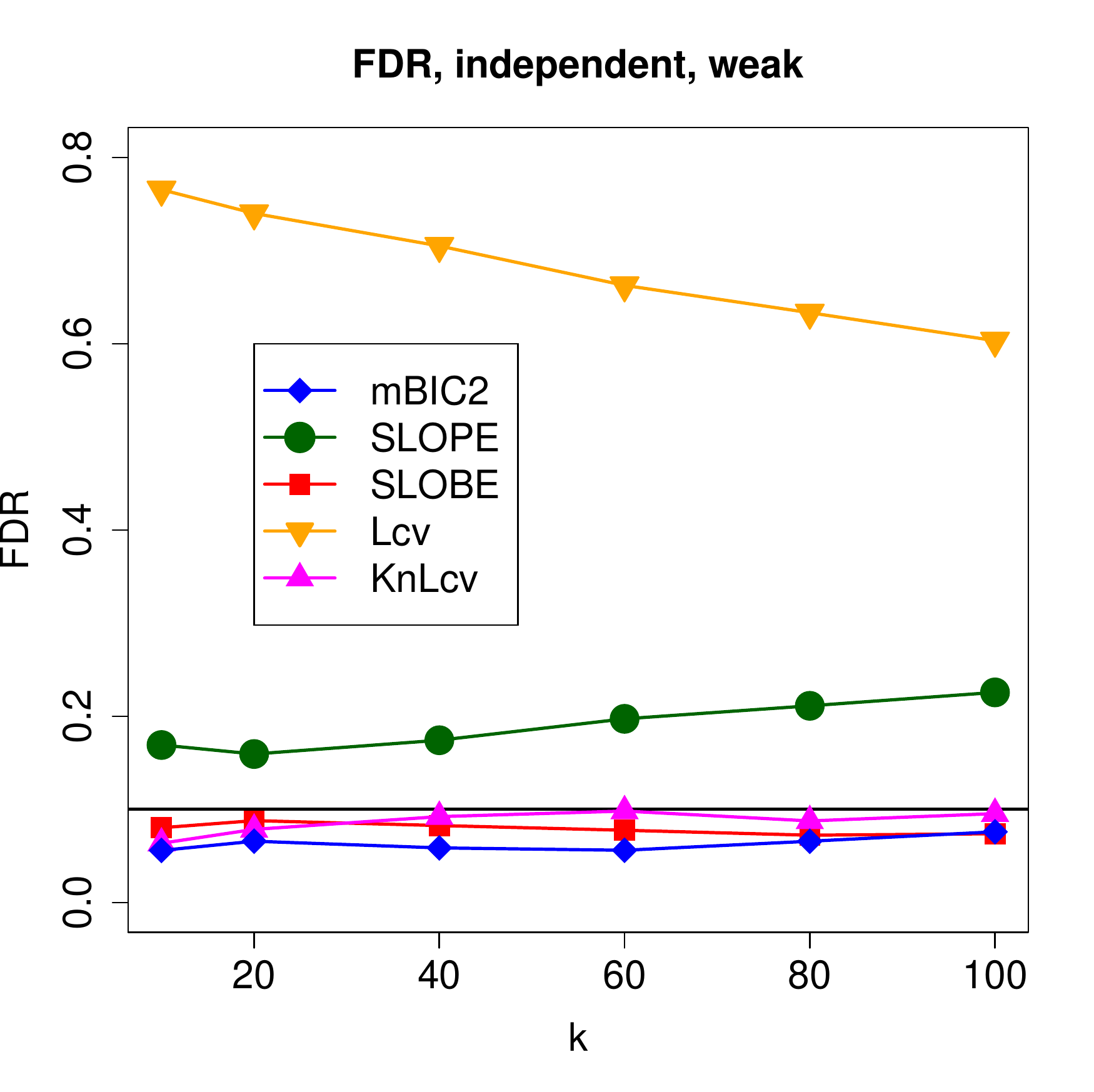}
	\end{minipage}
	\hspace{0.1cm}
	\begin{minipage}[t]{0.4\textwidth}
		\includegraphics[scale=0.3]{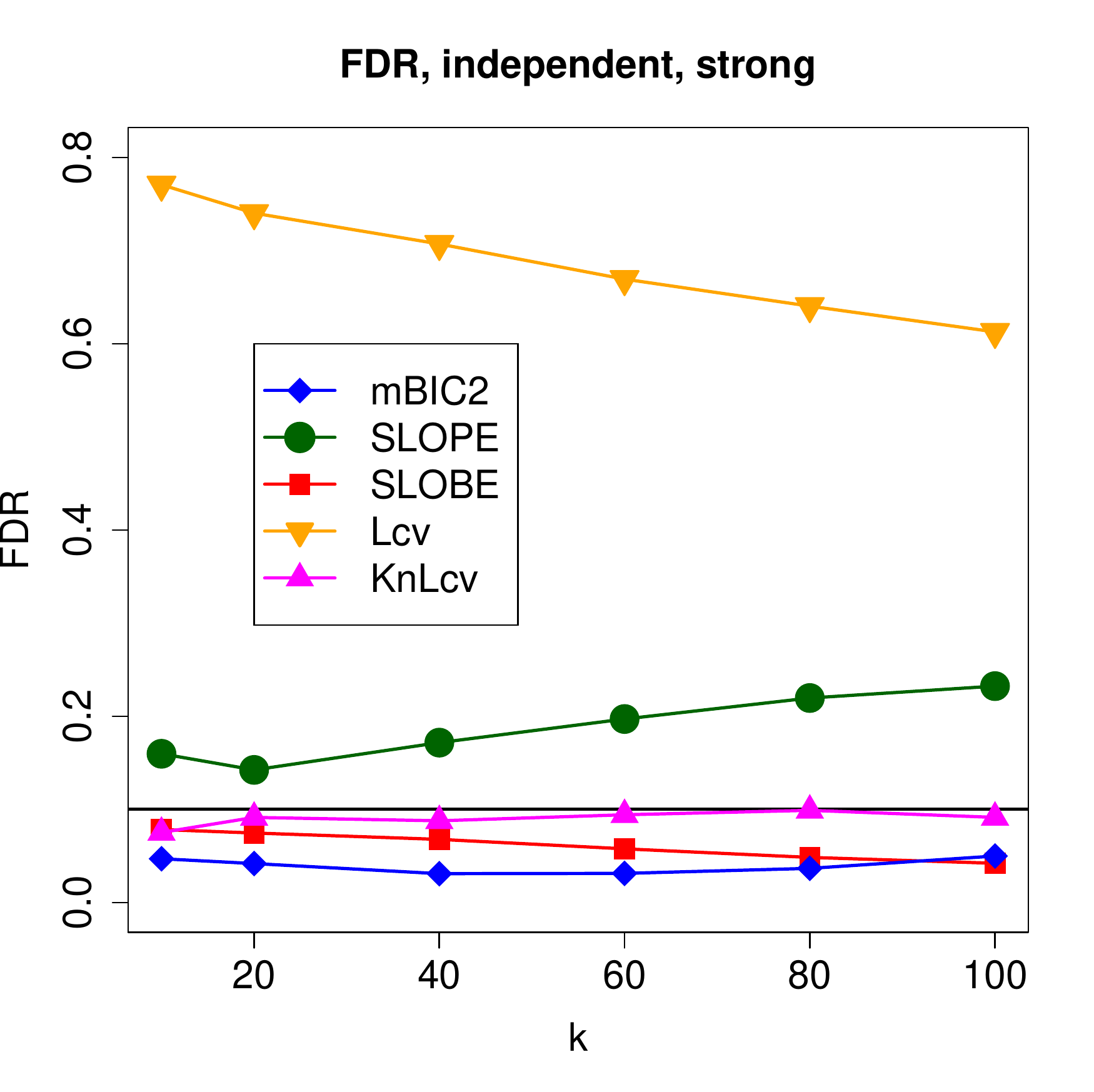}
	\end{minipage}
	\begin{minipage}[t]{0.4\textwidth}
		\includegraphics[scale=0.3]{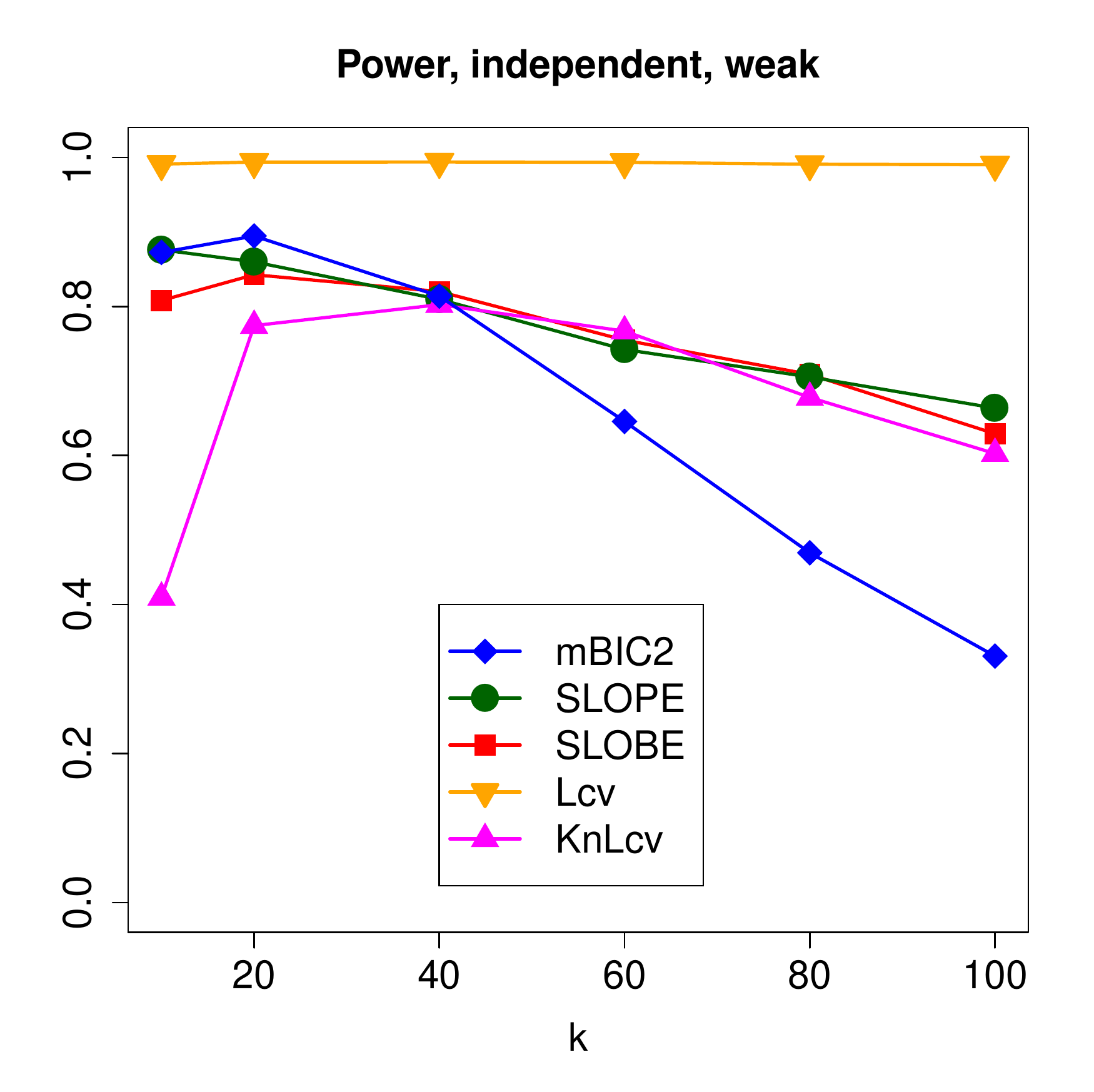}
	\end{minipage}
	\hspace{0.1cm}
	\begin{minipage}[t]{0.4\textwidth}
		\includegraphics[scale=0.3]{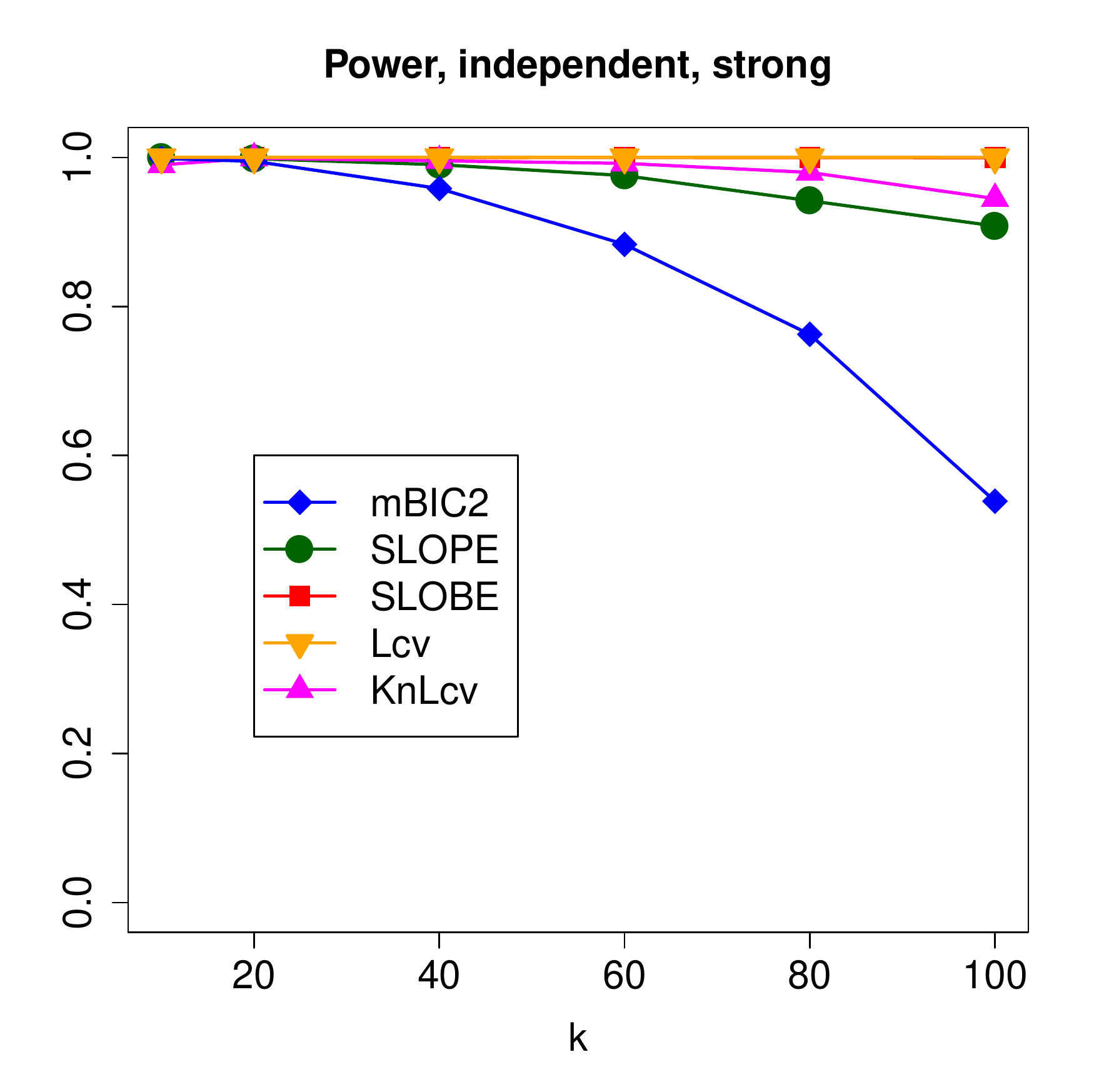}
	\end{minipage}
	\centering
	\begin{minipage}[t]{0.4\textwidth}
		\includegraphics[scale=0.3]{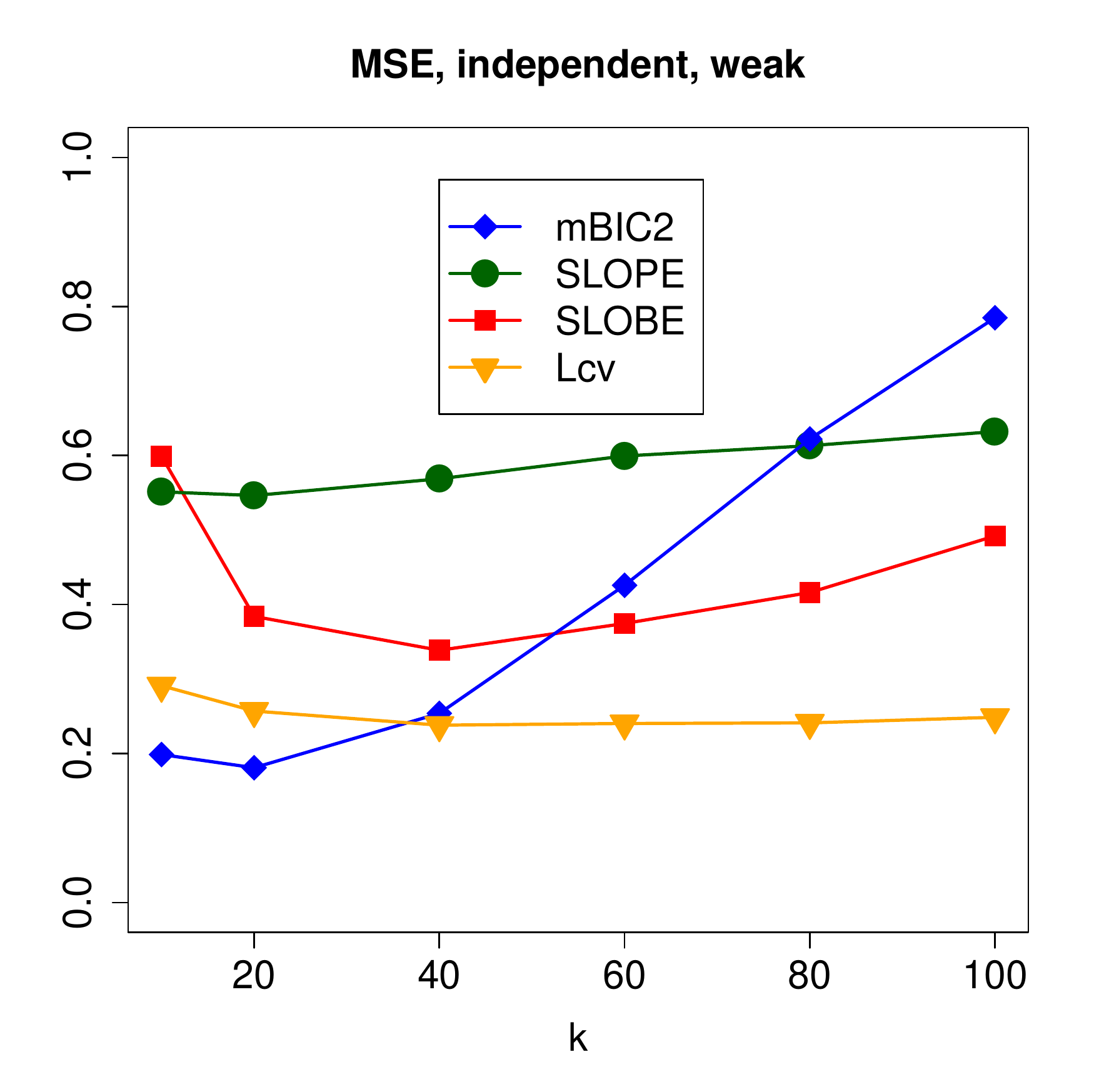}
	\end{minipage}
	\hspace{0.1cm}
	\begin{minipage}[t]{0.4\textwidth}
		\includegraphics[scale=0.3]{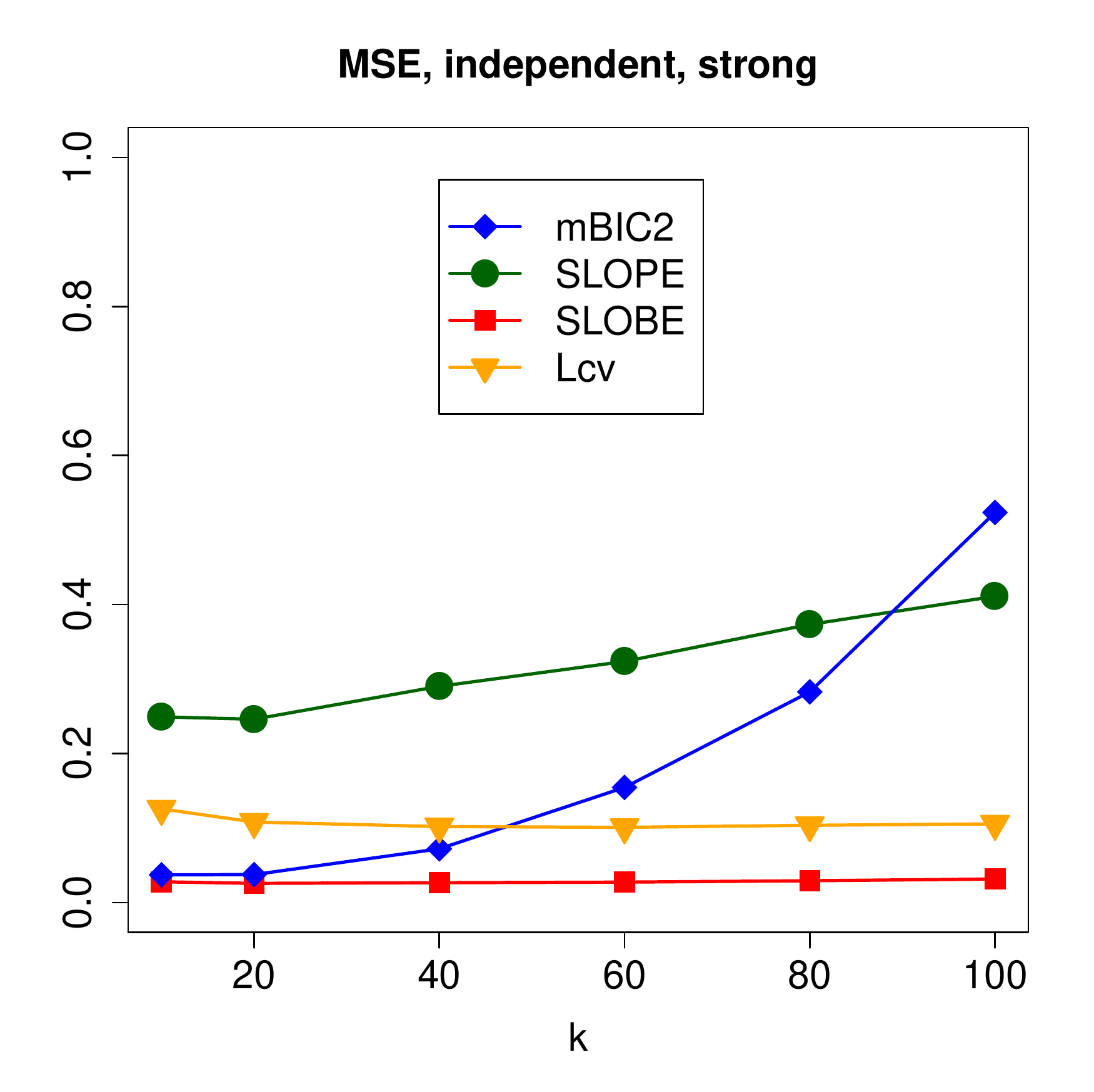}
	\end{minipage}
	\hspace{0.1cm}
	\begin{minipage}[t]{0.4\textwidth}
		\includegraphics[scale=0.3]{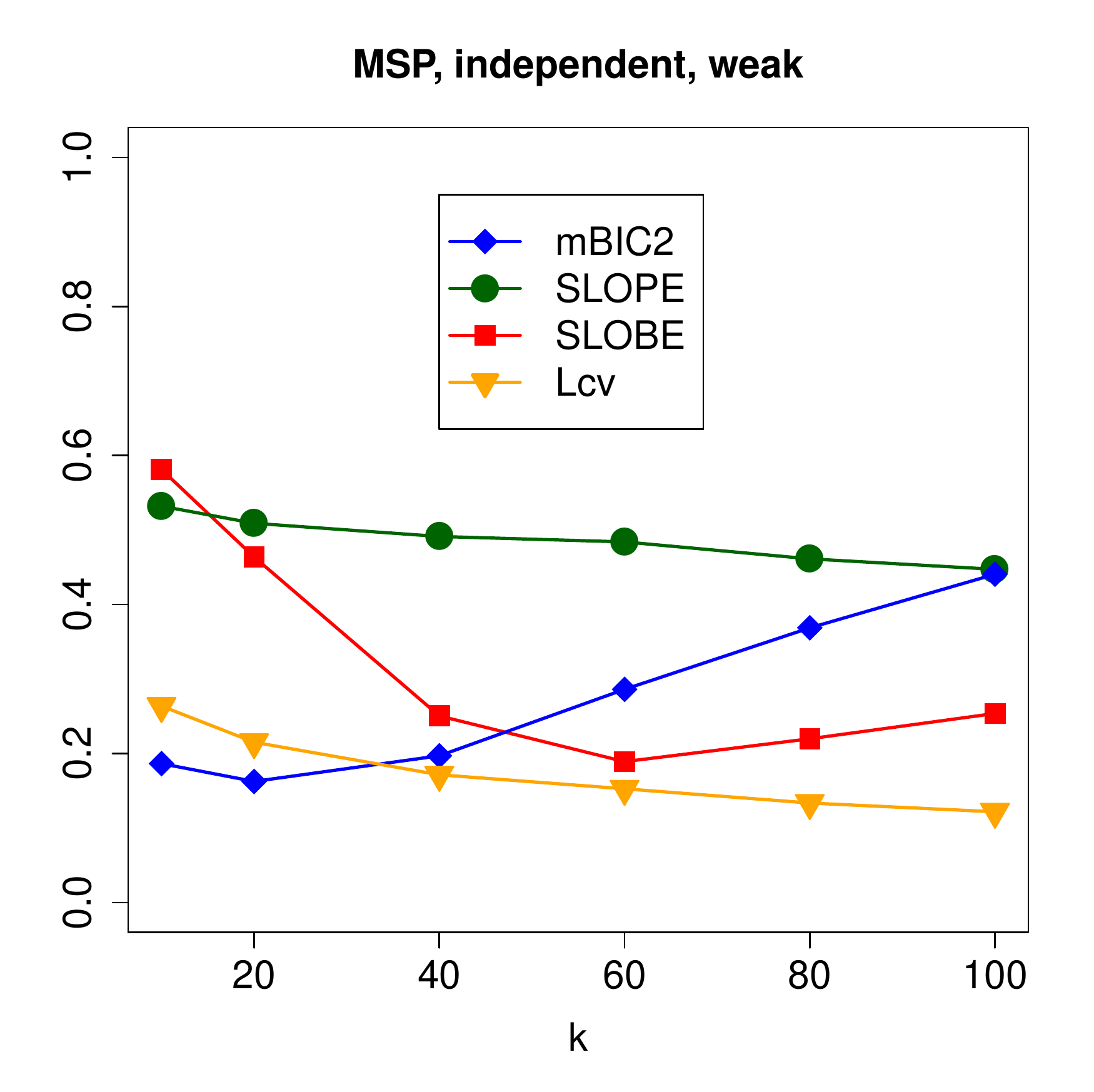}
	\end{minipage}
	\hspace{0.1cm}
	\begin{minipage}[t]{0.4\textwidth}
		\includegraphics[scale=0.3]{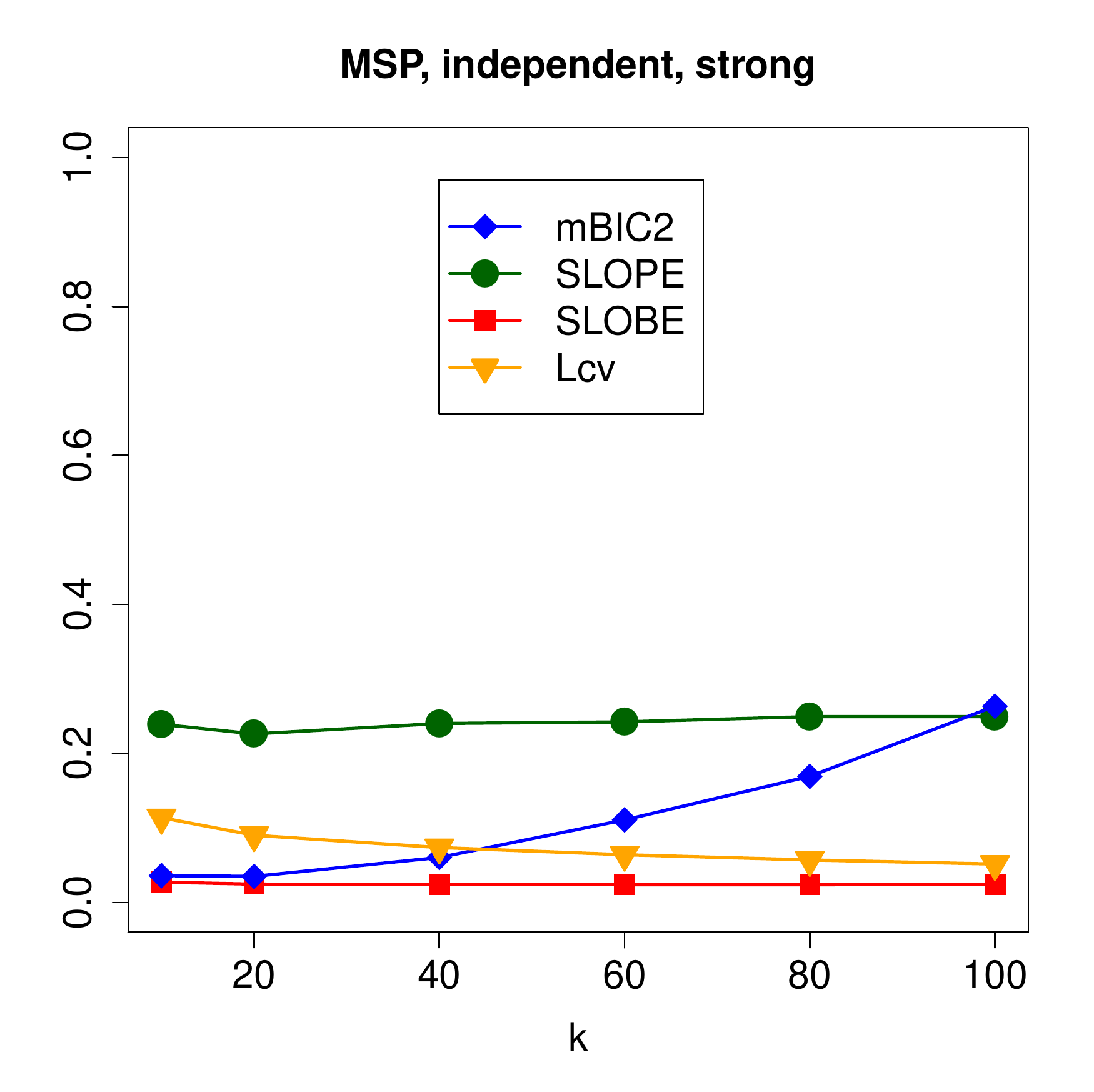}
	\end{minipage}
	\caption[Comparison of mBIC2 with different regularization methods. Simulations with independent regressors.]{Results for {\it independent} regressors and weak (\ref{weak}) and strong (\ref{strong}) signals.}
	\label{independent}
\end{figure*}

\begin{figure*}
	\centering
	\begin{minipage}[t]{0.4\textwidth}
		\includegraphics[scale=0.3]{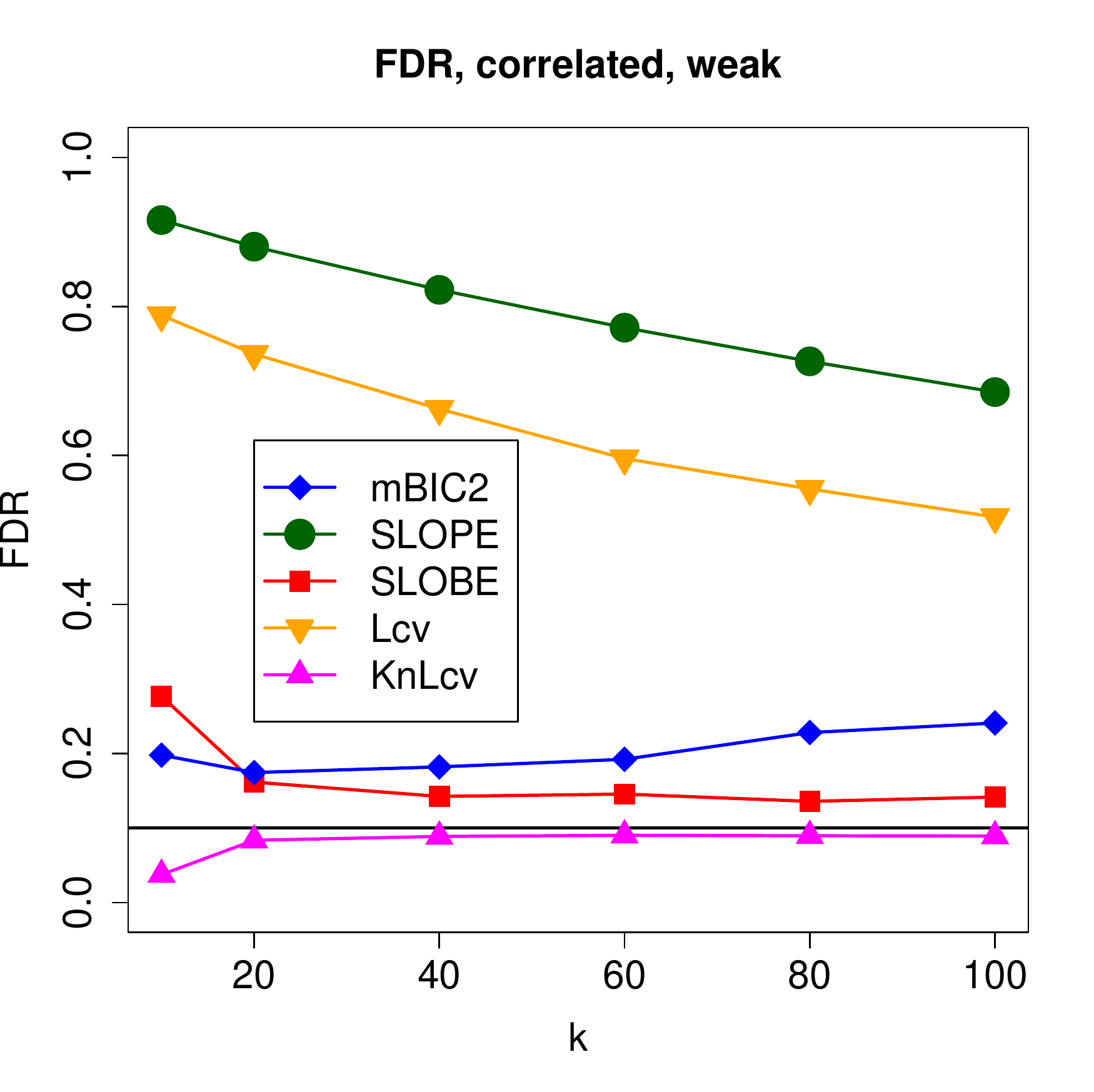}
	\end{minipage}
	\hspace{0.1cm}
	\begin{minipage}[t]{0.4\textwidth}
		\includegraphics[scale=0.3]{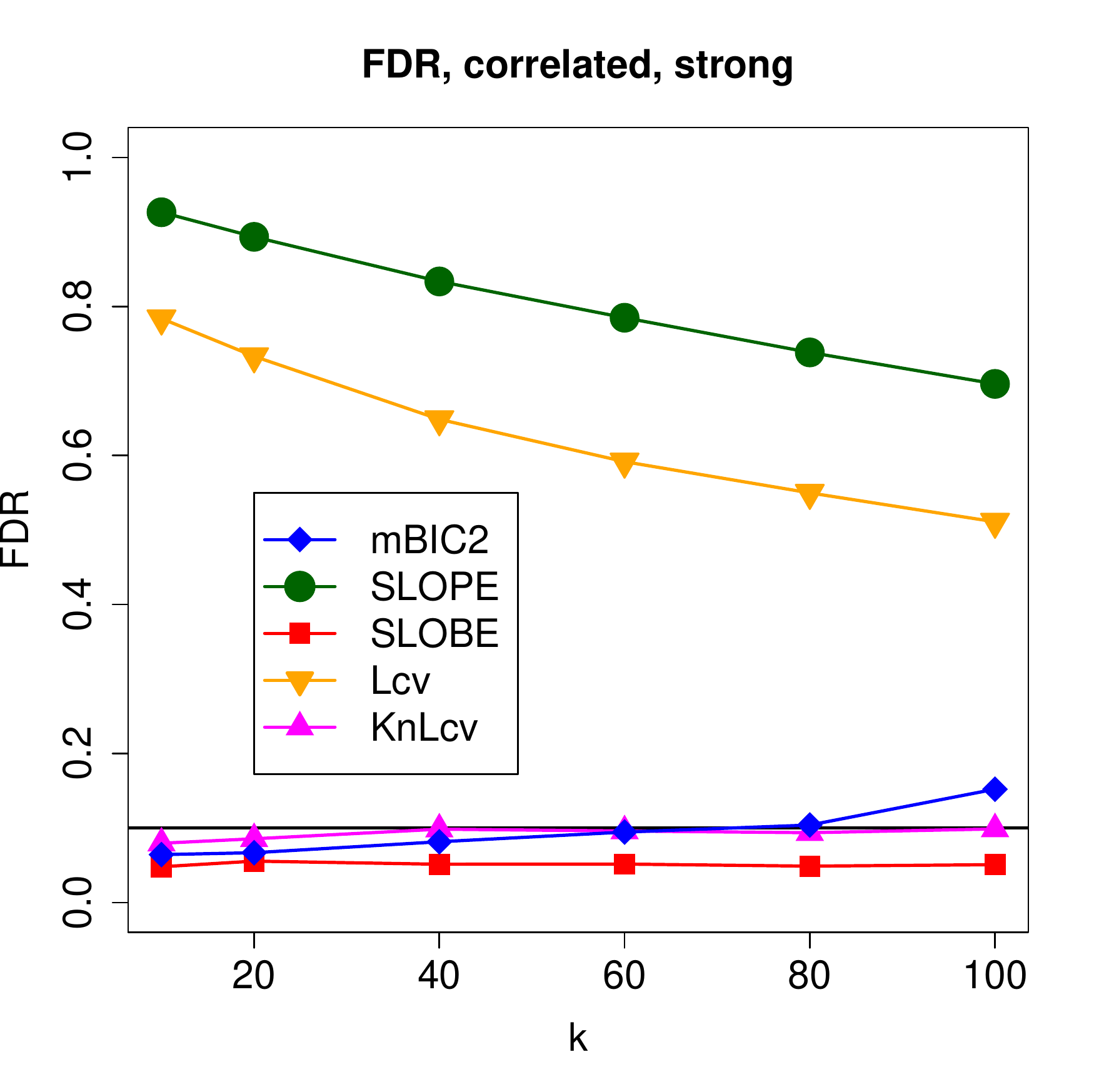}
	\end{minipage}
	\begin{minipage}[t]{0.4\textwidth}
		\includegraphics[scale=0.3]{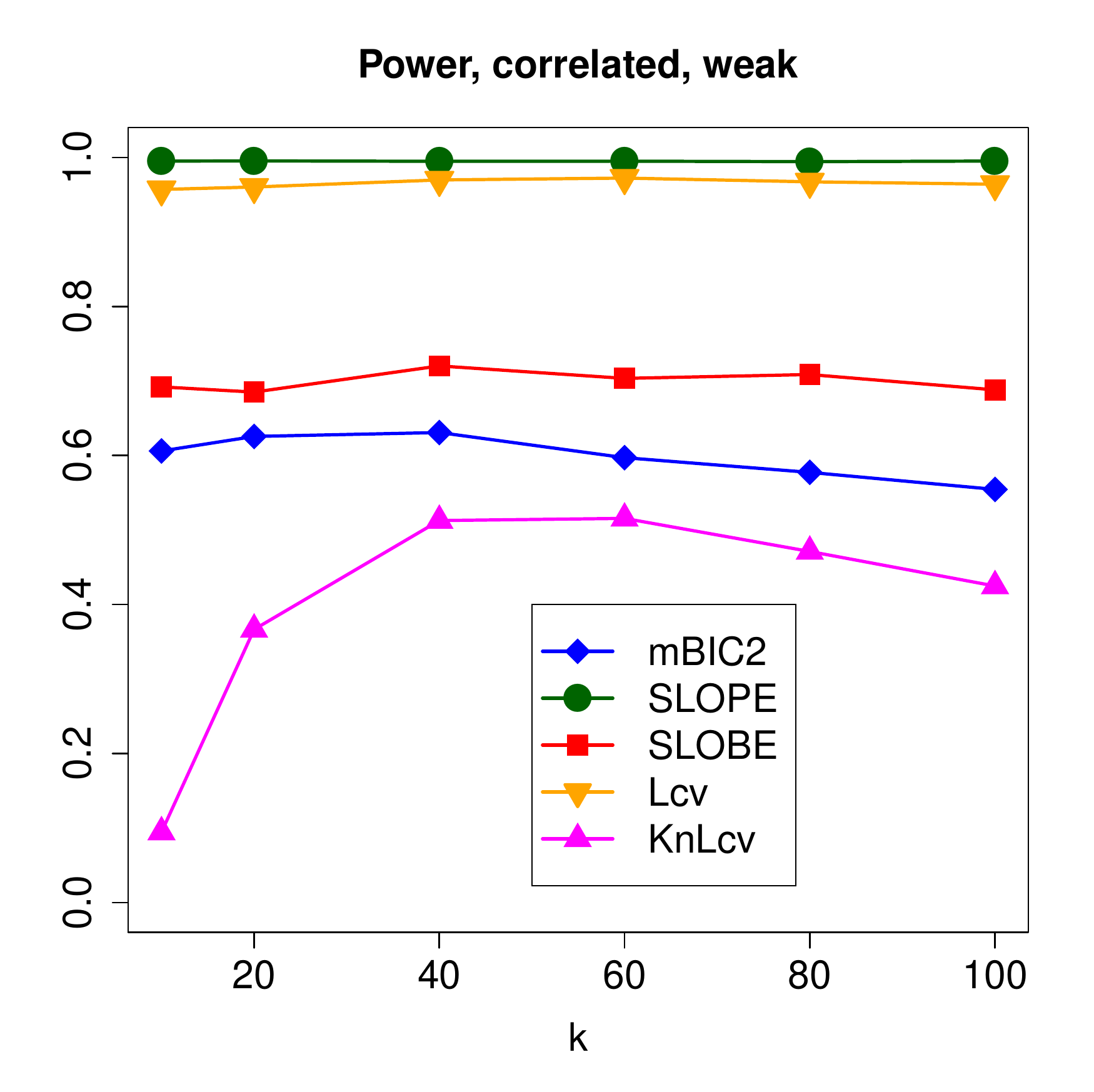}
	\end{minipage}
	\hspace{0.1cm}
	\begin{minipage}[t]{0.4\textwidth}
		\includegraphics[scale=0.3]{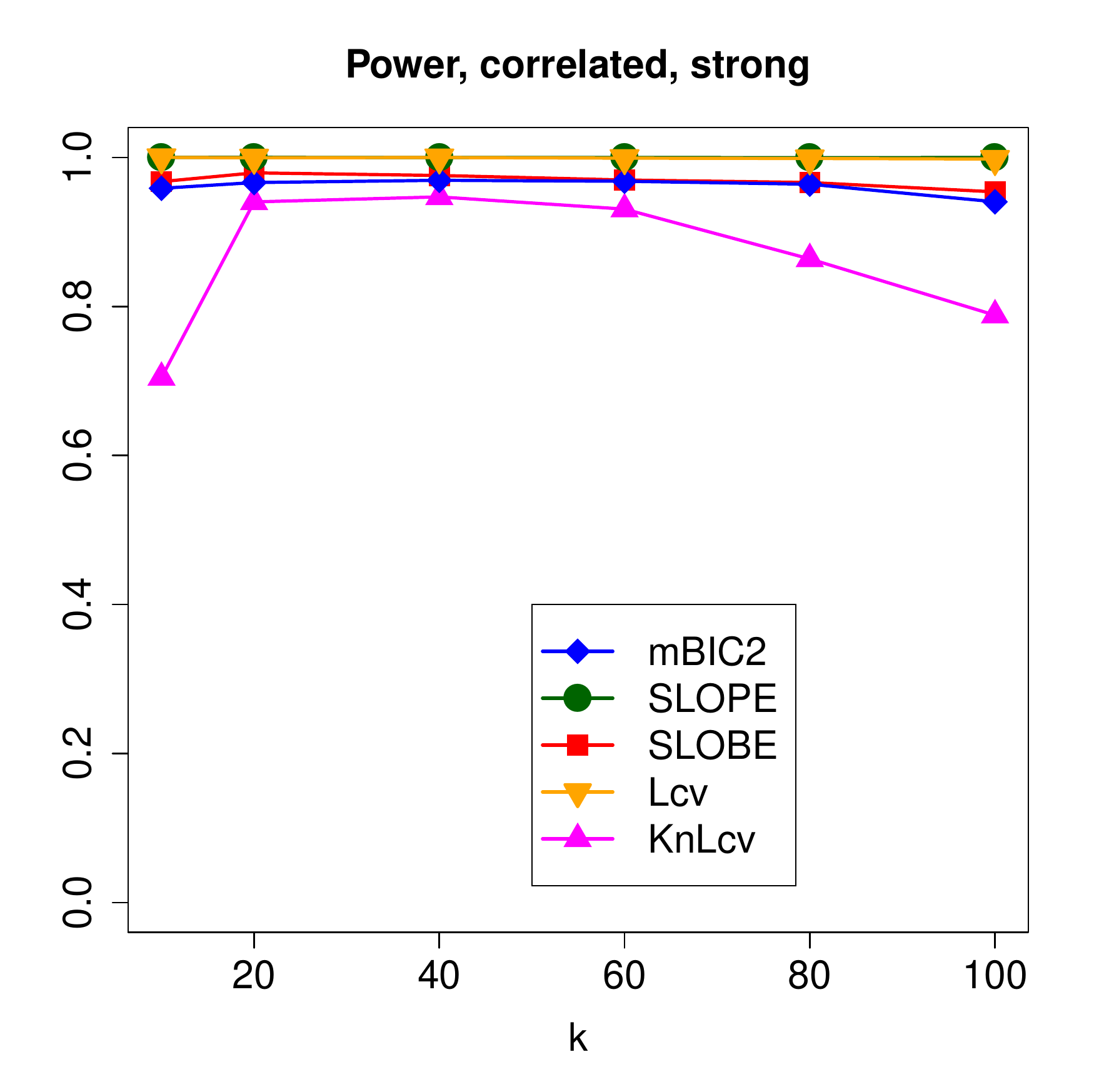}
	\end{minipage}
	\centering
	\begin{minipage}[t]{0.4\textwidth}
		\includegraphics[scale=0.3]{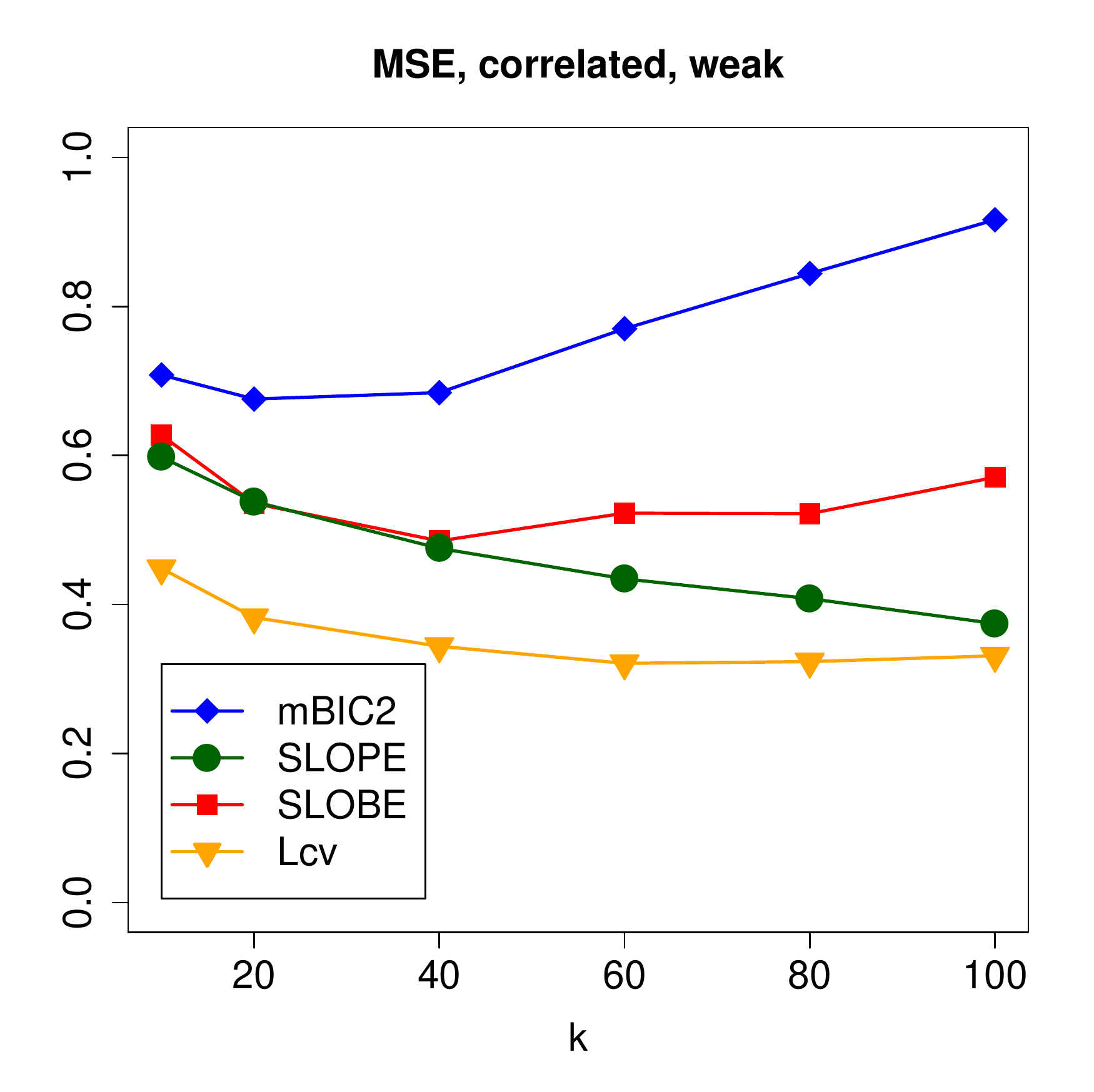}
	\end{minipage}
	\hspace{0.1cm}
	\begin{minipage}[t]{0.4\textwidth}
		\includegraphics[scale=0.3]{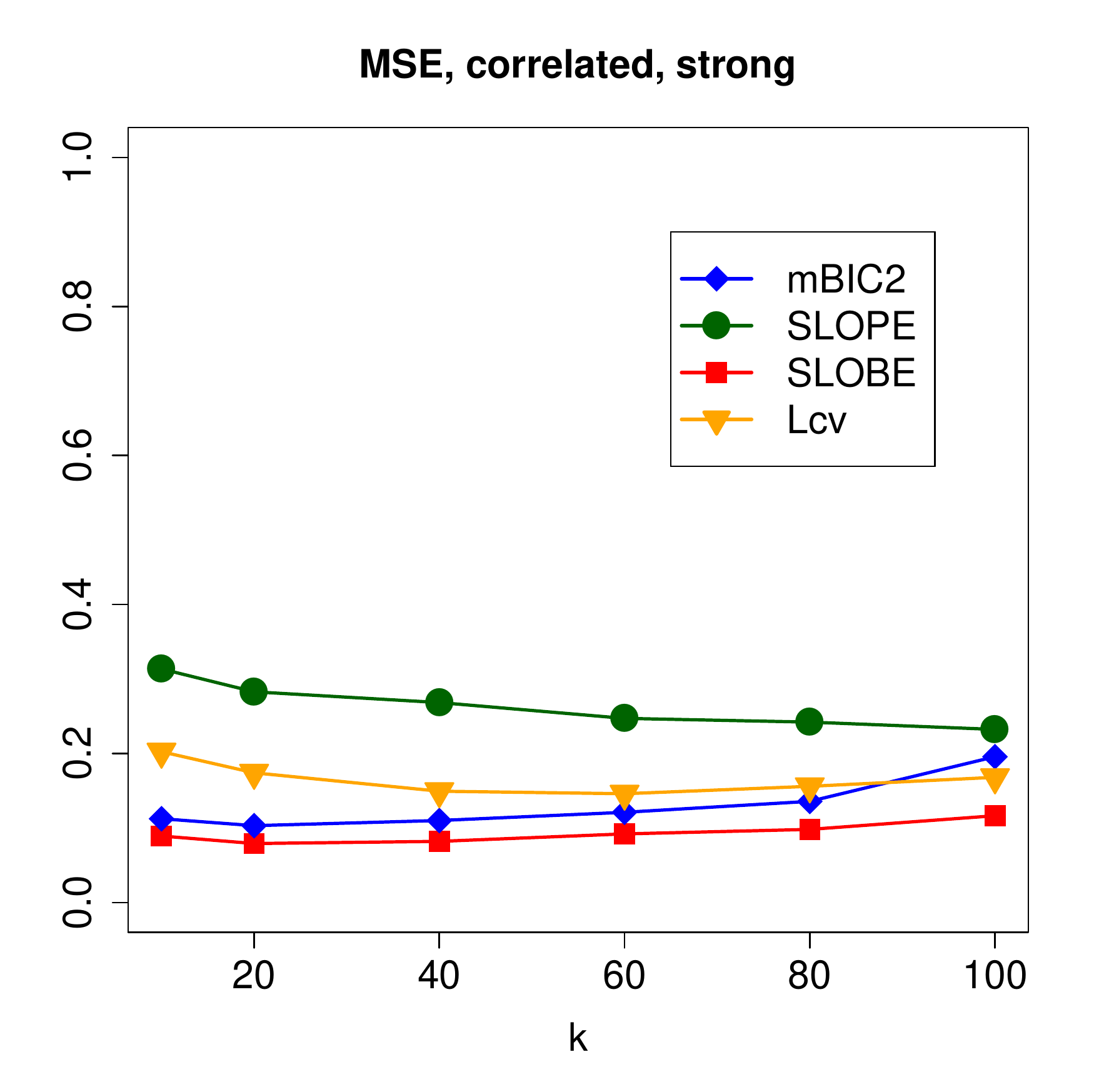}
	\end{minipage}
	\hspace{0.1cm}
	\begin{minipage}[t]{0.4\textwidth}
		\includegraphics[scale=0.3]{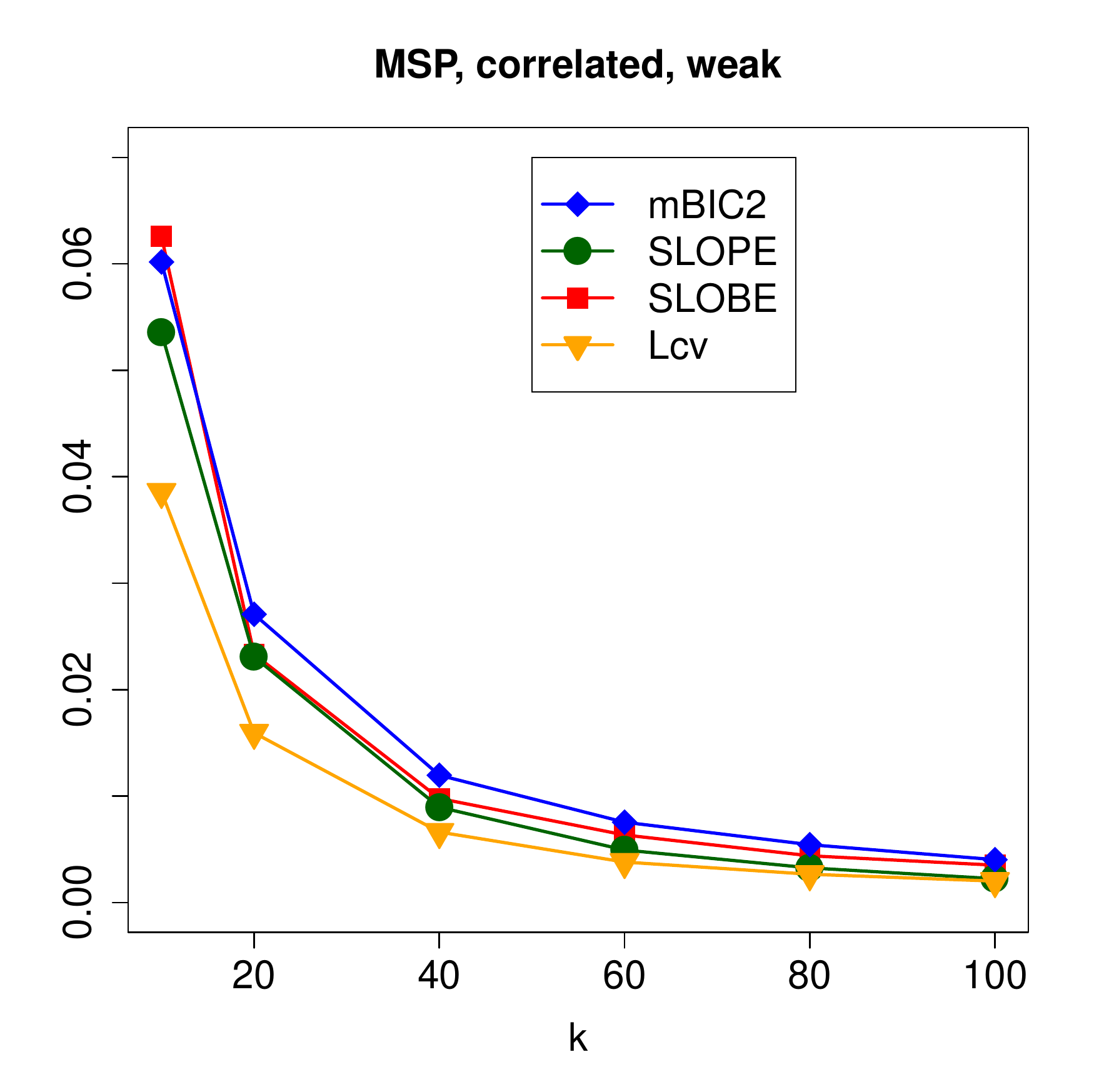}
	\end{minipage}
	\hspace{0.1cm}
	\begin{minipage}[t]{0.4\textwidth}
		\includegraphics[scale=0.3]{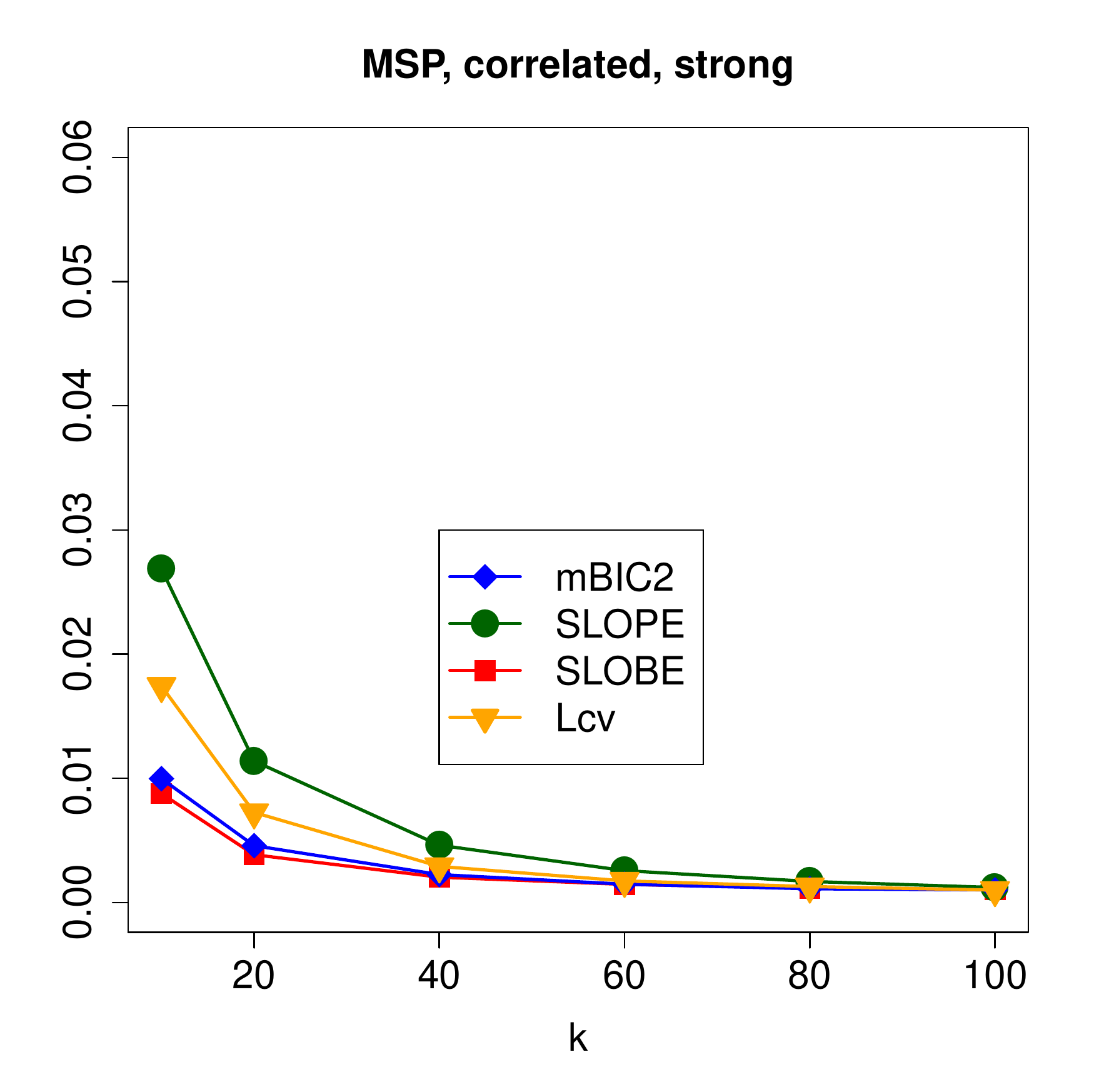}
	\end{minipage}
	\caption[Comparison of mBIC2 with different regularization methods. Simulations with correlated regressors.]{Results for {\it correlated} regressors and weak (\ref{weak}) and strong (\ref{strong}) signals.}
	\label{correlated}
\end{figure*}

\noindent {\bf Results:}

As illustrated in Figures \ref{independent} and \ref{correlated} mBIC2 controls FDR below 0.1 for the independent regressors.  It outperforms all other procedures when the regressors are independent and the signal is weak and sparse. However, it suffers from a loss of power and diminished estimation and prediction properties when the number of causal regressors increases. In case of the correlated design FDR of mBIC2 increases, particularly when the signals are  weak. Here, in most cases mBIC2 has a slightly larger FDR, smaller power and worse estimation and prediction properties than SLOBE. We believe that this is a drawback of the search procedure which has problems with identifying the optimal model when predictors are strongly correlated.

SLOPE based on the heuristic sequence of tuning parameters (\ref{heuristics}) keeps FDR close to the nominal level of $q=0.2$ when the regressors are independent. It also has a pretty good power but suffers from  relatively large estimation and prediction errors. This is due to excessive shrinkage by a large $\lambda$ sequence needed for FDR control. In case of correlated regressors SLOPE does no longer control FDR. It performs similarly to cross-validated LASSO but has a larger FDR and worse estimation and prediction properties. This is due to the specific sequence of tuning parameters, which is not ideal for the purpose of prediction. Interestingly, in the scenario with correlated predictors and weak signals  both crossvalidated LASSO and SLOPE have better estimation and prediction properties than the methods which control FDR. 

SLOBE, the adaptive version of SLOPE, exhibits very good properties under all considered scenarios. It nicely keeps FDR close to the nominal level 0.1, even when predictors are correlated. When the signals are strong SLOBE has superior predictive and estimation properties. For weak signals SLOBE outperforms mBIC2 when the predictors are correlated and when they are independent and the number of causal markers increases. 
However, it needs to be mentioned that the comparison between mBIC2 and SLOBE may depend on the actual values of $n$ and $p$. In the simulations reported in \cite{biomarkers} mBIC2 outperforms SLOBE when the predictors are independent and $n=1000$ and $p=2000$.  We will come back to this when discussing the results from analysing the first real data set in Section \ref{Sec:RealData}.

As expected, cross-validated LASSO  has a very large FDR, which allows to obtain a high power and good estimation and prediction properties  when signals are weak. When signals are strong then cross-validated LASSO has worse estimation and prediction properties than SLOBE, which nicely controls FDR. Due to the results reported in Section \ref{predSLOPE} we expect that the cross-validated version of SLOPE will outperform LASSO with respect to prediction properties. This however requires development of an efficient SLOPE algorithm, which is currently under development (see e.g. \cite{ bao2020, strongSLOPE}).  

As expected, knockoffs always control FDR. They provide large power and good estimation properties when regressors are independent and $k$ is sufficiently large. However, when $k$ is very small or regressors are strongly correlated,  knockoffs  have less power than other model selection methods. We believe that the lack of power for small $k$ is not necessarily an inherent property of the knockoff methodology and could be eliminated by applying some modifications. On the other hand, the lack of power for correlated regressors seems to result from increased variance of LASSO estimates after the design matrix is augmented with knockoff variables (i.e. $p$ changes from 500 to 1000). In case of some specific designs, like equicorrelated predictors, this could be solved for example by using the counting knockoffs of \cite{weinstein2017power} (see also \cite{weinstein2020}), which use less than $p$ knockoff copies. In other cases one could resort to the conditional randomization tests of \cite{knockrandom}, which however are much more computationally intensive.

\section{Real data examples}\label{Sec:RealData}

In this section the different model selection methodologies are applied to analyse two real data sets, one with a continous and one with a binary response. The {\it R} codes which illustrate how to use the different {\it R} packages to perform the  analysis are available as  Supplementary Material. 

\subsection{Continous response}\label{Sec:Continous}

The first data set is concerned with gene expression levels in lymphoblastoid cell lines of 210 unrelated HapMap individuals \cite{hapmap2005} from four populations (60 Utah residents with ancestry from northern and western Europe, 45 Han Chinese in
Beijing, 45 Japanese in Tokyo, 60 Yoruba in Ibadan, Nigeria) \cite{stranger2007}. This data set is available at {\it ftp://ftp.sanger.ac.uk/pub/genevar/} and was previously studied e.g. in \cite{bradic2011, fanfanbarut14, Wojtek}.  
The goal is to identify genes whose expression levels can be used  to predict the expression level of the gene CCT8, which appears within the Down syndrome critical region on human chromosome 21. Such analyses are performed to identify genes which regulate the expression of CCT8.

The original data set contains expression levels measured for 47293 probes. Following \cite{WangWuLi12, Wojtek} we preprocess the data by removing probes  for which the maximum expression level among the 210 individuals is smaller than the 25-th percentile of all measured expression levels and for which the range of expression levels among the 210 individuals is smaller than 2. After this preprocessing we are left with  $p=3220$ probes, which will be used to predict the expression level of CCT8.

We begin our analysis by identifying interesting explanatory variables using an advanced step-wise search strategy to minimize the mBIC2 criterion. Similarly as in the simulation study we first eliminate variables with a marginal p-value larger than 0.15, then perform a liberal forward selection with  BIC followed by  backward elimination using mBIC2, and finally a step-wise selection on the whole data set using mBIC2 as selection criterion.  These computations are performed with the R package {\it bigstep} \cite{bigstep}.

%\begin{lstlisting}
%data = prepare_data(Y, X,  verbose=FALSE)
%
%fit <- data \%$>$\%
%      reduce_matrix()\%$>$\%
%			fast_forward()\%$>$\%
%      multi_backward(crit=mbic2)\%$>$\% 
%      stepwise(crit=mbic2)
%
%\end{lstlisting}

This search strategy identified 5 important variables which are listed in Table \ref{Tab:RealData1}. Looking at the p-values when testing coefficients of the corresponding multiple regression model, one observes that all five variables are significant after Benjamini Hochberg correction at an FDR level $\alpha=0.05$, but V1354 would not be sifnificant after Bonferroni correction. The multiple $R^2$ of this model is equal to 0.581 and the value of the mBIC2 criterion equals 392.8249.

\begin{table}[t]
	%\noautomaticrules
	\tabletitle{Properties of the five variables selected with mBIC2 for the {\it Sanger} data.}
	\label{Tab:RealData1}
	\begin{tabular}{r|llccccl}
		\tch{Name}    &\tch{p val} &\tch{R(Y)} &\tch{R(V1004)} &\tch{R(V682)}  &\tch{R(V1370)} &\tch{R(V1354)}  &\tch{$\#(|R| > 0.65)$}  \\
		\hline 
		V1004 &  6.84e-07  & 0.59& & && &214\\
		V682 &   2.52e-08& 0.58& 0.64 & &&&160 \\
		V1370 & 7.94e-09 &0.51& 0.35&0.28&&&  1\\
		V1354 & 5.02e-05 &0.43& 0.26& 0.26&0.36& &0\\
		V206 &  8.79e-06 &0.25& 0.49&0.48& 0.37&0.26& 0
	\end{tabular}
\end{table}

Table \ref{Tab:RealData1} illustrates that variables V1004 and V682 are strongly correlated with the expression of CCT8 and represent a large cluster of at least 215 probes with strongly correlated expression levels. Probe V1370 has just one strongly correlated probe, while the other two selected probes are not strongly correlated to any other probes in the data set. Probe V206 is interesting since it is not that strongly correlated with CCT8 and would not be selected based on the marginal p-value. However, it is relatively strongly correlated with V1004 and V682 and is significant in the multiple regression model built by mBIC2.

When applying SLOBE for the analysis of our data set we obtain an empty model. This result is in agreement with the simulation results from \cite{biomarkers}, which show that SLOBE might have rather low power when $p \gg n$. The reason for this is quite well understood and has to do with how regularization techniques like LASSO operate.  Compared to the step-wise regression with mBIC2, which needs to estimate regression coefficients only in very small models, SLOPE, LASSO and SLOBE need to estimate all $p$ coefficients. This leads to excessive variance when $p \gg n$, which results in problems with identifying the optimal model. Therefore regularization techniques are usually  applied only after the number of variables has been substantially reduced by some screening procedure, like e.g. Sure Independence Screening \cite{SIS}. For our data set this technique was used previously by \cite{WangWuLi12, Wojtek}, who preselected 300 predictors based on their marginal correlations with CCT8. In our analysis we additionally include V206, which was selected by mBIC2 but has a relatively small marginal correlation with CCT8.

Applying the described {\it bigstep} mBIC2 procedure on the reduced data set yields 9 selected variables, which include all five variables selected by mBIC2 on the full data set. This model is likely to be too large since now the penalty in mBIC2 is adjusted only to  $p=301$ and does not take into account that the explanatory variables were preselected using marginal correlations with $Y$. After performing  backward elimination with mBIC2 adjusted to the number of variables in the full data set $p=3220$ we obtain exactly the same model as the one selected by {\it bigstep} on the full data set.

\begin{table}[t]
	%\noautomaticrules
	\tabletitle{Properties of the six variables selected with SLOBE for the  {\it Sanger} data.}
	\label{Tab:RealData2}
	\begin{tabular}{r|lccccccl}
		\tch{Name}    &\tch{p val} &\tch{R(Y)}&\tch{R(V980)} &\tch{R(V1370)} &\tch{R(3173)}  &\tch{R(V315)} &\tch{R(V1354)}   &\tch{$\#(|R| > 0.65)$}  \\
		\hline 
		V980  &  6.80e-06  & 0.51&      &     &     &    &&0\\
		V1370 &  4.25e-09  &0.51 & 0.34 &     &     &    &&1 \\
		V3173 & 3.69e-14   &0.47 & 0.30  &0.10  &     &    && 270\\
		V315 &  1.56e-05   &0.44 & 0.37 & 0.31& 0.14&    &&1\\
		V1354 & 4.95e-07   & 0.43&  0.19& 0.36& 0.10&0.19&& 0\\
		V206&4.53e-05      &0.25 &0.34  &0.37 &0.35 &0.30&0.26&0
	\end{tabular}
\end{table}

To perform  SLOBE on the reduced data set we use the estimator provided by the cross-validated LASSO as starting point
%\begin{lstlisting}
%Y1<-scale(Y);
%X1<-scale(X)/sqrt(n);
%Rcpp::sourceCpp('SLOBE_cpp_missing_corrected.cpp')
%objl<-cv.glmnet(X1,Y1,standardize=FALSE, intercept=FALSE)
%betal<-coefficients(objl, s='lambda.min');
%betal<-betal[2:302,1];
%
%obj<-SLOBE_ADMM_approx_missing(betal,X1,Y1,a_prior = 0.01*n, b_prior = 0.1*n,FDR=0.1).
%\end{lstlisting}
and identify 7 interesting variables: V2524, V980, V1370, V3173,  V315, 1354 and V206. Only three of these variables coincide with variables selected by mBIC2. Fitting a multiple regression model with these variables gives fairly large p-values for the variables V2524 (p = 0.003071) and V3173  (p = 0.000155). This is because of their strong correlation (R = 0.74). Backward elimination with mBIC2 removes V2524 and consequently the p-value for V3173 drops to 3.69e-14. This suggests that the large group of variables strongly correlated with V3173 contains some important predictors.  

As shown in Table \ref{Tab:RealData2} all p-values in the resulting model with 6 variables are small enough to be rejected by the Benjamini-Hochberg procedure adjusted to the number  of variables $p=3220$ in the original data set, whereas  V206 and V315 would not be rejected by the Bonferroni procedure (V315 is pretty much at the decision boundary).  The multiple $R^2$ for this model is equal to 0.6116 and the value of mBIC2  is equal to  392.0696, which is smaller than the value of mBIC2 for the model selected by the extended step-wise procedure from {\it bigstep}. Hence this model would also be preferred according to mBIC2 but could not be identified with the initial search strategy.

\begin{table}[t]
	%\noautomaticrules
	\tabletitle{Correlation between mBIC2 and SLOBE variables for the  {\it Sanger} data.}
	\label{Tab:RealData3}
	\begin{tabular}{r|llccccl}
		\tch{Name}    &\tch{R(V980)} &\tch{R(V1370)} &\tch{R(3173)}  &\tch{R(V315)} &\tch{R(V1354)}  &\tch{R(V206)}  \\
		\hline       
		V1004 & 0.51& 0.36& 0.64 & 0.59& 0.26 & 0.49\\
		V682  & 0.43& 0.28& 0.69 & 0.31& 0.26 & 0.48 \\
		V1370 & 0.34& 1.00& 0.10 & 0.31& 0.36 & 0.37 \\
		V1354 & 0.19& 0.36& 0.10 & 0.19& 1.00 & 0.26 \\
		V206  & 0.34& 0.37& 0.35 & 0.30& 0.26 & 1.00 \\
	\end{tabular}
\end{table}

Comparing the models from Table \ref{Tab:RealData1} and Table \ref{Tab:RealData2} we observe that they have three variables in common: V206, V1354 and V1370. Probe V3173, which is the strongest predictor in the SLOBE model, replaced  V1004 and V682 selected by the stepwise procedure. The two remaining variables in the SLOBE model V980 and V315 are also strongly correlated with V1004 (their marginal correlations with V1004 exceed 0.5) and still somewhat correlated with V682.
Hence the SLOBE model substituted two strongly correlated variables which were also strongly correlated with V206 by three less correlated variables which are also slightly less correlated with V206
The six variables in the reduced SLOBE model have a maximal pairwise correlation that does not exceed 0.4, whereas in the model selected by the stepwise procedure three pair-wise correlations exceed 0.45.

The above comparison illustrates the difficulties that emerge when one wants to identify  the ''best'' model among a set of predictors which are strongly correlated.
Both mBIC2 and SLOBE tend to select only small subsets of correlated groups and depending on the context it might or might not be desirable to have only few representatives for clusters of correlated predictors.
One solution to this problem is provided by SLOPE, which tends to select larger subsets of correlated variables. 

%\begin{lstlisting}
%
%obj<-trainSLOPE(X1,Y1)
%obj1<-SLOPE(X1,Y1, alpha= obj\$optima[2.2,2] ) 
%
%\end{lstlisting}

To verify the performance of SLOPE on the reduced data set we used  SLOPE combined with cross-validation as implemented in the  R package \textit{SLOPE}.
SLOPE identified 50 variables which contained all variables selected by mBIC2 and SLOBE apart from variable V1004, which was previously selected by the step-wise procedure although it had only weak marginal correlation with CCT8.  SLOPE selected 21 variables from the cluster of correlated variables related to V1004, V682 and V3173, which reflects the importance of this cluster. Compared to the models selected by mBIC2 and SLOBE, cross-validated SLOPE included additional 23 variables whose marginal correlation with CCT8 varied between 0.57 and 0.43. Based on the results of simulations we would expect that many of these additional findings are false positives. 
Cross-validated LASSO with the {\it glmnet} package identified 40 variables, including 18 variables which represent the large cluster of correlated probes. Furthermore LASSO included 17 variables where marginal correlation with CCT8 was ranging between  0.43 and 0.57. Again, we would expect that most of these additional detections are false positives.

\begin{figure}
	\begin{center}
		\includegraphics[scale=0.7]{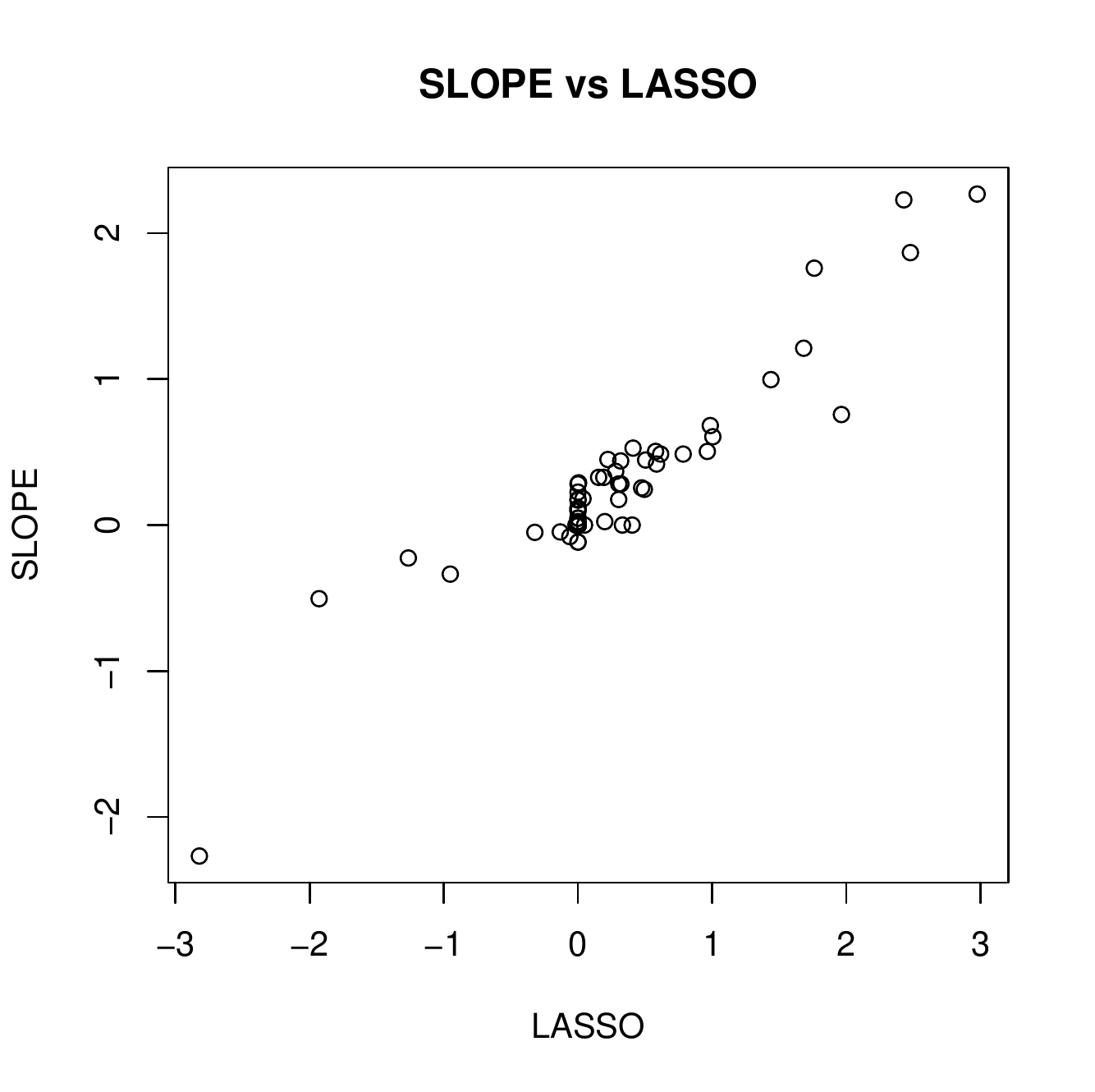}
	\end{center}
	\caption[Scatter plot of SLOPE and LASSO estimates for the Sanger data.]{Slope estimates vs LASSO estimates for the Sanger data}\label{fig:slopevsLASSO}
\end{figure}

The scatter plot from Figure \ref{fig:slopevsLASSO} compares  regression coefficients estimated by SLOPE and LASSO. The 8 variables which have coefficients with the largest absolute value coincide for both methods. These include the 6 variables V1370, V206, V1354, V3173, V980 and V315 which were previously identified by mBIC2 or SLOBE and two other variables which are strongly correlated  with V3173 (R$>$0.71). The only large negative coefficient corresponds to V206, the probe which is only weakly correlated with CTT8 but which is an important regressor in all models built by the advanced methods used in this section.  The scatter plot shows  also that the shrinkate to zero is stronger for SLOPE coefficients than for LASSO coefficients. This is due to the inclusion of a larger number of correlated variables by SLOPE. We can also see that non-zero SLOPE estimates have a tendency to cluster around horizontal lines, because SLOPE shrinks similar coefficients towards each other. The simulation results from Section \ref{predSLOPE} indicate that these shrinkage properties potentially improve the prediction properties of SLOPE, but this assertion remains to be verified based on more extensive comparisons on other real data sets.

\subsection{Binary response variable}\label{Sec:Binary}

As an example of a data set with a binary response we consider the  ARCENE data set, which is available at the UCI Machine Learning Repository \cite{UCI}. The dataset was used for the NIPS 2003 challenge on feature selection \cite{guyon2008feature}. The website of the workshop provides details about the challenge and further description of the data sets \cite{ARCENE_data}.

The ARCENE data set consists of mass-spectrometry data for cancer patients and for controls. For the NIPS 2003 challenge there was a training set and a validation set, each with 44 cancer patients and 56 controls. Aditionally there was also a test set of 310 cases and 390 controls. This is however not available at the UCI repository. There are 10000 metric explicatory variables, most of them measuring the abundance of proteins of a given mass but some of them were created randomly. The task of the challenge was to predict cancer cases using the mass-spectrometry features. Here we want to focus  on the question which features are actually related to cancer. In other words we want to illustrate the behaviour or our methods in terms of model identification and not so much in terms of prediction. 
%The R script to perform this analysis is available at {\bf WEB LINK}.

In a preprocessing step 39 variables are removed which have only zero measurements and 139 variables are removed which yield quasi-complete separation. For the remaining 9822 features we perform logistic regression analysis considering the following methods for model selection.

\begin{itemize}
	\item  \textbf{Multiple testing}: Testing individual features with simple logistic regression and applying Bonferroni correction
	\item $\mathbf{L_0}$ \textbf{penalties}: Model selection with the original BIC and with its modifications discussed in Section \ref{Sec:mXIC} using the R package \textit{bigstep}
	\item \textbf{LASSO}: Considering the LASSO search path and models obtained with cross validation  using the R package \textit{glmnet}
	\item \textbf{SLOPE}: Considering the SLOPE search path and models obtained with cross validation  using the R package \textit{SLOPE}
\end{itemize}

%%%%%%%%%%%%%%%%%%%%%%%%%%%%%%%%%%%%%%%%%%%%%%%%%%%%%%

After Bonferroni correction 478 features are significant at a nominal significance level $\alpha = 0.05$.
The analysis of the ARCENE data set is challenging due to the large amount of correlation between explicatory variables, particularly among those which are significantly associated with cancer status. Pairwise correlations among the 478 significant features are larger than 0.5 in 84.4\%, and larger than 0.8 in 22\% of all cases. 

This has an interesting consequence when performing model selection with the four modifications of BIC we have introduced before. These $L_0$ penalties  tend to select a small number of representatives of clusters of highly correlated features.  For the sample size $n = 200$ mAIC is the most conservative choice and the \textit{stepwise} function of bigstep selectes only two features. For the other three criteria (mAIC2, mBIC, mBIC2) stepwise selects three features.  However, a plain \textit{stepwise} search is not guaranteed to find the best model because there is the danger of getting stuck in local optima. For the criterion mBIC2 we performed some very simple strategies to escape the potential local minimum from the stepwise procedure. Adding two forward steps with the milder criterion BIC and then swiching back to stepwise with mBIC2 gave an improved model with five features. Two other strategies we tested did not further improve this model.

\begin{table}[t]
	%\noautomaticrules
	\tabletitle{Properties of the five variables selected with mBIC2 for the {\it Arcene} data.}
	\label{Tab:RealDataArcene1}
	\begin{tabular}{r|lccccll}
		\tch{Name}    &\tch{p val} &\tch{R(V3365)} &\tch{R(V729)}  &\tch{R(V698)} &\tch{R(V6584)}  &\tch{$\#(|R| > 0.5)$}  &\tch{$\#(|R| > 0.8)$} \\\hline 
		V3365 & 3.54 e-08 & & & & &1389 & 207\\
		V729 &  0.0043  & -0.29 & & & & 785 & 23\\
		V698 &  6.12 e-08 & 0.49 & -0.20 & &  & 1638 & 34\\
		V6584 &  0.00085 & -0.19 &  0.34 & -0.01 & & 160 & 9\\
		V3161 &  0.0062 & -0.05  &  0.11 & -0.08 & 0.14 & 0 & 0
	\end{tabular}
\end{table}

Table \ref{Tab:RealDataArcene1} provides some information about the five selected variables. Interestingly only the first and the third variable have very small marginal p-values, whereas particularly the second and the fifth feature are marginally not that strongly associated with cancer status. The strongest correlation is between the first and the third variable  ($R = 0.49$). Particularly the fifth variable is not correlated to any of the others. The first three variables have a large number of strongly correlated features. Specifically the first variable represents more than 200 other variables with an absolute correlation  larger than 0.8. Only the fifth variable has no strong correlation with any other features. Knowing about the data generation for the challenge this indicates that it might be a random variable which was added to the set of features.  Among the 478 features which were significant after Bonferroni 95\% have a correlation larger than 0.5 with at least one of the first three features. This illustrates how well the first three features selected by the stepwise procedure represent the marginally significant variables. The two features which were additionally selected with the multiple forward steps are not strongly correlated to the marginally significant features, having a maximum correlation of 0.31 and 0.13, respectively.

%%%%%%%%%%%%%%%%%%%%%%%%%%%%%%%%%%%%%%%%%%%%%%%%%%%%%%

To present the results of LASSO we initially focus on the beginning of the LASSO search path. The first variables to enter are V3365 and  V5005, followed by V7748, followed by V5005 and  V7748, followed by  V9215 and  V9585. In this initial phase no variables are removed along the search path. Here are the correlations between the 5 variables selected by mBIC2 and the first seven variables along the LASSO search path (in the order they have entered).    \\[3mm]

\begin{table}[t]
	%\noautomaticrules
	\tabletitle{Correlations between variables selected by mBIC2 and the first seven variables on the LASSO path for the {\it Arcene} data.}
	\label{Tab:RealData}
	\begin{tabular}{l|lllllll}
		&V1936 &V3365& V3629& V4973& V7748& V9215& V9585 \\
		V3365 &-0.48 & 1.00& -0.43&  0.52&  0.50& -0.05&  0.52 \\
		V729  &-0.03 &-0.29& -0.04& -0.15& -0.20&  0.37& -0.15 \\
		V698  &-0.32 & 0.49& -0.33&  0.70&  1.00&  0.15&  0.70 \\
		V6584 & 0.04 &-0.19& -0.02& -0.09& -0.02&  0.89& -0.09 \\
		V3161 & 0.08 &-0.05&  0.03& -0.08& -0.07&  0.11& -0.08 \\ 
	\end{tabular}
\end{table}

\ \\

What we see is quite typical for the LASSO search path. The first three variables to enter are strongly correlated with V3365, the next two and the last are strongly correlated with V698. LASSO selects more correlated variables than the $L_0$ penalties do. Due to shrinkage each variable in the model tends to explain less variance than it would do in  an ordinary regression model. This allows to include several strongly correlated predictors which typically stabilizes and enhances the prediction properties and to some extend prevents loss of important predictors due to their replacement by some strongly correlated regressors. On the other hand there is the danger that false positives may be included in the model.

% In the end this can have the result that false positives may enter the search path already relatively early \cite{FDR_LASSO}.

A rather tricky question is which model to choose along the LASSO search path. In terms of prediction the most common answer is to make use of cross validation. For model identification however this approach has some serious drawbacks. First of all the model one obtains depends on the random selection of subsamples for cross validation and the effect on the selected model can be dramatic. In our example the model sizes obtained in different cross validation runs varies between 40 and 80. 
In the R script we look in more detail at a representative model of size 50. This model includes three of the five features found with mBIC2 (V3365,  V729, V3161) and there are two more features (V7748, V9275), which are strongly correlated with the remaining two features from the mBIC2 model ($|R| > 0.95$). Furthermore there are quite a number of additional features correlated with features from the mBIC2 models, but also many which are uncorrelated. 
It is more than likely that this model includes many false positives. A viable strategy is to perform some model selection among the features obtained with LASSO cross validation based on our modifications of BIC. Using mBIC2 this approach results in a model with the five features mentioned above. This model is almost identical to the previous mBIC2 model but has a slightly lower criterion.

%%%%%%%%%%%%%%%%%%%%%%%%%%%%%%%%%%%%%%%%%%%%%%%%%%%%%%

\begin{figure}
	\begin{center}
		\subfigure[\label{Fig:LASSO_path}]{\includegraphics[angle=90,width=7cm,angle=-90]{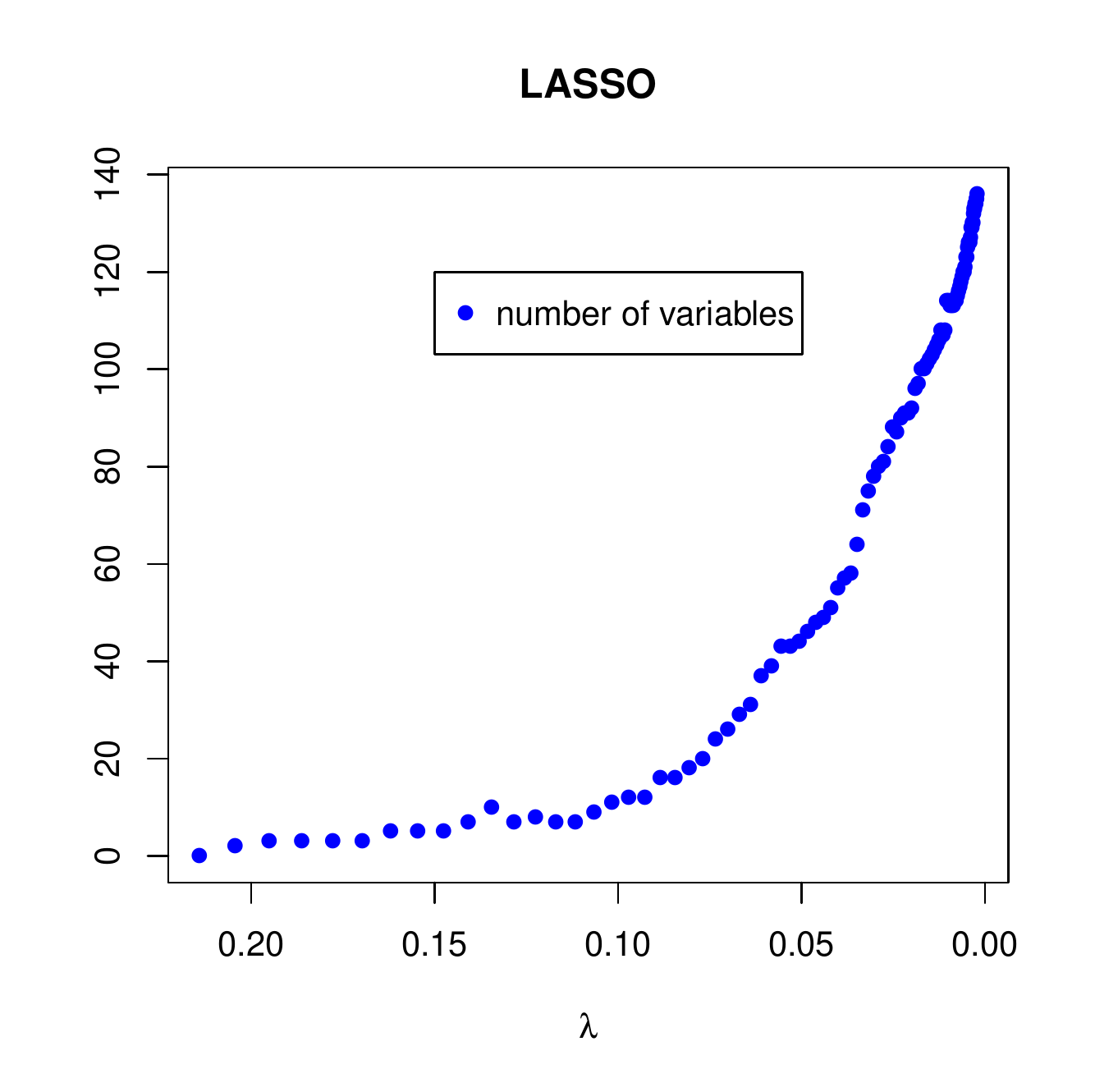}} 
		\subfigure[\label{Fig:SLOPE_path}]{\includegraphics[angle=90,width=7cm,angle=-90]{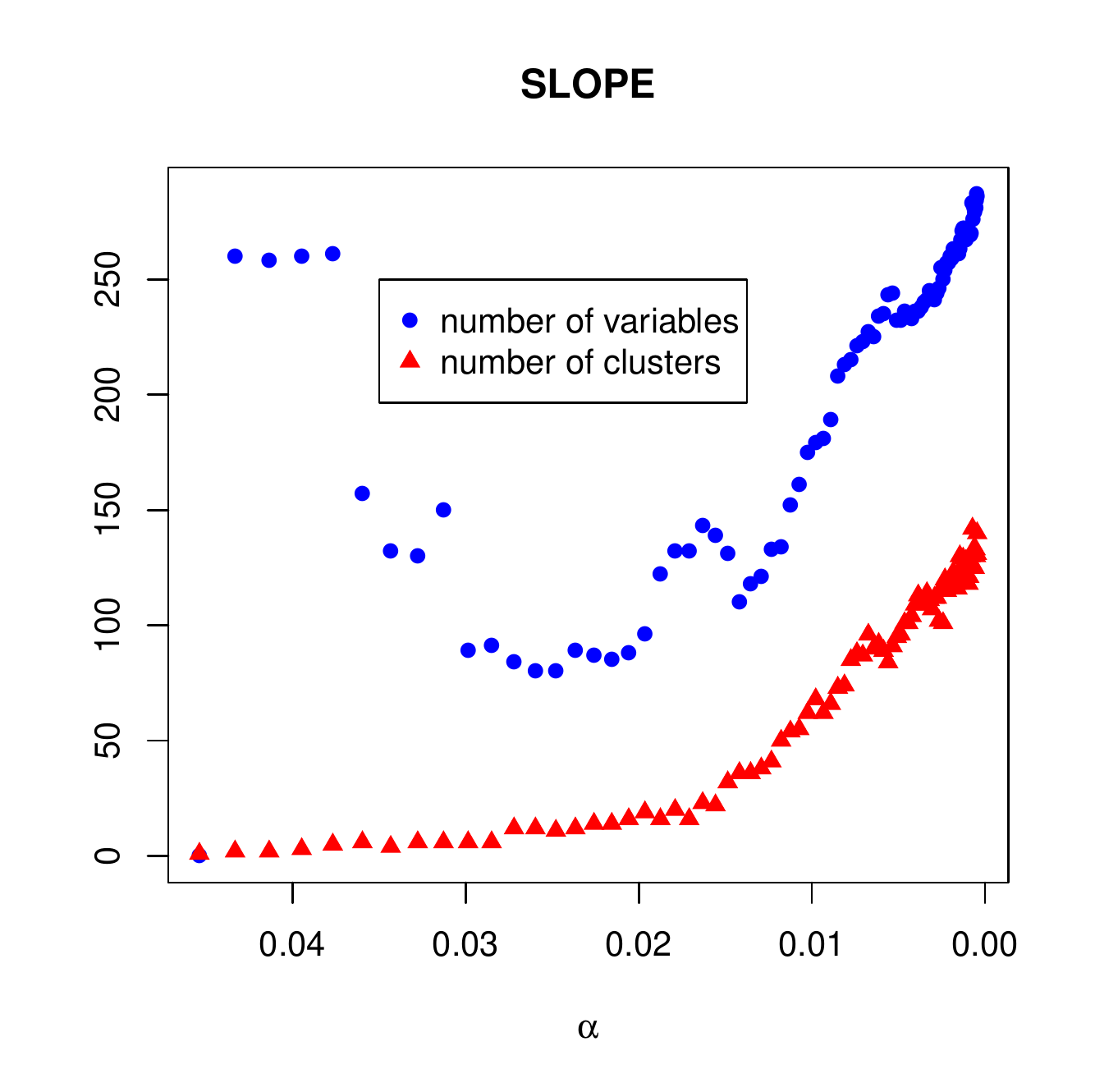}}
	\end{center}
	%\caption[Search path]{Search path for LASSO and for SLOPE.}\label{Fig:Searchpath}
	\caption[Number of variables along the LASSO path and numbers of variables and clusters along the SLOPE path for the Arcene data.]{Number of variables along the LASSO path and numbers of variables and clusters along the SLOPE path. In case of LASSO clustering does not occur.}\label{Fig:Searchpath}
\end{figure}

Finally we want to discuss the results obtained by SLOPE. The first observation is that the SLOPE search path has a very different initial behaviour from the LASSO search path. %(see Figure \ref{Fig:Searchpath}).
The LASSO search path starts with very small models which are incrementally increased and only relatively rarely features from the search path are removed when decreasing the penalty 
In contrast, Figure \ref{Fig:Searchpath}) illustrates that the SLOPE search path immediately starts with a rather large number of features which are then thinned out before the model size increases again. The large number of non-zero coefficients at the beginning of the SLOPE path results from  clustering of similar regression coefficients. This behaviour is more pronounced when there exist large gaps between consecutive elements of the SLOPE sequence. The SLOPE path is obtained by multiplying the basic SLOPE sequence (BH with $q=0.2$) with a constant $c$. Large values of $c$ lead to larger gaps between the elements of $\lambda$. Thus, at the beginning of the path SLOPE has a tendency to replace single predictors with clusters of predictors. The corresponding regression coefficient all have almost the same (very small) value. Thus, the initial 260 features are in fact grouped in just one cluster with the same value of regression coefficient. The clusters tend to become smaller and their number tends to increase along the SLOPE path but at the end of the path the number of clusters is still two times smaller than the number of nonzero coefficients.

%#############################################################
%
%   POTENTIAL FIGURE WITH TWO GRAPHS (ALSO FOR LASSO)
%
%\begin{figure}
%	\begin{center}
%	\subfigure[\label{Fig:LASSO_path}]{\includegraphics[angle=90,width=7cm,angle=-90]{LASSO_path.pdf}} 
%	\subfigure[\label{Fig:SLOPE_path}]{\includegraphics[angle=90,width=7cm,angle=-90]{clustering.pdf}}
%	\end{center}
%	\caption[Search path]{Number of variables along the LASSO path. Number of variables and number of clusters along the SLOPE  path.}\label{Fig:Searchpath}
%\end{figure}
%

\begin{figure}
	\begin{center}
		\subfigure{\includegraphics[angle=90,width=7cm,angle=-90]{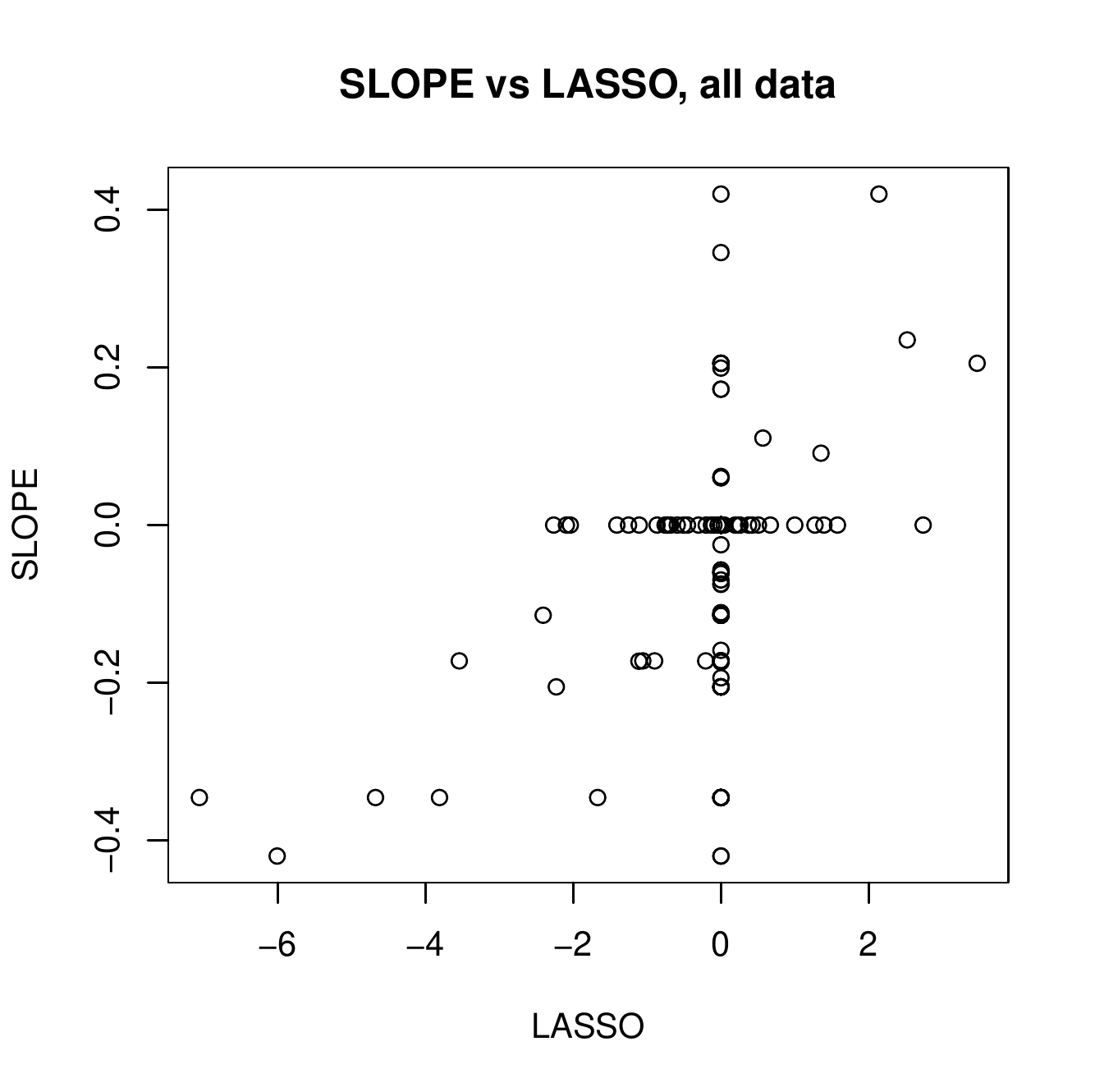}} 
		\subfigure{\includegraphics[angle=90,width=7cm,angle=-90]{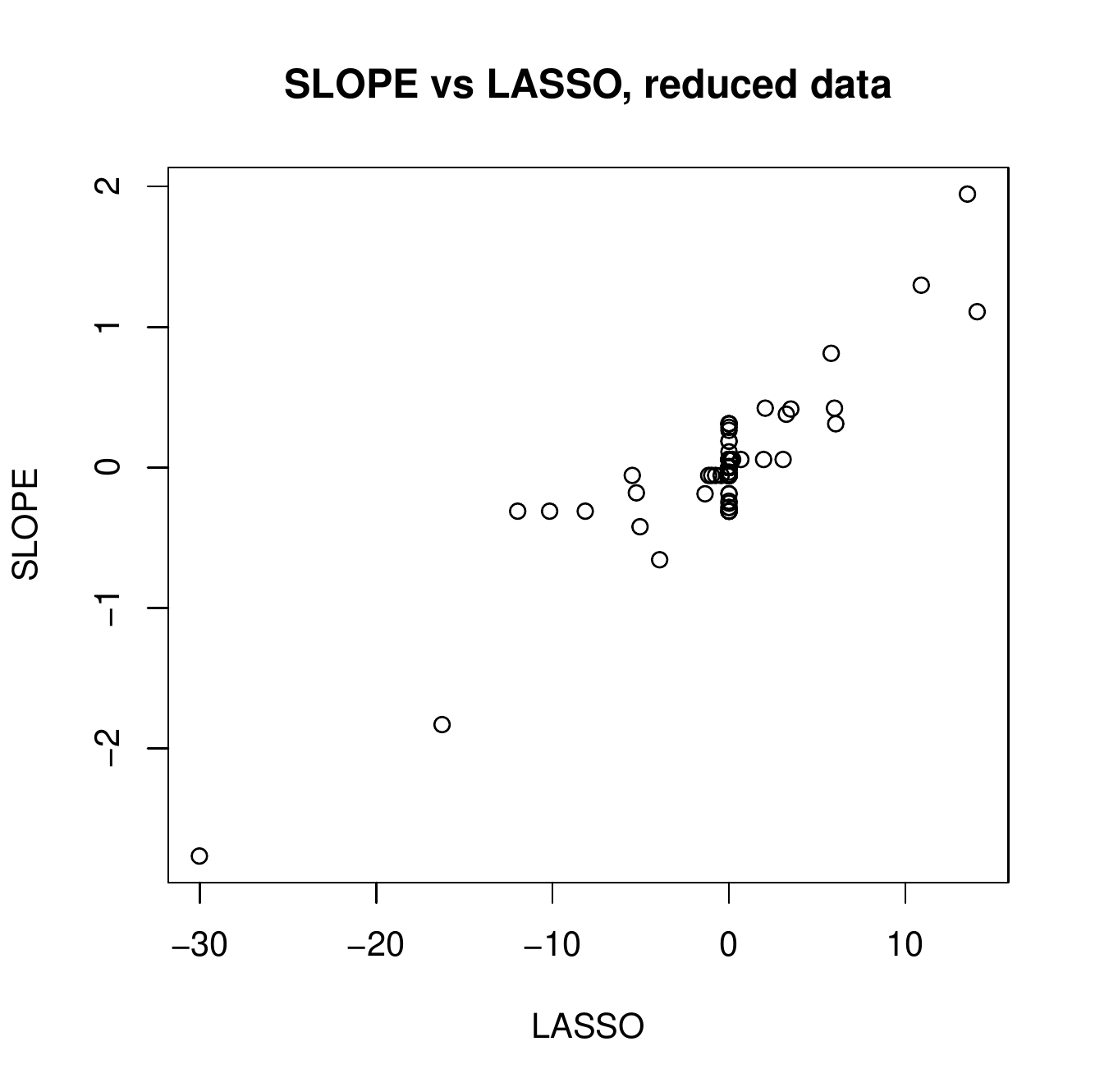}}
	\end{center}
	\caption[Scatter plot of SLOPE and LASSO estimates for the Arcene data.]{Slope estimates vs LASSO estimates for the Arcene data.}\label{slopevsLASSObin}
\end{figure}

The left panel of Figure \ref{slopevsLASSObin} presents a scatter plot of the cross-validated LASSO estimates vs the SLOPE estimates, where we considered that model on the SLOPE path with the smallest number of nonzero regression coefficients. There is only a fairly small correlation between SLOPE and LASSO estimates which is due to the rather large variance of estimators for both methods when $p>n$. The right panel of Figure \ref{slopevsLASSObin} provides the scatter plot for a reduced dataset, where  300 variables with smallest marginal p-value were pre-selected. These were augmented with three variables selected by mBIC2 although they had relatively large marginal p-values. Here the estimates of SLOPE were obtained using cross-validation as implemented in the {\it SLOPE} package. After reduction of the number of features the correlation between SLOPE and LASSO estimates increases. The five most important predictors obtained with both methods coincide, among them the three variables selected by mBIC2 with large marginal p-values (V729, V6584 and V3161) and two other variables (V7734 and V2804) from the cluster strongly correlated with V3665. The resulting model including these five top variables selected by LASSO and SLOPE has only slightly larger residual deviance than the model identified by mBIC2. Comparing SLOPE to LASSO we note that SLOPE selects 40 variables from the cluster related to V3665 while LASSO includes only 7 variables from this cluster. Like for the first data example with a continous response SLOPE estimates are again smaller and more shrinked towards each other than LASSO estimates. It remains to be validated if this goes along with an improvement of the prediction/classification accuracy. 

\section{Summary}\label{Sec:Summary.Simulation}
The theoretical results reported in this chapter as well as the results from simulations and real data analysis illustrate that there does not exist a uniformly optimal model selection method for high dimensional regression problems. The choice of the methodology depends on the study purpose, on the actual values of $n$ and $p$ and on the amount of correlations between explanatory variables. 

When $n$ is larger or comparable to $p$ then the convex optimization methods like LASSO, SLOPE or their extensions like SLOBE perform very well. Here SLOBE or knockoffs based on cross-validated LASSO (Lcv) can be used for preventing false discoveries, with Lcv knockoffs providing exact FDR control  but having lower power than SLOBE when signals are very sparse or when the explanatory variables are strongly correlated. Our simulations illustrate that SLOBE also keeps FDR close to the nominal level under a wide range of scenarios and yields very good estimation and prediction properties when the signal is strong. Cross-validated versions of LASSO and SLOPE include many false positives but yield very good prediction properties when  the signal is weak. As shown in the real data analysis this goal is accomplished by selecting many representatives from a set of correlated predictors. Our simulation study from Section \ref{predSLOPE} suggest that the prediction properties of SLOPE are better than those of LASSO for a wide range of scenarios, particularly when there is a larger number of weak and correlated predictors. Here SLOPE has a tendency to include more of the correlated predictors than LASSO, which is related to its ability to cluster regression coefficients of variables which have a similar influence on the loss function \cite{kremer2019,schneider2020}. This property brings the potential for identification of low dimensional models, where some regression coefficients are equal to each other, which remains an interesting topic for a further exploration. Concerning model selection properties of LASSO it is important to be aware of the fact that selecting variables according to the order in which they appear on the LASSO path is definitely not an optimal solution (e.g., see \cite{weinstein2020}). Concerning model selection properties it is much better to order variables according to the magnitudes of their estimates from the cross-validated LASSO and then threshold them using some version of the knockoff methodology or using some model selection criteria, like mBIC2. 

The major drawback of the above mentioned regularization techniques is that for $p\gg n$ they loose their superior estimation and model selection properties because the respective estimators have too large variance due to a large number of estimated parameters. Model selection based on modifications of the Bayesian Information Criterion, like mBIC2, is based on the least squares fitting of the compared sub-models  and does not suffer much from increased $p$ when the signal is sufficiently sparse to guarantee a low variance of least squares estimators.   Our simulations reported here as well as in \cite{GWAS,DBF14,admixtures,biomarkers} show that selection based on mBIC2 allows to control FDR under a wide range of scenarios. Furthermore already relatively simple extensions of the stepwise selection strategy implemented in {\it bigstep} \cite{bigstep} can discover representatives of important clusters of variables even when $p\gg n$ and when predictors are correlated. Since mBIC2 has a tendency to include only very few correlated variables, the step-wise selection strategy has the potential to include predictors which have small marginal correlations with the response and which tend to be missed by simple screening strategies like Sure Independence Screening (SIS, \cite{SIS}). This happened in both our real data examples, where the final optimal models selected by SLOBE or LASSO used variables identified by mBIC2 but missed by SIS. 

The main limitation of performing model selection with mBIC2 for $p\gg n$ is that if $p$ is getting too large then the optimization problem is getting extremely challenging and heuristic search methods based on stepwise selection strategies might no longer be efficient enough. However, there is currently a lot of interesting research going on to overcome these limitations (e.g. see \cite{Rahul1}).
Finally our results illustrate that the analysis of high dimensional data might always remain a complex task. The best results are often obtained by the combination of different methods. Ideally such an analysis will lead to a coherent and consistent descriptions of the underlying mechanisms but one has to be aware of the limitations of what one can expect from modeling of phenomena in a high dimensional setting.

\section{Acknowledgments} 
We would like to thank the editor and the reviewers for suggestions which helped to improve the presentation of this chapter. We would also like to thank Dominik Nowakowski for performing simulations for Section \ref{predSLOPE} and to Wojciech Rejchel for suggesting and preprocessing the data set used in Section \ref{Sec:Continous}. 
M.Bogdan acknowledges support of the Polish National Center of Science with grant Nr 2016/23/B/ST1/00454. 

\section{Glossary}
\begin{Glossary}
	\item[FWER] Family Wise Error Rate
	\item[FDR] False Discovery Rate
	\item[GLM] Generalized Linear Models 
	\item[MSE] Mean Squared Error
	\item[LASSO] Least Absolute Shrinkage and Selection Operator
	\item[SLOPE] Sorted L-One Penalized Estimation
\end{Glossary}

%\part{Applications in Medicine}

%\part{Further Topics}

\bibliographystyle{plain}
\bibliography{bibtex_BogFrom_new}

\printindex

\end{document}